\documentclass[a4paper]{article}
\pdfoutput=1

\usepackage{jcappub}
\usepackage{graphicx}
\usepackage{rotating}
\usepackage{color}
\usepackage{epsfig}
\usepackage{bm}
\usepackage{multirow}
\usepackage{url}
\usepackage{hyperref}

\newcommand{\be}{\begin{equation}}
\newcommand{\ee}{\end{equation}}

\newcommand{\GeV}{{\rm GeV}}

\newcommand{\TeV}{{\rm TeV}}

\def\Rea{{\rm{Re}\,}}

\newcommand{\ds}{{\sffamily DarkSUSY\,}}

\newcommand{\pyt}{{\sffamily PYTHIA\, }}

\def\a{\alpha }
\def\b{\beta }
\def\g{\gamma }

\def\c{\chi }

\begin{document}

\begin{flushright}
\large \tt FERMILAB-PUB-14-010-A, TUM-HEP 927/14
\end{flushright}

\vskip 0.2in

\title{Indirect Detection Analysis: Wino Dark Matter Case Study}

\author[a]{Andrzej Hryczuk}
\author[b]{Ilias Cholis}
\author[c,d]{Roberto Iengo}
\author[e]{Maryam Tavakoli}
\author[c,d]{Piero Ullio}

\affiliation[a]{Physik Department T31, Technishe Uniersit\"{a}t M\"{u}nchen, James-Franck-Str.~1, 85748 Garching, Germany}
\affiliation[b]{Center for Particle Astrophysics, Fermi National Accelerator Laboratory, Batavia, IL 60510, USA}
\affiliation[c]{SISSA, Via Bonomea, 265, 34136 Trieste, Italy}
\affiliation[d]{INFN, Sezione di Trieste, Via Bonomea 265, 34136 Trieste, Italy}
\affiliation[e]{School of Astronomy, Institute for Research in Fundamental Science (IPM), P.O. Box 19395-5531, Tehran, Iran}
\emailAdd{andrzej.hryczuk@tum.de}
\emailAdd{cholis@fnal.gov}
\emailAdd{iengo@sissa.it}
\emailAdd{maryam.tavakoli@desy.de}
\emailAdd{ullio@sissa.it}

\date{\today}

\abstract{
We perform a multichannel analysis of the indirect signals for the Wino Dark Matter, including one-loop electroweak and Sommerfeld enhancement corrections. We derive limits from cosmic ray antiprotons and positrons, from continuum galactic and extragalactic diffuse $\gamma$-ray spectra, from the absence of $\gamma$-ray line features at the galactic center above 500 GeV in energy, from $\gamma$-rays toward nearby dwarf spheroidal galaxies and galaxy clusters, and from CMB power-spectra. Additionally, we show the future prospects for neutrino observations toward the inner Galaxy and from antideuteron searches. For each of these indirect detection probes we include and discuss the relevance of the most important astrophysical uncertainties that can impact the strength of the derived limits.  We find that the Wino as a dark matter candidate is excluded in the mass range bellow $\simeq$ 800 GeV from antiprotons and between 1.8 and 3.5 TeV from the absence of a $\gamma$-ray line feature toward the galactic center. Limits from other indirect detection probes confirm the main bulk of the excluded mass ranges.
}

\maketitle
\section{Introduction}
\label{sec:Intro}

Dark Matter (DM) composes 85$\%$ of the total matter in the Universe  \cite{Hinshaw:2012aka, Ade:2013zuv} but its nature remains 
one of the main questions in cosmology and high energy physics. In addition Supersymmetry has for long been a favorable theory for the extension of the Standard Model providing a natural candidate, the lightest supersymmetric particle (LSP) for a weakly interacting massive particle (WIMP).
In the Minimal Supersymmetric Standard Model (MSSM), the lightest out of the four neutralinos, each of which is a linear combination of the R=-1 neutral Wino, Bino and Higgsinos, is the DM LSP. In this paper we concentrate on the pure Wino case, for which at the thermal scenario, the correct relic density is achieved at $m_{\chi}$ being at the order of 3 TeV \cite{Hisano:2006nn,HryczukPHD}. Much lighter Winos have to be produced non-thermally, while heavier give too large relic abundance, and one has to invoke for example a dilution via late entropy production or the decay of the Wino into a lighter state, such as a gravitino or an axino.

TeV scale Wino evades all current direct detection and collider bounds. At such a high mass scale the direct detection experiments lose sensitivity for a purely kinematical reason: the nuclei mass sets the characteristic scale for sensitivity, which for all working direct detection experiments is optimal around $\mathcal{O}(100 {\,\rm GeV})$. A substantial upgrade of the technology is needed, which might be provided by the DARWIN project \cite{Baudis:2012bc}, and other proposed ton-scale dark matter direct detection experiments. Also for a pure Wino, the elastic scattering on a nucleus vanishes, because the coupling of the neutralino to $Z$ or Higgs bosons scales with gaugino-higgsino mixing. On the collider front, even the LHC at 14 TeV is not enough for a discovery of $2\text{--}3$ TeV weakly interacting particle like Wino, see e.g. \cite{Chattopadhyay:2006xb}. Thus the only sensitive probe for the Wino is through its indirect signals; either through cosmic rays (CRs), $\gamma$-rays, microwaves or neutrinos, which is the aim of this work.

On the indirect detection, signals can come from $\chi \chi \longrightarrow W^{+}W^{-}$ at tree level, with $\chi$ being the neutral member of the $SU(2)_{L}$ triplet. This gives CRs, $\gamma$-rays and neutrinos from the decay of the $W^{\pm}$ and the subsequent hadronization of their products.  As CRs propagate in the Galaxy or at far away galaxies, they produce additional $\gamma$-rays in their interactions with the local interstellar medium; or microwaves from synchrotron emission; providing additional possible signals of Wino annihilations. These energetic CRs can even impact the CMB at the recombination epoch. 

In this setup the higher order corrections are crucial for making robust predictions for the indirect detection. It is especially important, because this is the only feasible way of excluding (or detecting) Wino DM, at least in the near future. Monochromatic $\gamma$-rays, potentially giving a smoking-gun signature of DM annihilation, can only be produced at the loop level. Additionally, three body final state processes $\chi \chi \longrightarrow W^{+}W^{-}\gamma$, $ W^{+}W^{-}Z$ and $\chi \chi \longrightarrow \chi^{+} \chi^{-} \longrightarrow W^{+}W^{-}\gamma$, $ W^{+}W^{-}Z$ have to be included, as they modify the final spectra and total annihilation cross section.

Wino dark matter has been studied on its indirect detection prospects already in the past, see e.g. \cite{Ullio:2001uq,Baer:2005zc,Chattopadhyay:2006xb,Grajek:2008jb,Belanger:2012ta}. Most of these works were interested in the low mass region, at most a few hundreds of GeV. The reason is that at a tree level such Wino can have large cross sections possibly giving interesting signals. On the other hand, when one goes beyond tree level approximation, and in particular includes the Sommerfeld effect, also TeV scale Wino starts to have an interesting phenomenology. 
Sommerfeld effects become important for $m_{\chi} \gg m_{W}$ enhancing the current epoch (at $\sim 10^{-3}$c velocities) annihilation cross section and modifying the final annihilation products injection spectra. 
This was already noticed in \cite{Hisano:2005ec}, where positron and antiproton signals were discussed, especially inspired by the HEAT cosmic ray results. After \textit{PAMELA} reported the positron fraction rise people were suggesting heavy DM as a possible explanation and this model was also advocated as one of the possibilities \cite{Grajek:2008pg,Kane:2009if}. However, none of these works considered electroweak corrections and all concentrated on only one or two detection channels. 

In this work we discuss  all possible channels, which is essential for making robust claims on the exclusion or detection. More specifically, in section~\ref{sec:SWino} 
we present the effects of the Sommerfeld enhancement and the electroweak corrections on the CR, neutrino and $\gamma$-ray spectra produced from the DM annihilations, as well as  our general methodology for calculating the end product signals (after galactic propagation of CRs). In section~\ref{sec:InDM}, we show our results for a variety of indirect detection probes, that can provide more or less strong limits on Wino DM with masses anywhere between 0.5 and 3.2 TeV.  We study the impact that DM annihilations can have on CR antiprotons and positrons, galactic and extragalactic diffuse $\gamma$-rays, in $\gamma$-ray signals from nearby dwarf spheroidal galaxies and from galaxy clusters. We also include current limits from observations of the CMB angular temperature and polarization power spectra and also future perspectives from neutrinos from the galactic center and from antideuterons. For each of these indirect detection probes, we study not just a reference case, but include the most important astrophysical uncertainties that can impact the strength of the derived limits. Those uncertainties can be related to the background production and propagation assumptions of the CRs (for the CR probes), to the local DM density, to the 
DM density profile of our Galaxy, to the impact of the DM substructures, or to the target selection (see more details in the individual subsections of section~\ref{sec:InDM}). Such a study allows us a comparison on the strength between the various indirect detection methods. Finally, in section~\ref{sec:concl} we discuss on the combination of these indirect detection probes and conclude.

\section{The Sommerfeld effect and Wino DM}
\label{sec:SWino}

The Sommerfeld enhancement is a non-relativistic effect, resulting in correcting the annihilation cross section due to presence of some ``long range force'' between the particles in the incoming state. It can be described as an effect of distorting the initial wave function of the incoming two-particle state by a non-relativistic potential. This potential is taken to be Yukawa or Coulomb, as the force arises due to exchange of massive or massless boson.\footnote{In the cosmological setting in reality the potentials are always Yukawa type, since in the thermal background due to the plasma screening there are no strictly massless modes.} For a review of this effect in the dark matter context we refer the Reader to e.g. \cite{Finkbeiner:2010sm,Hryczuk:2010zi} and references therein. Here, we would only like to stress the main implications of this effect for the Wino DM model. At the tree level and in the leading order Born approximation, which in the following we will simply refer to as the "tree", the annihilation cross section for pure Wino is given by:
\begin{equation}
\sigma v|_{\c\c\rightarrow W^+W^-} = \frac{8\pi \a_2^2}{m_\c^2}\frac{(1-m^2_W/m^2_\c)^{3/2}}{(2-m^2_W/m^2_\c)^2},
\end{equation}
where $m_\c$ and $m_W$ are the Wino and W boson masses, respectively, and $\a_2=g_2^2/(4\pi)$ with $g_2$ being the weak coupling constant.

At this level one can distinguish two distinct phenomenologically relevant mass windows: \textit{i)} low mass ($m_\c\lesssim 500\;\GeV$) giving possibly measurable indirect detection signals, but too low thermal relic density and \textit{ii)} large mass (TeV scale) with relic abundance in accordance with thermal production, but with very weak experimental signatures. However, this picture is significantly altered when higher order corrections, and in particular the Sommerfeld effect, are taken into account. 

In the low mass regime, the main difference comes from electroweak corrections, with Sommerfeld effect being negligible. The total cross section is changed only at a \% level, but the spectrum becomes softer, changing the predictions for the indirect detection signals and in consequence the experimental bounds on this scenario. On the other hand, if the Wino mass is at the TeV scale the Sommerfeld effect starts to dominate. The total cross section can be enhanced by more than an order of magnitude and in particular, if the Wino mass happens to be around $m_\c\approx 2.4\;\TeV$, this enhancement acquires a strong resonance behavior.\footnote{The resonance appears when the annihilation takes place through a loosely bound state formed due to the interactions between the neutralino and chargino pairs and mediated by the exchange of the $W$ boson. This happens when the Bohr radius of the Wino pair matches the interaction range, i.e. $1/(\alpha_2m\c) \approx 1/m_W$.} The cross section is then strongly boosted and can potentially lead to observable signals. 

Throughout this work we will assume that the Wino constitutes the whole DM, meaning that for most of the considered possible values of $m_\c$, the main production mechanism in the early Universe has to be non-thermal. Nevertheless, let us note that the considered higher order effects also introduce corrections to the thermal relic density, which are however significantly milder  since the Sommerfeld effect is weaker at higher velocities \cite{Hisano:2006nn,Hryczuk:2010zi} (see also \cite{Zavala:2009mi}).

\subsection{Implications for the present-day DM annihilation}

The consequences of these effects for the indirect detection are encoded in the change of the annihilation spectrum and the total cross section. If by $A_{\chi\chi\to {\rm SM}}$ we call the perturbative annihilation amplitude into generic SM states, then the full (Sommerfeld enhanced) amplitude is given by\footnote{See also recent works with different, effective field theory approach for including the Sommerfeld effect in case of multiple channels \cite{Beneke:2012tg,Hellmann:2013jxa}.}
\begin{equation}
\label{sdef}
A^{{\rm SE}}_{\chi^0\chi^0\to {\rm SM}} = s_0 A_{\chi^0\chi^0\to {\rm SM}}+s_\pm A_{\chi^+\chi^-\to {\rm SM}},
\end{equation}
where the $s_0$ and $s_\pm$ are called the Sommerfeld factors and are in general complex functions of the relative velocity $v$ and $m_\c$. They are obtained by summing the contributions of the ladder diagrams or, equivalently, solving the appropriate Schr\"{o}dinger equations (see \cite{Hryczuk:2011vi} for details). In this approach, the amplitudes $A$ can be computed at any given order of perturbation theory, in our case $\mathcal{O}(g^6)$, while the Sommerfeld factors are treated as being non-perturbative. The cross section is then obtained by integrating the modulus square of Eq. (\ref{sdef}) over the phase space.\footnote{In the phase space integration we neglect the mass difference between $\c^\pm$ and $\c^0$.} For masses $m_\c \sim m_W$ we have $s_0\approx 1$ and $s_\pm\approx 0$ thus the total result is the perturbative one, while for $m_\c\gg m_W$, $|s_0|\gg |s_\pm|\sim \mathcal{O}(1)$. Therefore compared to the perturbative result the one with Sommerfeld effect introduces three modifications: \textit{i)} enhances the value of the cross section, \textit{ii)} opens up the annihilation channels $ZZ$, $Z\gamma$ and $\gamma\gamma$ (without the SE they are of higher order), and finally \textit{iii)} modifies the spectra.

\begin{figure}[t]
\centering
	\includegraphics[scale=0.8]{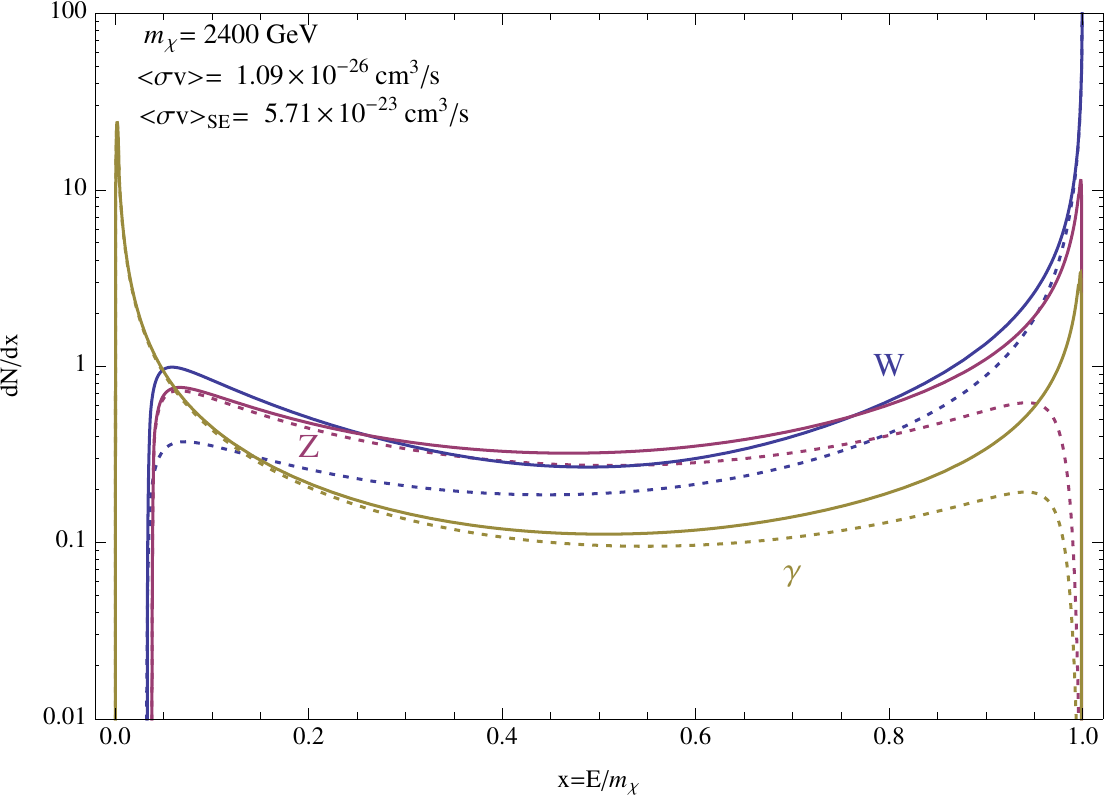}
	\caption{Initial annihilation spectra of $W^\pm$, $Z$ and $\gamma$ for the case near the resonance, with $m_\c=2.4$~TeV. The solid lines correspond to full Sommerfeld enhanced result, while dotted to the $\mathcal{O}(g^6)$ one. Notice the difference in normalization for $dN/dx$ in these two cases, see Eq.~\eqref{eq:setotnorm}.}
    \label{fig:spectini}
\end{figure}

On Fig.~\ref{fig:spectini} we show the primary annihilation spectra for an example case of $m_\c=2.4$ TeV.\footnote{The method of doing the computation of the spectra is discussed in the Appendix \ref{app:spectrum}.} It is chosen such to be near the resonance, where the impact of the Sommerfeld effect is most clearly visible. First of all note that the perturbative result, given just by the standard two- plus three-body annihilation process (dotted lines), is normalized differently than the full Sommerfeld one (solid lines). The ploted spectra are \textit{per annihilation}, i.e. normalized such that integrated over $x$ give the total number of produced primary particles. Therefore,
\begin{equation}
\label{eq:setotnorm}
\frac{dN_{tot}}{dx}=\frac{1}{\sigma_{tot}}\frac{d\sigma_{tot}}{dx}, \qquad \frac{dN_{\text{SE}}}{dx}=\frac{1}{\sigma_{\text{SE}}}\frac{d\sigma_{\text{SE}}}{dx}.
\end{equation}
This is why these lines are close to each other, even though for this mass the SE enhances the cross section by nearly four orders of magnitude.

For massive gauge bosons the spectrum posses a visible threshold at $x=m_{W,Z}/m_\c$. Photons on the other hand are regulated by introducing an effective photon mass $m_\gamma$, which physically is an effect of the energy resolution below which one cannot distinguish the $W^+W^-\gamma$ state from $W^+W^-$ one. In collider experiments the IR regulator is set by the known energy resolution of the detector. In our case however, it is not that simple, because of the effects of propagation before the signal actually reaches the detector. Having this in mind, we choose the energy resolution to be of 1\%. To take it into account in the spectrum, we have followed the philosophy of \cite{Ciafaloni:2010qr}: one subtracts the part of the $\gamma$ spectrum below $E_{\rm res}$ and adds it to the $W$ one. This is done in such a way, that including the virtual corrections one obtains correct total cross section after integrating over whole range of $x$.
However, instead of flat distribution as in \cite{Ciafaloni:2010qr}  we have chosen, a more physical Gaussian one. Moreover, we applied this procedure consistently to whole spectrum, by introducing a smearing, i.e. for every $x$ we convolute the initial spectra with a Gaussian distribution
\be
G(x_{\rm obs},x)=\frac{1}{\sqrt{2\pi}x_{res}} \exp\left(\frac{(x_{obs}-x)^2}{2x_{res}^2}\right),
\ee
where $x_{\rm res}=E_{\rm res}/m_\c$ and $x_{\rm obs}$ is the observed value. This effectively makes the change:
\be 
D(x)\rightarrow D(x_{\rm obs})=\int_0^1 dx D(x) G(x_{\rm obs},x),
\ee
where $D(x)$ is the initial annihilation spectrum and $D(x_{\rm obs})$ the corrected one. Indeed, such a smearing simulates the physical process that makes the spectrum and cross section finite in the IR. In our computations we used a small regulator $m_\gamma \ll E_{\rm res}$. This means that after the smearing it is $E_{\rm res}$ which plays the role of a regulator, as if $m_\gamma$ was actually chosen to be zero. 

The impact of the Sommerfeld enhancement is most visible for large $x$, where in fact it starts to dominate and produces a strong signal. This comes mainly from the large enhancement of the two-body processes giving line components, but also from the amplification of the bump just below the upper threshold for $\gamma$ and $Z$. This bump appears, because not only soft, but also the collinear (even hard) gauge boson emission is logarithmically enhanced.

\subsection{Fluxes at production}
\label{sec:fluxes}

The $W$ and $Z$ bosons produced in the annihilation process will subsequently decay into quarks and leptons. Quarks then undergo hadronization producing mesons and even baryons, which can be stable, like protons and antiprotons, or fragment into leptons and photons. Additionally, the particles produced in these process can have very high invariant masses, i.e. the primary particle can be off-shell with large vitruality. The resulting process is then not a decay, but a \textit{splitting} and the whole process produces a shower of final particles. In particular, the primary gauge bosons produced in the annihilation are very energetic: their invariant mass is of the order of the mass of the neutralino, which in our case of interest is at the TeV scale.

All this processes can be described by the Dokshitzer-Gribov-Lipatov-Altarelli-Parisi (DGLAP) evolution \cite{Gribov:1972ri,Altarelli:1977zs,Dokshitzer:1977sg}, which relies on the fact that the branching probabilities in the soft/collinear approximation are universal. They depend only on the virtuality $\mu^2$ and the so-called splitting functions, which one can derive given field content and interactions. 

This approach is very well know and exploited a lot in collider physics, especially in the simulations of jets. Because of this, since many years robust numerical codes taking care of all the splitting/hadronization/fragemantation processes exist. Two most widely used are \pyt \cite{Sjostrand:2006za} and {\sffamily HERWIG} \cite{Corcella:2002jc}, from which in our work we used the former one, as it is already implemented inside \ds \cite{Gondolo:2004sc}. Unfortunately, they are optimized for the high energy collisions and not non-relativistic annihilations. Moreover, these codes concentrate mostly on the QCD jets and not EW processes. They have also some peculiarities, e.g. \pyt does include the photon bremsstrahlung from fermion states, but not from $W^\pm$. One thus has to be careful when using them for the DM annihilation. Nevertheless, this can be done, as we will describe below. We follow the approach of \cite{Ciafaloni:2010ti} to include the additional electroweak splitting functions, that are missing in \pyt\!\!\!. However, in contrary to what was done there and then used in the PPPC 4 DM ID code \cite{Cirelli:2011fk}, for the photon and $W/Z$ bremsstrahlung we use our full $\mathcal{O}(g^6)$ computation. In this way we take the advantage of our model specific treatment, for which we have computed the whole loop corrections, with the Sommerfeld effect included. This allows to study possible spectral features (depending on the virtual internal bremsstrahlung (VIB) and not the final state radiation (FSR) or related to the $x \approx 1$ region, in which the collinear approximation fails), as well as includes also non-logarithmically enhanced corrections. 

The first step is to compute the spectra of $f=\gamma,\nu,e^+,\bar p,\bar d$ at production per annihilation, i.e. the quantity:
\begin{equation}
\frac{dN^f_{tot}}{dx}=\frac{1}{\sigma_{tot}}\frac{d\sigma^{\chi\chi\to X\to f}_{tot}}{dx},
\end{equation}
where $x=E_f/m$, $E_f$ is the kinetic energy of particle $f$, $\sigma_{tot}$ is the total annihilation cross section (summed over all possible annihilation channels), and $\sigma^{\chi\chi\to X\to f}_{tot}$ denotes the sum of cross sections for all processes giving rise to particle $f$ (with all multiplicities etc. included).\footnote{This implicitly assumes, that:
\be
\sigma_{tot}\approx \int_0^1 \frac{d\sigma^{\chi\chi\to X\to f}_{tot}}{dx}dx,
\ee
i.e. that the subsequent production of particle $f$ from decay/fragmentation of primary annihilation products does not change the total cross section. This is clearly justified, since all those additional contributions are of a higher order.}
 
We start from the final spectra of $f=\gamma,\nu,e^+,\bar p$ (for antideuterons see a separate discussion in Sec.~\ref{sec:IDantid}). In the total spectrum, including the order $\mathcal{O}(g^6)$ terms, we have possible initial states $I=W,Z,\gamma$. To get final spectra one has to convolute the initial ones with the fragmentation tables \cite{Ciafaloni:2010ti}:\footnote{Note that the final $f$ spectra are vs. the kinetic energy $x=E_{k}/m$, while in the formula $z$ is the total energy fraction carried by a given primary channel particle (e.g. $W$); that is why $x\leq z \leq 1$. The same applies to the splitting functions, where $z=E/m$.}
\begin{equation}
   \frac{dN^f_{tot}}{dx} (M,x) = \sum_{I=W,Z,\gamma} \int_x^1 dz \,S_I\, D_{I}(z)\,
   \frac{dN^{\rm MC}_{I\to f}}{dx}\left(zM,\frac{x}{z}\right)\,,
   \label{eq:spectra}
\end{equation}
where
\begin{equation}
D_I(z)=BR_I \frac{dN_I}{dz}
\end{equation}
is the spectrum of $I$ (splitting function) and the symmetry factors  are $S_W=1$, $S_Z=S_\gamma=1/2$.

\begin{figure}
\centering
	\includegraphics[scale=0.6]{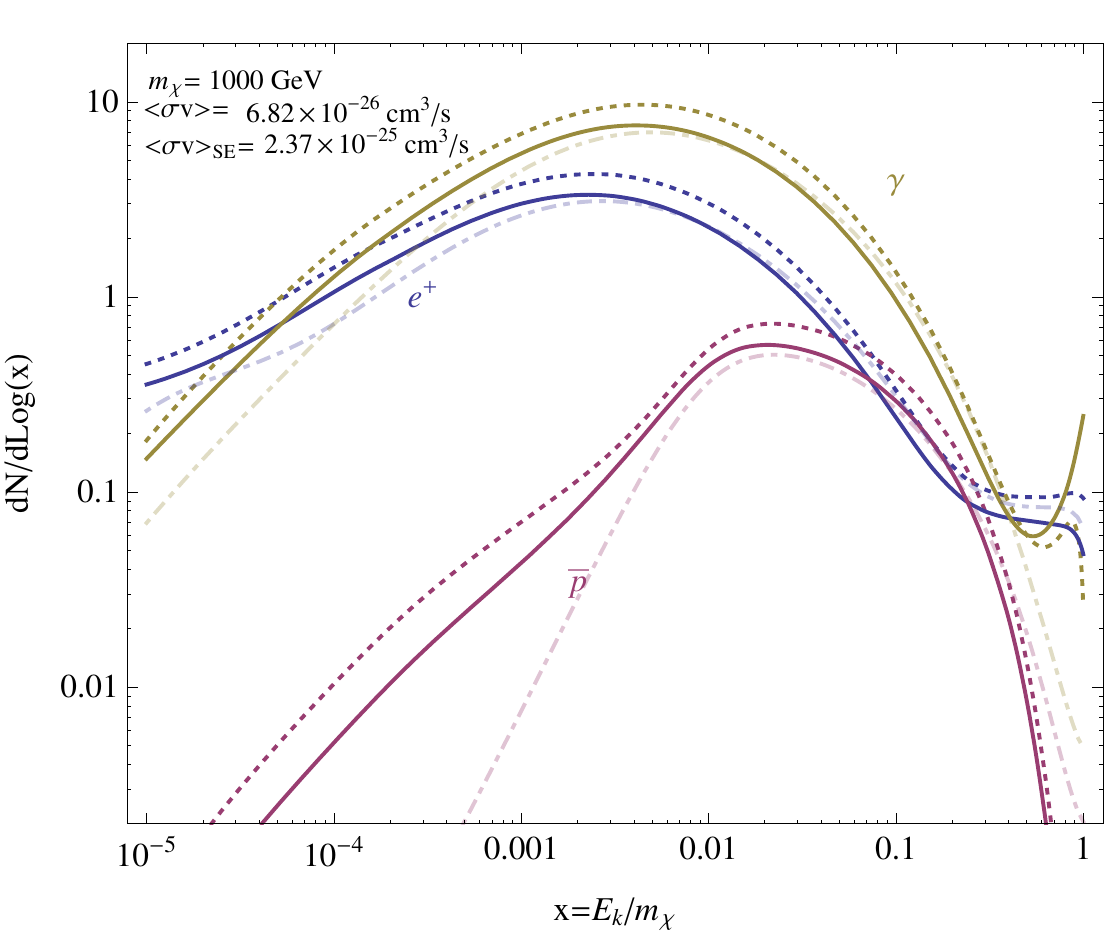}
		\includegraphics[scale=0.6]{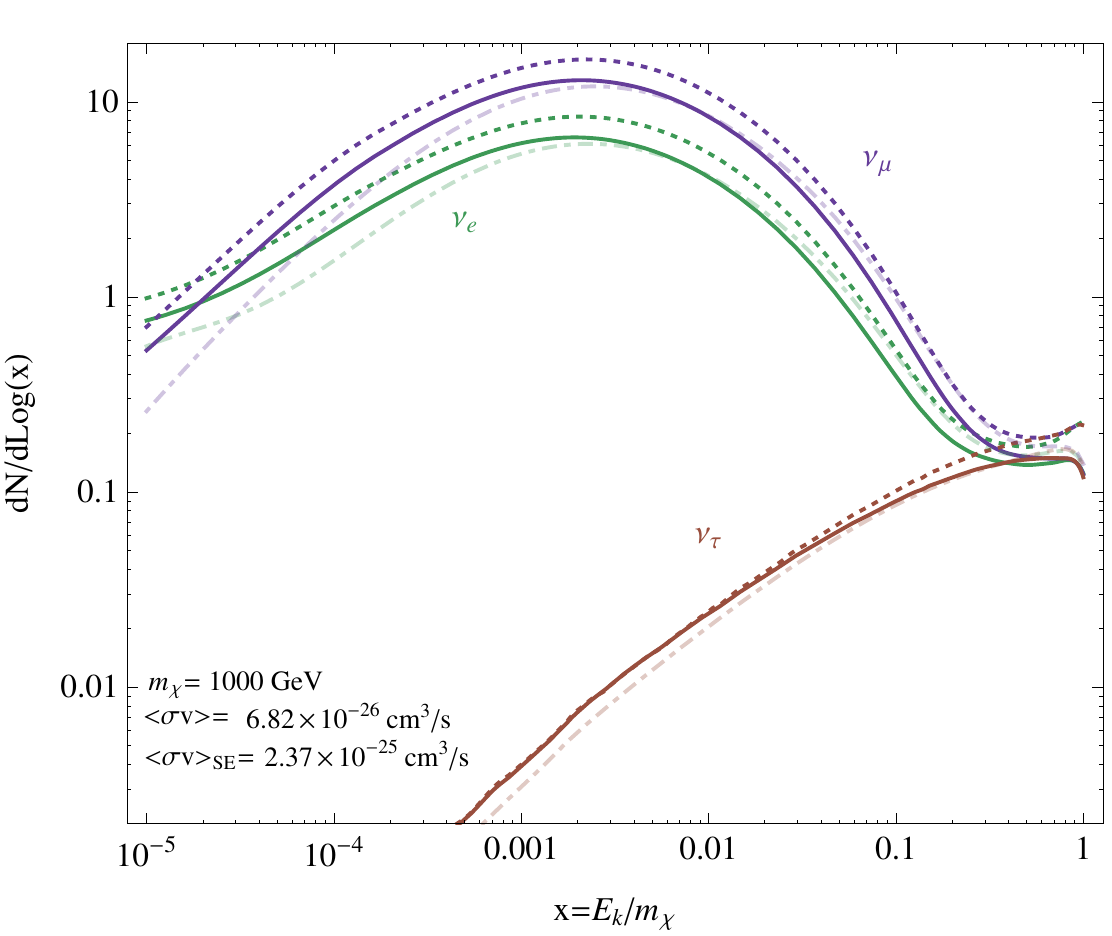}
			\includegraphics[scale=0.6]{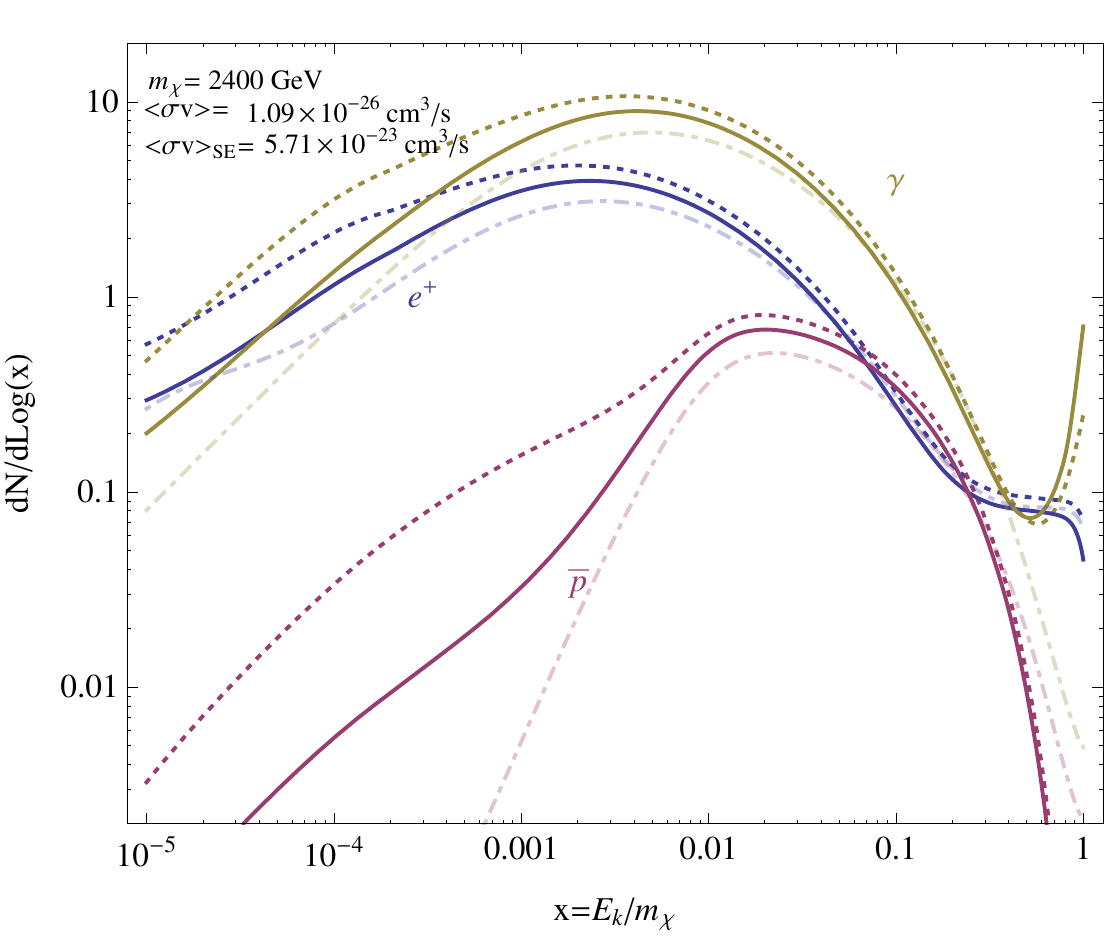}
		\includegraphics[scale=0.6]{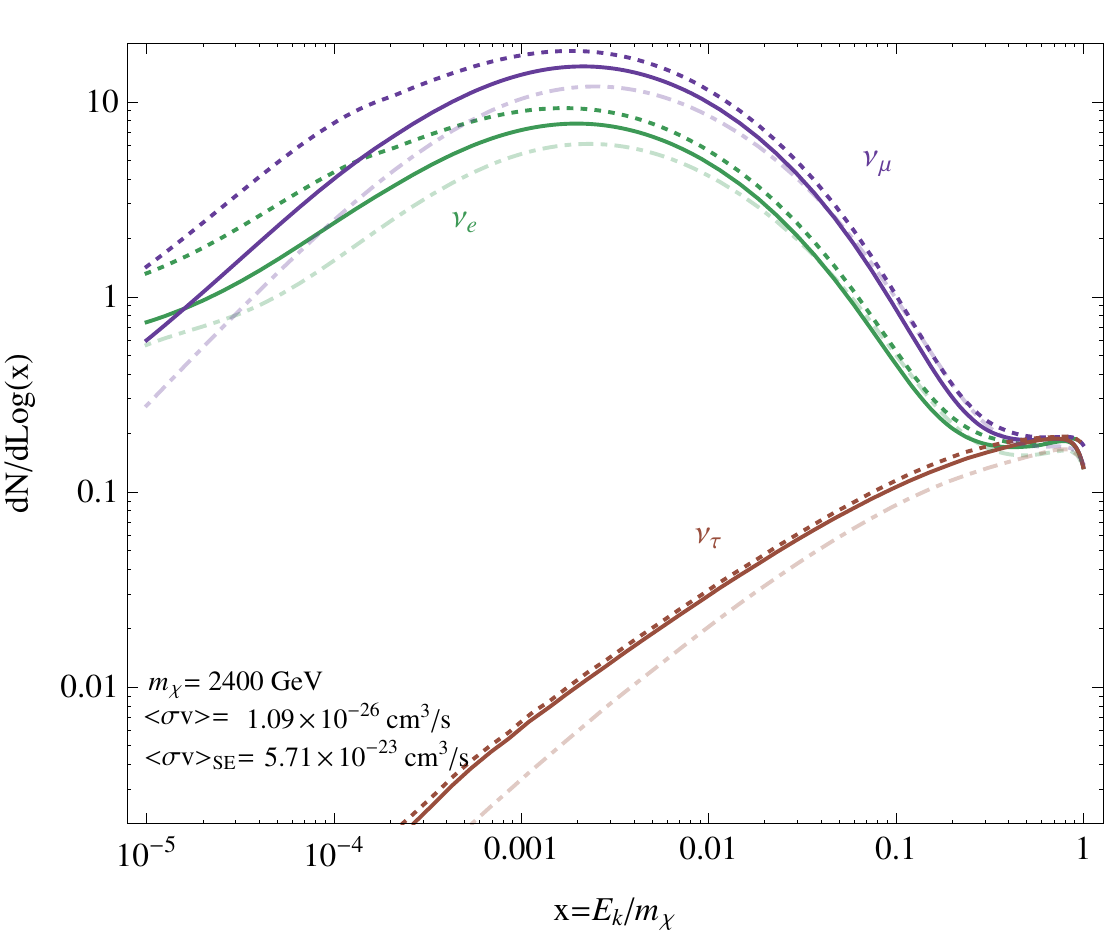}
	\includegraphics[scale=0.6]{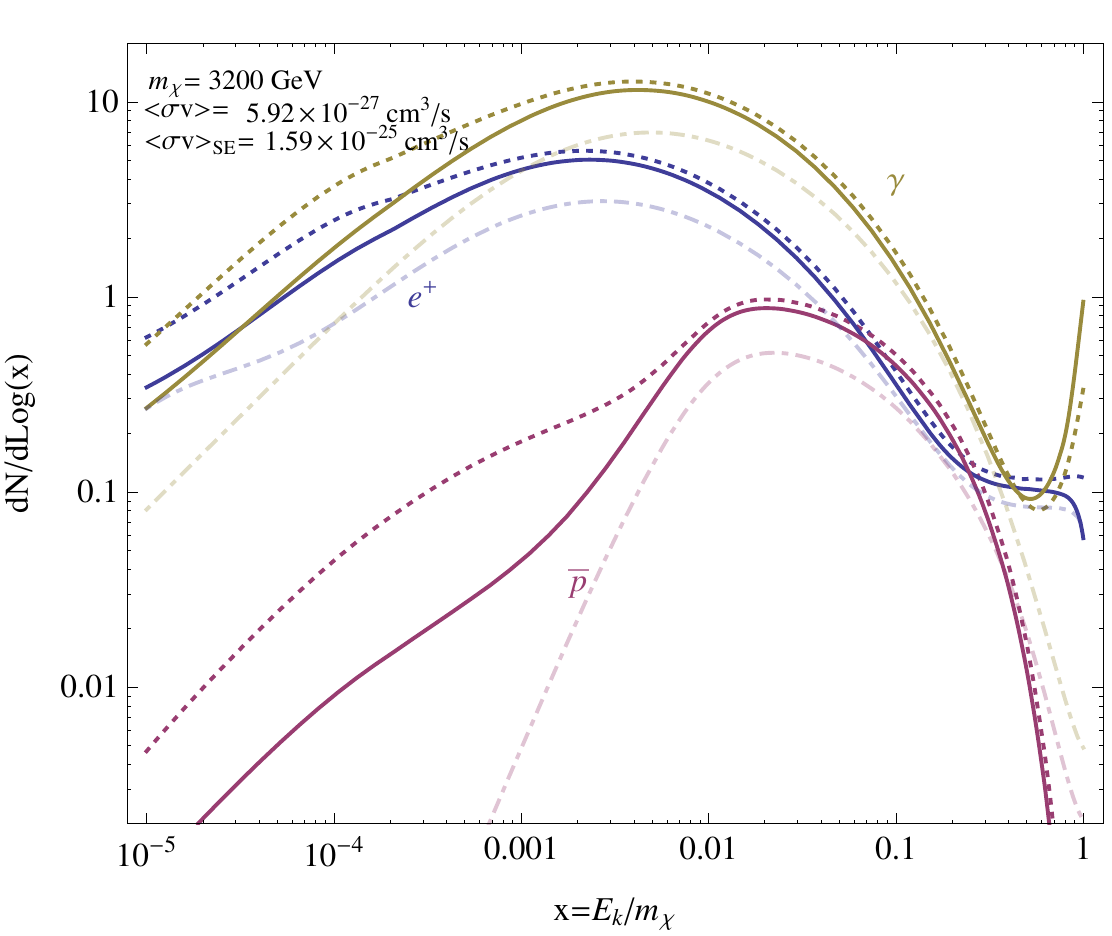}
		\includegraphics[scale=0.6]{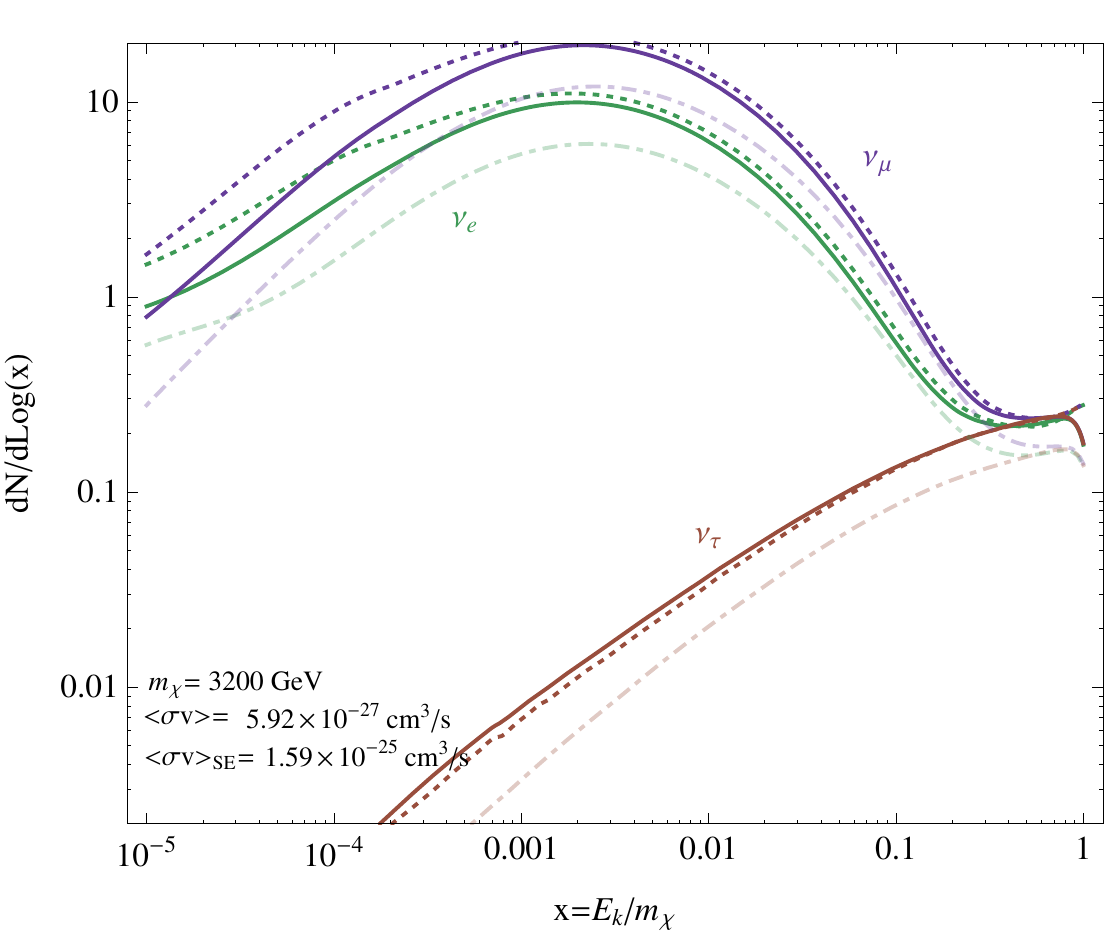}
	\caption{Number of final stable particles per annihilation for there representative Wino masses: typical $m=1$ TeV (top), near resonance $m=2.4$ TeV (middle) and giving correct thermal relic abundance $m=3.2$ TeV (bottom). The chained lines show the tree level result, dotted the EW corrected, while the solid the full Sommerfeld enhanced one.}
    \label{fig:spectfin}
\end{figure}

The results are given on Fig.~\ref{fig:spectfin} for three representative masses. In the left column we show the $e^+$, $\bar p$ and $\gamma$ spectra, where on the right the neutrino ones. The chained line represents the result one would obtain for the tree level annihilation process. At the TeV scale it nearly does not change with the Wino mass, because we consider only one $W^+W^-$ annihilation channel and thus $m_\c$ affects essentially only the total cross section and not the spectrum. The dominant final state are soft photons, coming mainly from production and then decay of $\pi^0$s. Electrons are produced in direct $W$ decay or splitting and also by charged pions. Finally, antiproton production is suppressed in most energies, but nevertheless prove to be very useful in putting constraints.

The results do change considerably with the inclusion of electroweak and Sommerfeld corrections plotted as solid lines. For the soft part, when the Wino mass is relatively low the Sommerfeld effect is rather mild and nearly entire modification of the spectrum comes from radiative corrections. As advocated before, a clear enhancement of the very low energetic final states is visible. On the other hand in the higher end of the spectrum additional hard $\gamma$ component arises, to which both radiative and Sommerfeld corrections contribute. The former mainly due to logarithmic enhancement of the collinear photon FSR, while the latter also amplifies a monochromatic $\gamma$-ray line.

When looking at higher masses, these effects are becoming slightly stronger. Notice however, that the near resonance case of $m_\c=2.4$ TeV does not introduce much stronger deformations of the spectra than the generic case of large $m_\c$. The reason is that the main effect of the Sommerfeld resonance is anyway the enhancement of the total cross section, while the spectrum change is rather mild.

In total, the amplification of the signal is visible in both the very soft part (electroweak corrections) and the hard $\gamma$-ray component (Sommerfeld), while the total cross sections is also strongly enhanced (Sommerfeld). 

In the case of neutrinos the overall behavior is similar. The tree level result gives weaker signals and is nearly mass independent. The $\nu_\tau$ has a completely different spectrum than $\nu_e$ and $\nu_\mu$, because of difference in their production mechanisms. Neutrinos arise mostly due to pions' decays, through:
\be
\pi^+\to \mu^+ \nu_\mu \to e^+ \nu_e \bar{\nu}_\mu  \nu_\mu \quad {\rm and} \quad
\pi^-\to \mu^- \bar{\nu}_\mu \to e^- \bar{\nu}_e \nu_\mu \bar{\nu}_\mu \;.
\ee
This is due to the helicity suppression of a pion decay, favoring muons as the heaviest kinematically allowed final states: the $\tau$ is heavier than $\pi^\pm$. Our full result again shows some enhancement, here mainly in the soft part. What is however most important phenomenologically in this case, is the effect on the total cross section amplifying the dark matter component vs. background in very the high energetic neutrinos (see Sec.~\ref{sec:neutrinos}). Note also, that these are results at production, so no oscillation effects were included.  

\subsection{General propagation of CRs}
\label{sec:DRAGON}

The propagation theory which is most widely accepted is the diffusion model with possible inclusion of convection \cite{Ginzburg:1990sk}. It was tested to provide the most adequate description of CR transport in our Galaxy. Within this framework the general CR propagation equation can be written as:
\begin{align}
\label{eq:propag}
\frac{\partial N^i}{\partial t}-\vec{\nabla}\cdot\left(D_{xx} \vec{\nabla}-\vec{v_c}\right) N^i+\frac{\partial}{\partial p}\left(\dot p-\frac{p}{3}\vec{\nabla}\cdot \vec{v_c}\right) N^i -\frac{\partial}{\partial p}p^2 D_{pp} \frac{\partial}{\partial p}\frac{N^i}{p^2} &=&   \\
Q^i(p,r,z)+\sum_{j>i} c\b n_{\rm gas}(r,z)\sigma_{ij}N^j - c\b n_{\rm gas}\sigma_{{\rm in}}(E_k) N^i &-\sum_{j<i} \frac{N^{i}}{\tau^{i \rightarrow j}} + \sum_{j>i} \frac{N^{j}}{\tau^{j \rightarrow i}}\;, \nonumber
\end{align}
where $N^i(p,r,z)$ is the number density of the $i$-th particle species with momentum $p$ and velocity $v=c\b$. 

We use DRAGON \cite{DRAGONweb,Evoli:2008dv} to solve the propagation equation at steady state, assuming cylindrical symmetry in space, with the galacto-centric radius $r$, the height from the Galactic disk $z$ and rigidity $R$ as the 2+1 dimensions of our numeric grid. The diffusion coefficient is expressed by:
\be
D(R,r,z)=D_0 \beta^\eta\left(\frac{R}{R_0}\right)^\delta e^{\left(|z|/z_d\right)}e^{\left((r-r_{\odot})/r_d\right)},
\label{eq:diffcoeff}
\ee
where the free parameters are the diffusion coefficient normalization $D_0$, spectral indices $\eta$ and $\delta$, parameters setting the thickness $z_d$ and radial scale $r_d$ of diffusion zone. The $R_0$ is the point in rigidity to which we fix the normalization. The diffusion coefficient grows with $r$, since it is proportional to the diffusion length which gets larger with $r$ as the large scale galactic magnetic field gets weaker far away from the galactic center.

The form of the source term is:
\be
Q(R,r,z)=f(r,z)\left(\frac{R}{R_i}\right)^{-\gamma_i},
\ee
where $f(r,z)$ is a function reflecting the spatial distribution of supernova remnants (SNRs) and $\gamma_i$ is the injection spectral index for species $i$. For electrons and positrons one adds also an exponential cut-off with energy, $e^{-E/E_c}$, with $E_c$ being set to a few TeV. The physical reason is that leptons loose energy very efficiently, thus very energetic ones need also to be very local. On the other hand we do not see nor expect many local sources of TeV scale leptons.

The gas distribution we used in our work is the one recently derived in \cite{Tavakoli:2012jx,Pohl:2007dz}. It is a new, and arguably most accurate available model for three dimensional distribution of atomic hydrogen gas in our Galaxy, reproducing the global features of the gas distribution such as spiral arms. It was derived using the 21cm Leiden-Argentine-Bonn survey data \cite{Kalberla:2005ts}, which is the most sensitive 21cm line survey up to date with the most extensive spatial and kinematic coverage.

To describe the effect of solar modulation we adopt a standard way using the force field approximation \cite{Gleeson:1968zza}. In this approximation for a given CR with mass $m$, atomic number $Z$ and mass number $A$ the modulated spectrum $\Phi_{mod}(E_k)$ is related to the unmodulated one $\Phi(E_k)$ by a formula:
\be
\Phi_{mod}(E_k)=\frac{\left(E_k+m\right)^2-m^2}{\left(E_k+m+\frac{Z|e|}{A}\phi\right)^2-m^2}\Phi\left(E_k+\frac{Z|e|}{A}\phi\right),
\ee
where $\phi$ is the modulation potential. For electrons and positrons the same formula holds but with $Z/A=1$. It captures effectively the effect of the scattering on the solar magnetic fields. The value of $\phi$ is typically determined by fitting the CR spectra at very low energies for a given propagation model. Note, that although theoretically the same value of this potential should be used for different CR species, in practise this is not the case. One always confront the model with the observational data, and these were taken by different experiments in different years. In particular, since now we are in the vicinity of the solar maximum, in the recent years the time dependence of the modulation was expected to be rather strong. This suggests, that what one should in fact do, is to not include the modulation as an effect on the propagation, but rather use it to "demodulate" the data. Effectively, however, this also boils down to using different values of $\phi$ in order to make the low energy CR data consistent, see e.g. \cite{Putze:2010fr}.

Therefore, in the results discussed below we adopt a different values for the modulation potential, which we fix by fitting to the B/C, proton, electron and total $e^++e^-$ data.

\subsection{General methodology}
\label{sec:methodology}

We identified 11 benchmark models with varying diffusion zone thickness, from $z_d=1$ kpc to $z_d=20$ kpc. The lower limiting value comes from the fact, that the galactic disk itself is extending to few hundreds of parsecs. The latter is chosen such, to be sure to enclose all the region with non-vanishing magnetic field, which is known to extend at least to few kpc. We also chose to fix spectral index $\delta=0.45$ motivated by fit to the CR data and the radial scale of $r_d=20$ kpc. The precise value of the latter do not introduce relevant effect and again comes from the radial scale of Milky Way. The convection was neglected, essentially because it is never a dominant effect in kpc scales and precisely for this reason it is not yet well understood on a quantitative level.

All the other parameters were fitted to the data. They are given in Table \ref{tab:propagation} together with the reduced $\chi^2$ values of the fits to B/C, protons, antiproton flux and $\bar p/p$ ratio. The data sets we used were \textit{i)} for B/C \textit{AMS-02} \cite{AMSsite} and \textit{PAMELA} \cite{PAMELABtoC}, \textit{ii)} for protons and helium \textit{PAMELA} \cite{Adriani:2011cu} and \textit{iii)} for antiprotons and $\bar p/p$ ratio \textit{PAMELA} \cite{2010arXiv1007.0821P}. In the case of proton and helium fluxes we additionally compared with the \textit{AMS-02} data \cite{AMSsite} and found good agreement when allowed for a different value of the modulation potential. The modulation potential was always treated as a free parameter of the fit, as advocated before. 

{\tabcolsep=0.16cm
\begin{table}
\centering
\footnotesize{
\begin{tabular}{|c c c | c c c|c  c|c c c c c|}
\hline \multicolumn{ 3}{|c|}{Benchmark} & \multicolumn{ 3}{c|}{Fitted} & \multicolumn{ 2}{c|}{Fitted} & \multicolumn{ 5}{c|}{Goodness} 
\\ \cline{ 1- 13}
 $z_d$ & $\delta$ & $r_d$ & $D_0\times 10^{28}$  & $v_A$ & $\eta$ & $\gamma^p_1/\gamma^p_2/\gamma^p_3$ & $R^p_{0,1}$ & $\chi_{B/C}^2$ & $\chi_p^2$ & $\chi_{\bar p}^2$ & $\chi_{\bar{p}/p}^2$ & $\chi_{\rm tot}^2$ \\ 
 
 [kpc]  & &  [kpc] &  [${\rm cm}^2{\rm s}^{-1}$] &  [${\rm km\; s}^{-1}$] & & & GV &  & & & & \\ \hline


 1 & 0.45 & 20 & 0.47  & 15.0 & -0.57 & $2.12/2.36/2.3$ & 14.5 & 0.38  & 0.31  & 0.66  & 0.79  & 0.55  \\ \hline

 1.4 & 0.45 & 20 & 0.70 & 15.0 &-0.57  & $2.12/2.36/2.3$ & 14.5 & 0.39  & 0.26 & 0.59  & 0.94  & 0.63  \\ \hline
 
 1.7 & 0.45 & 20 & 0.89 & 16.8 & -0.57 & $2.12/2.36/2.3$ & 14.5 & 0.42  & 0.24 & 0.58 & 0.71 & 0.52  \\ \hline

 2 & 0.45 & 20 & 1.12 & 18.8 & -0.57 & $2.12/2.36/2.3$ & 14.5 &  0.57 & 0.4 & 0.57 & 0.54 & 0.53  \\ \hline

 3 & 0.45 & 20 & 1.65  & 18.0 & -0.57 & $2.18/2.37/2.3$ & 14.0 & 0.44  & 0.49 & 0.54 & 0.52 & 0.55  \\ \hline

 4 & 0.45 & 20 & 2.2 & 18.0 & -0.57 & $2.20/2.37/2.3$ & 14.0 & 0.54   &  0.41 & 0.47 & 0.47  & 0.52   \\ \hline

 6 & 0.45 & 20 & 3.08 & 19.0 & -0.57 & $2.18/2.37/2.3$ & 14.0 & 0.65  & 0.47 & 0.45 & 0.47 & 0.52  \\ \hline

 8 & 0.45 & 20 & 3.6 & 18.5 & -0.57 & $2.18/2.37/2.3$ & 14.0 &  0.51 & 0.50 & 0.48 & 0.47 & 0.53  \\ \hline

 10 & 0.45 & 20 & 4.1 & 19.5 & -0.57 & $2.10/2.35/2.2$ & 15.5 &  0.62 & 0.90 & 0.47 & 0.50 & 0.69   \\ \hline

 15 & 0.45 & 20 & 4.6 & 18.5 &-0.57  & $2.10/2.35/2.2$ & 15.5 &  0.56 & 0.95 & 0.50 & 0.51 & 0.72  \\ \hline

 20 & 0.45 & 20 & 5.0 & 17.5 & -0.57 & $2.10/2.34/2.2$ & 14.2 & 0.56  & 0.53 & 0.46 & 0.47 & 0.53  \\ \hline

\end{tabular}
}
\caption{Benchmark propagation models. Everywhere the convection is neglected $v_c=0$. The second break in the proton injection spectra is always 300 GV. For primary electrons we use a broken power-law with spectral indices $1.6/2.59$ and a break at 7 GV. For He and heavier nuclei we assumed one power-law with index 2.3 and 2.25, respectively. The propagation parameters were obtained by fitting to B/C, proton and He data, while the primary electrons were obtained from the measured electron flux. The values  of antiproton and $\bar p/p$ $\chi^2$s are then predictions. The total $\chi^2_{\rm tot}$ has been obtained by combining all the channels. See the text for more details.}
\label{tab:propagation}
\end{table}
}

All considered propagation models give a very good fit to the CR data. As a second step, for these models we calculate the predicted diffuse $\gamma$-ray sky maps and check for consistency with the \textit{Fermi} data. We note also, that our probing of the diffusive zone of the diffusive zone thickness is dense enough to be able to make an interpolation of the result for any $1 \;{\rm kpc}\leq z_d \leq 20 \;{\rm kpc}$.

\begin{figure}
\centering
	\includegraphics[height=6.5cm,width=7.3cm]{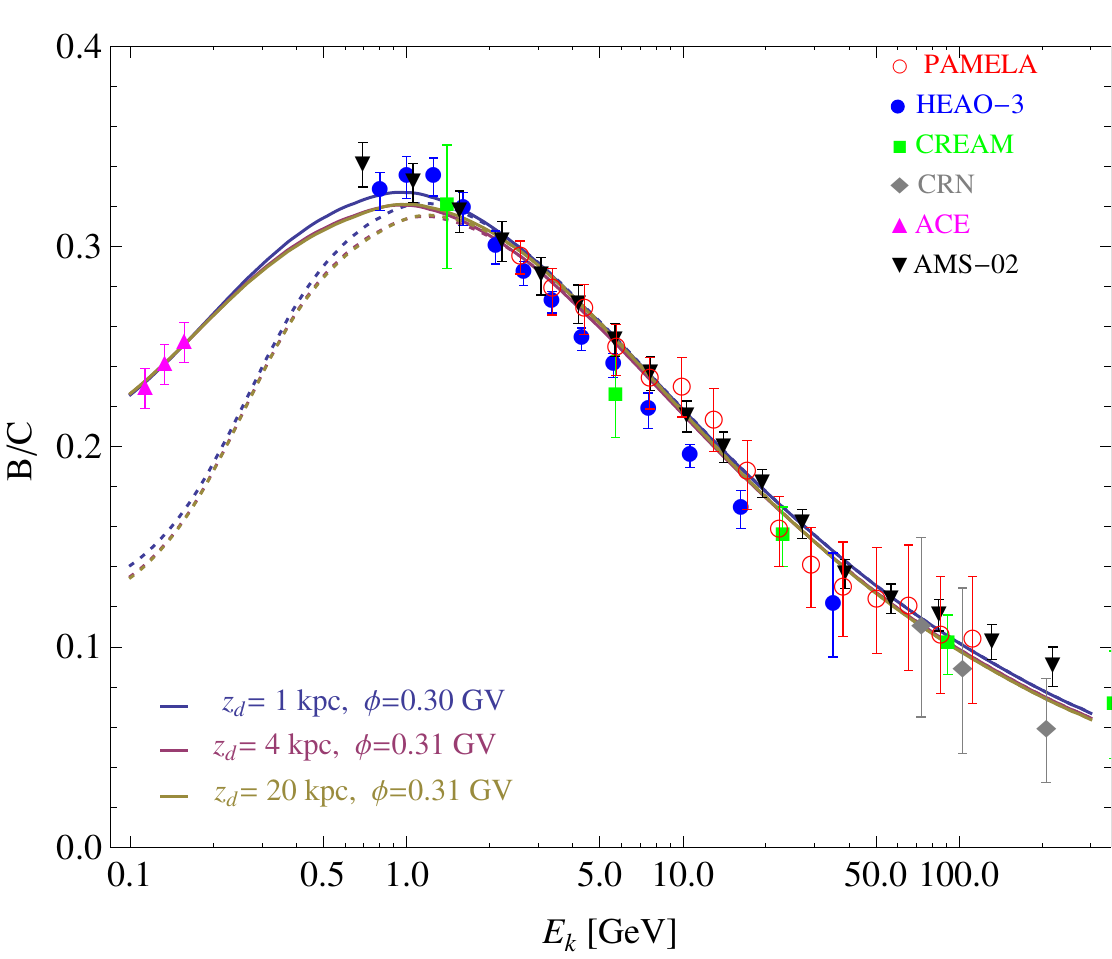}
		\includegraphics[height=6.5cm,width=7.5cm]{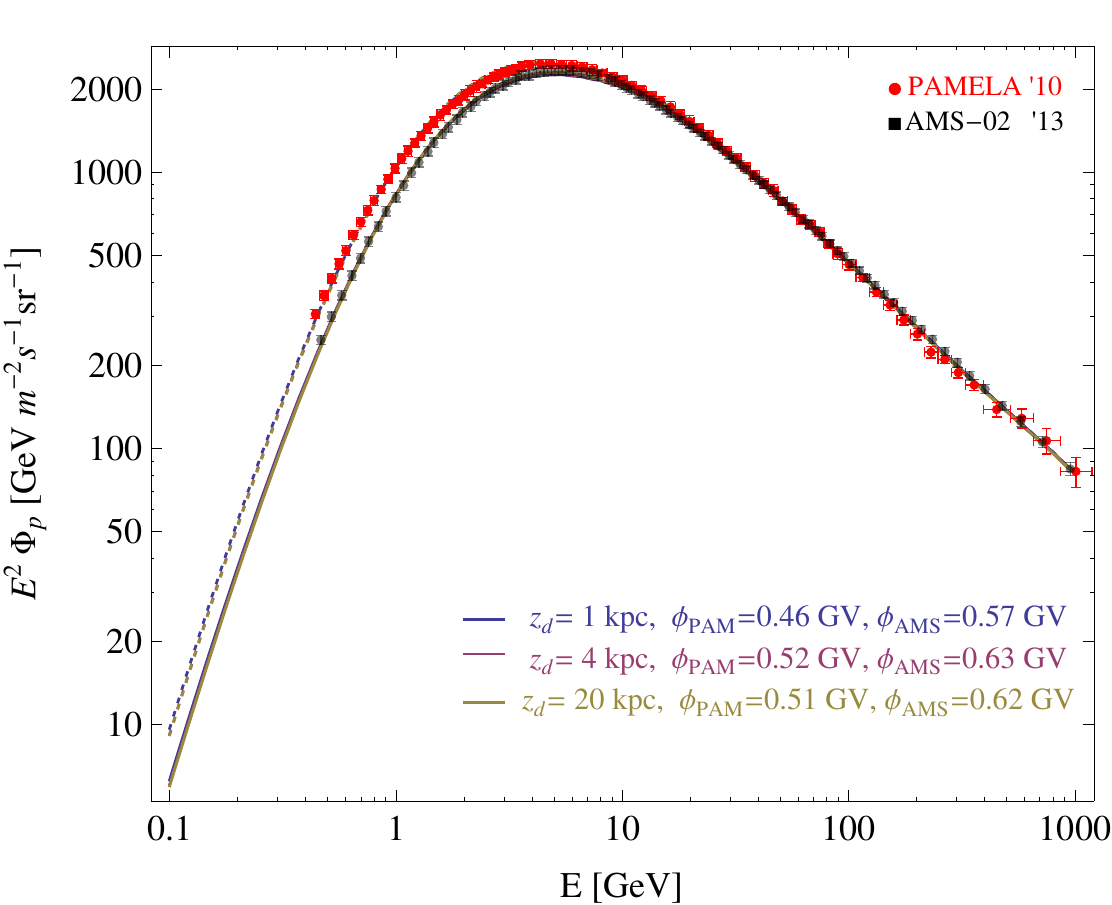}
		\includegraphics[height=6.5cm,width=7.5cm]{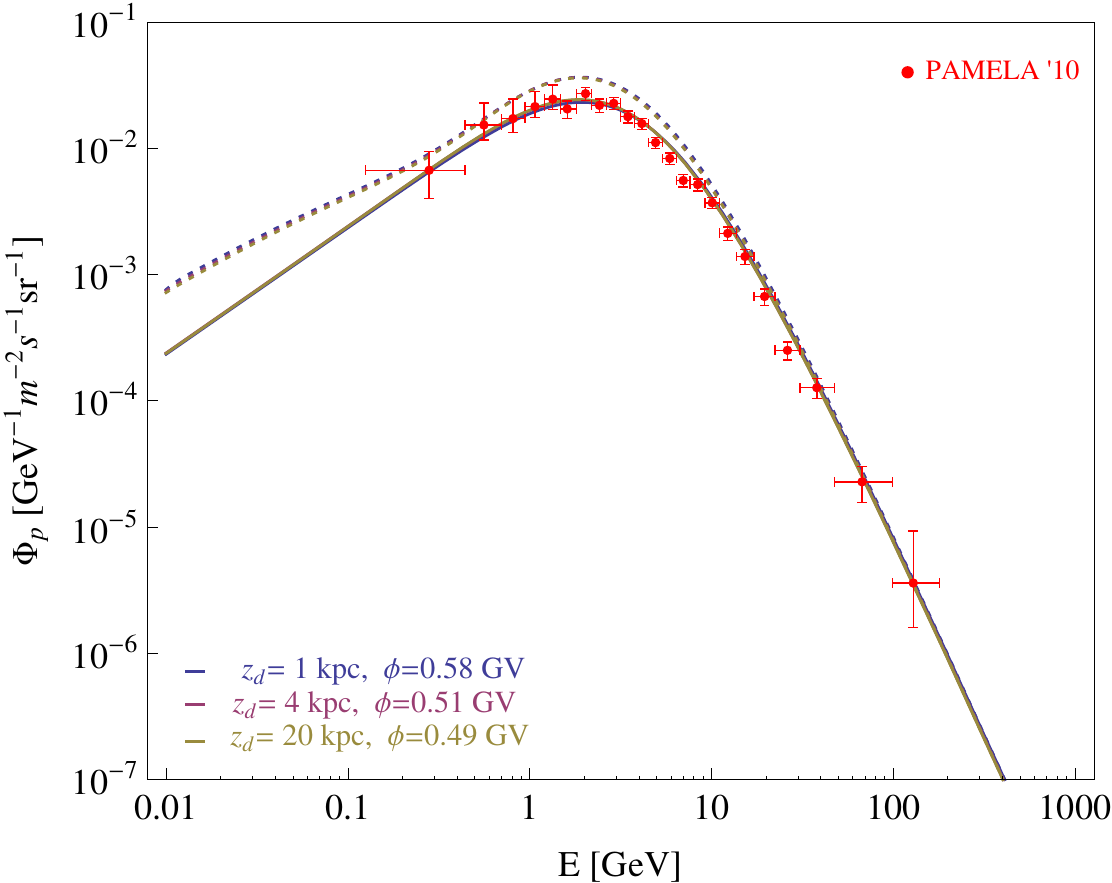}
	\includegraphics[height=6.5cm,width=7.5cm]{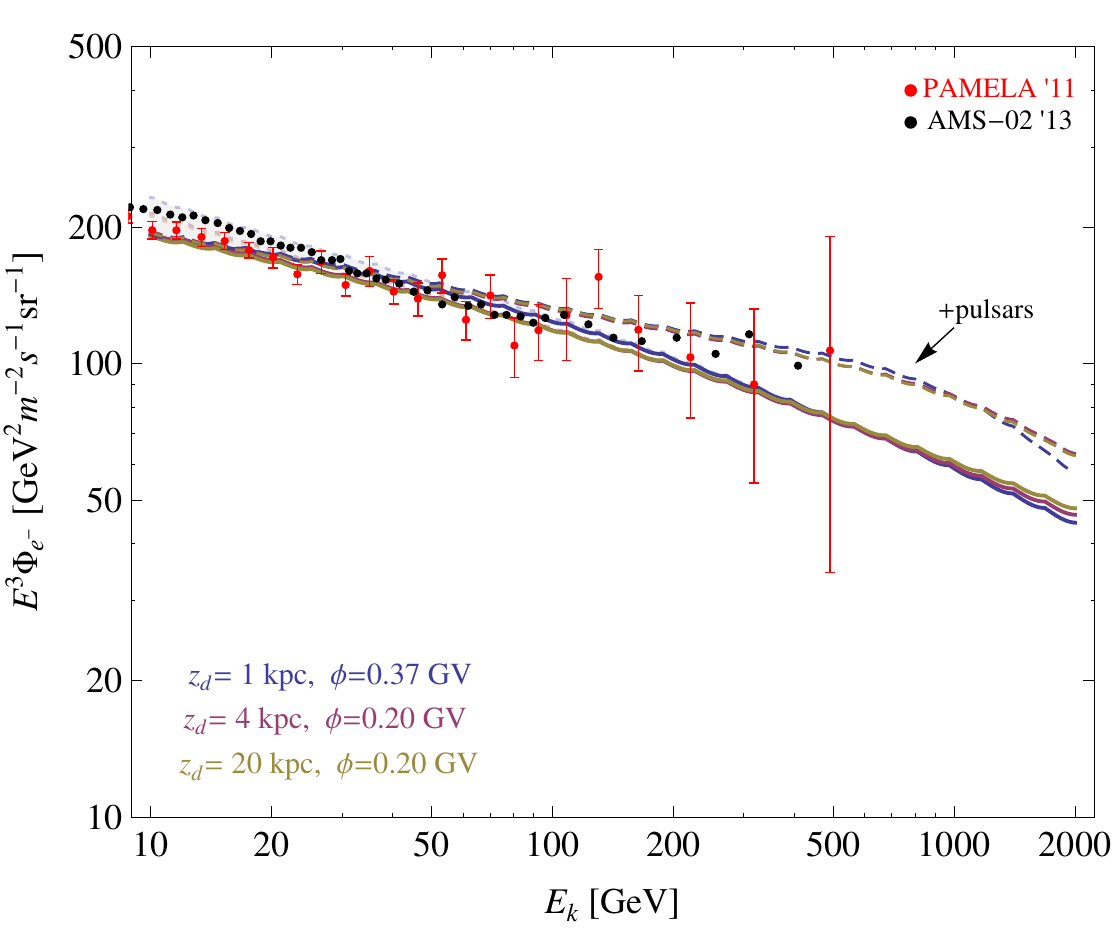}
	\caption{Comparison of propagation models. The solar modulated results are given by the solid, while for comparison also unmodulated spectrum is shown with dotted lines. \textit{Top left:} B/C data, \textit{top right}: protons, \textit{bottom left}: antiprotons and \textit{bottom right}: electrons. All the benchmark models give very good fit. In the case of electrons the dashed lines show the result with also the pulsar component included.	}
    \label{fig:propag1}
\end{figure}

As an example of how our models match the observational data, on the Fig.~\ref{fig:propag1} we show the fit of the thin $z_d=1$ kpc, medium $z_d=4$ kpc and thick $z_d=10$ kpc cases. In the B/C and protons one can see the strong solar modulation effect at low energies (with doted lines everywhere corresponding to unmodulated result), even for rather moderate values of the potential $\phi$. 

In the case of electrons, the simple force field approximation is insufficient in predicting the correct spectra at energies below few GeV. Therefore,  we chose to take into account only the data with $E>10$ GeV, due to the lack of full understanding of the solar modulation and also the precise values of parameters for secondary production mechanisms at such low energies. Moreover, what seem to be more robust choice is to insist on good agreement with the \textit{Fermi} diffuse $\gamma$-ray data and not the low energy electrons, since the backgrounds are much better understood in this case. Therefore, our prediction at energies below 10 GeV do not fit well the electron  data, but gives much better agreement with the diffusive $\gamma$-rays. The dashed lines on bottom right of Fig.~\ref{fig:propag1} give the total spectrum including the background and pulsar components. They are shown in order to convey that they are not very sensitive to variation of the propagation model and that they also improve the agreement with the electron data. For more discussion of these contributions see Sec.~\ref{sec:leptons}.

\begin{figure}
\centering
	\includegraphics[scale=0.65]{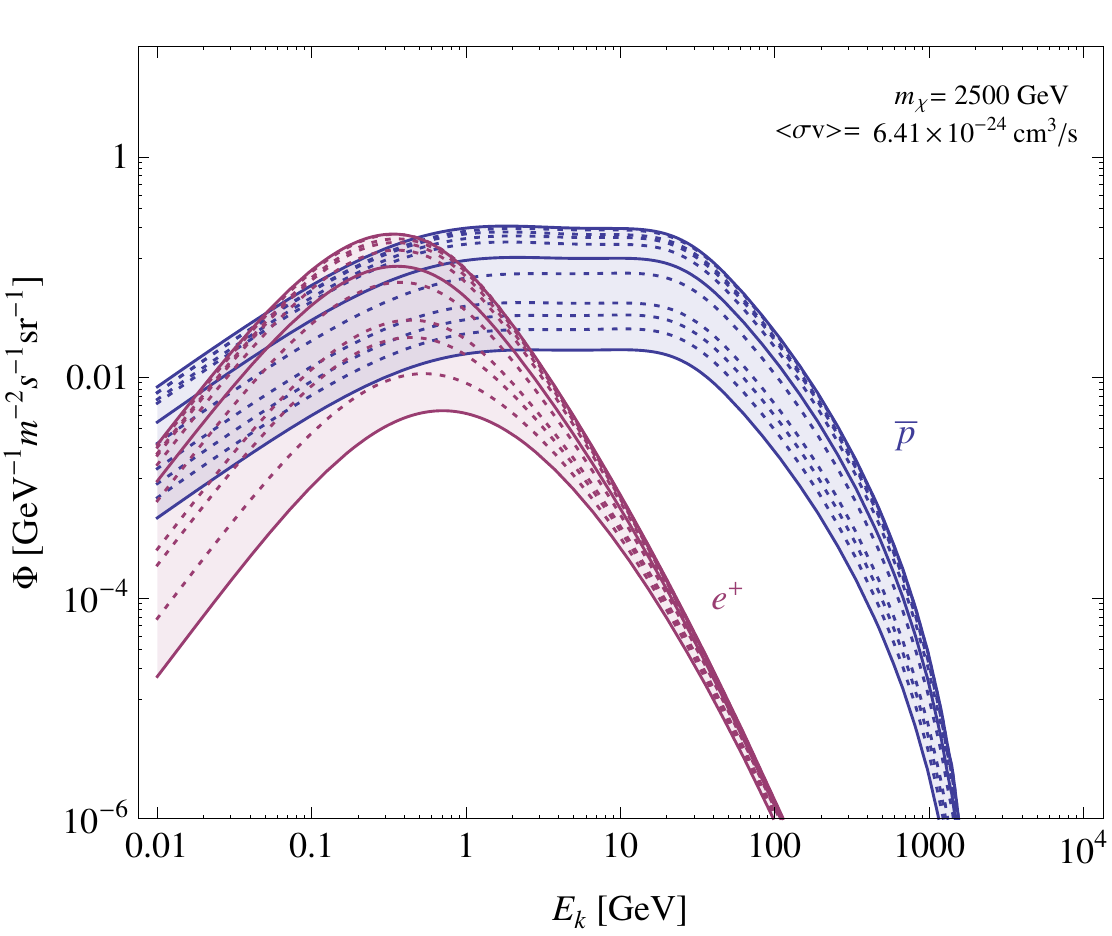}
	\includegraphics[scale=0.65]{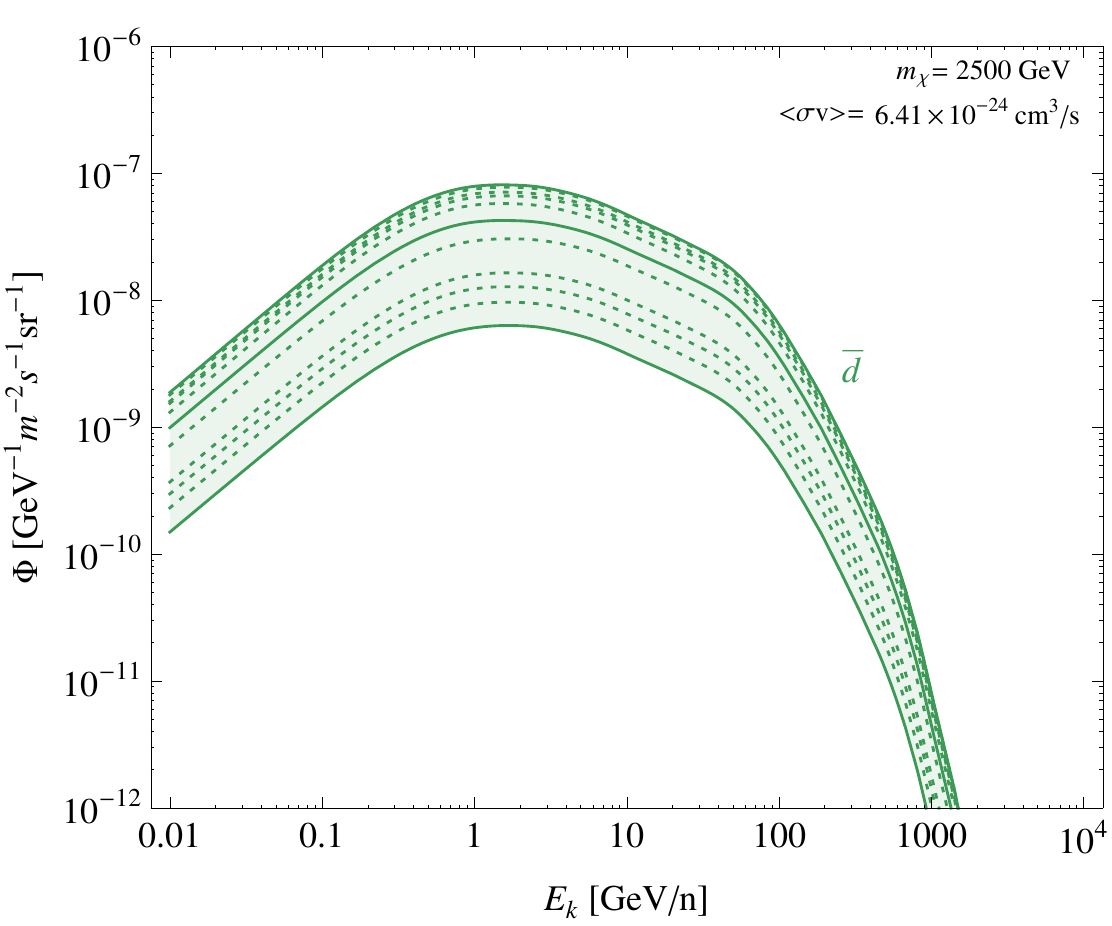}
	\caption{Effect of different propagation models onto indirect detection signal from  Wino with $m_\c=2.5$ TeV. \textit{Left:} antiproton (blue) and positron (violet) fluxes. \textit{Right:} antideuteron flux. The dotted lines correspond to our benchmark models, while solid ones single out the $z_d=20,\;4,\;1$ kpc cases (from top to bottom).} 
    \label{fig:propag2}
\end{figure}

Anticipating the discussion of the dark matter originated fluxes, on Fig.~\ref{fig:propag2} we show how they are affected by varying the propagation model. The dotted lines correspond to our benchmark models, while solid ones single out the thin, medium and thick cases. As expected, the uncertainty associated with the propagation model is less important when going to higher energies, but even then it remains substantial. Indeed, in the Wino model the phenomenologically most important effect of this uncertainty is the variation of high energy $\bar p$ fluxes originating from the dark matter, as we discuss below.

\section{Search Channels for Indirect Wino DM detection}
\label{sec:InDM}

In the following we will discuss what are the indirect detection signals for the Wino dark matter simultaneously in several channels. The questions we are going to answer are: \textit{i)} for what range of masses Wino is already excluded as a dark matter candidate, \textit{ii)} what is the impact of various uncertainties and how can they affect the exclusion limits, and finally \textit{iii)} does a configuration exist in which this model can explain the positron fraction rise and also  be consistent with the existing constraints from other indirect DM search channels.

The strategy is the following. Since in the Wino DM model the main annihilation channel is into gauge bosons, one can expect the antiproton constraints to be one of the strongest for masses less than a TeV given the current $\textit{PAMELA}$ data (see also \cite{Cirelli:2008pk,Donato:2008jk,Cholis:2010xb,Evoli:2011id}). The biggest uncertainty in this channel is the diffusion of the DM originated antiprotons. The astrophysical background is fixed by the requirement of fitting the proton data and the data on a secondary to primary ratio in CRs such as boron over carbon (B/C), because essentially all observed antiprotons are secondaries generated by interaction of primary CR protons. The significance of the diffusion effect can be parametrized by the scale of the diffusion zone $z_d$, see Eq.~\eqref{eq:diffcoeff}. Therefore, we start from determining the propagation parameters for a given value of $z_d$ by fitting the obtained CR spectra to the data: the B/C, protons, He and afterwards electrons. For this set of models we compute the DM signal in antiprotons and confront with the observations, putting limits on the Wino model. 

Having settled the propagation properties, we use them to determine the signals in leptons, due to experimental results chosen to be the total $e^++e^-$ flux and positron fraction defined as $e^+/(e^++e^-)$. In this case, diffusion is somewhat less relevant, since leptons loose energy much faster and the locally measured flux stems from much more local sources than for protons. What is then most uncertain is the production mechanism of secondary leptons and their exact energy losses during the propagation. However, as we will see, in our case the precise distinguishing between astrophysical and DM components in this channel proves to be rather difficult. Nevertheless, determining the total fluxes is important for the determination of the total diffuse $\gamma$-ray spectrum, which partially comes from the inverse Compton scattering of leptons on the radiation fields and the bremsstrahlung processes. The data sets we used for deriving the limits were \textit{i)} for electrons \textit{PAMELA} \cite{Adriani:2011xv},  \textit{ii)} for $e^++e^-$ flux \textit{AMS-02} \cite{AMSsite}, \textit{Fermi} \cite{Ackermann:2010ij}, HESS \cite{Aharonian:2008aa, Aharonian:2009ah} and MAGIC \cite{BorlaTridon:2011dk} with energies $E \geq 50$ GeV, and \textit{ii)} for positron fraction \textit{PAMELA} \cite{Adriani:2008zr} and \textit{AMS-02} \cite{Aguilar:2013qda} with energies $E \geq 20$ GeV. These energy cuts were motivated by the poor agreement between different data sets in the lower energies.

After obtaining lepton spectra, for all our propagation models and different Wino masses, we compute the $\gamma$-ray sky and compare with \textit{Fermi} data. From this we \textit{a posteriori} deduce which of our initial propagation models are the best ones, that is giving best fit to the data, and what are the uncertainties there. Having all this information, we can already put some more robust bounds on the Wino model. Finally, we close the whole picture by discussing the signals coming from dwarf spheroidal galaxies (dSph), neutrinos and antideuterons.

We did not include any effect coming from substructures. The reason is that although simulations tend to favor rather non-negligible amount of substructures in the Galaxy halo, they are still rather far from being conclusive \cite{Moore:1900zz} (see also \cite{1984ApJ...284..544G}). On the other hand, overdensities would amplify the DM signal.\footnote{Additionally, substructures are colder, with much lower velocity dispersions, see e.g. \cite{Strigari:2006rd}. Note however, that in contrast to the "dark force" Sommerfeld models, in our case this does not introduce any effect: recall that in the two channel version of the SE, below the threshold the effect is independent of the velocity.} Therefore, in order to give conservative limits or prospects for DM searches, we decided not to include substructures, and consider only a smooth halo profile, which we take to be the NFW \cite{Navarro:1995iw}:
\be
\label{eq:NFWprofile}
\rho(r)=\rho_{NFW} \frac{R_{c}}{r}\frac{1}{(1+ r/R_{c})^{2}}.
\ee
with $R_{c} = 20$ kpc.

The overall normalization of the density profile is obtained via the determination of the local dark matter density. This quantity is again not known exactly. We adopt as reference a value of $\rho_{DM}(r_0) = 0.4 {\,\rm GeV/ cm}^{3}$ in agreement with \cite{Catena:2009mf, Salucci:2010qr}. 

In the results below we always assume that the Wino accounts for the whole dark matter. This in most of the choices of $m_\c$ requires it to be of non-thermal origin, except for the masses of around 3.2 TeV. It is worth to emphasize that this is not at all an \textit{ad hoc} scenario, as non-thermal Wino arises naturally in many well motivated theories, see e.g. \cite{Moroi:1999zb,Acharya:2007rc}.

\subsection{Antiprotons}
\label{sec:antiprotons}

In the cosmic rays antiprotons are far less abundant then the protons. They are believed not to be produced in astrophysical sources and hence the observed flux is secondary coming from interactions of protons (and to a certain extent also heavier nuclei) with the interstellar gas composing mostly of hydrogen (atomic and molecular) and helium. These produced antiprotons then propagate and interact with the gas by themselves, sometimes annihilating and sometimes scattering inelastically and loosing energy. The latter process introduces softening of the spectrum and is commonly taken into account by treating all the inelastic collisions as annihilating $\bar p$s and replacing them by the so-called \textit{tertiary} source.

Note, that for all these processes, the gas distribution plays an important role. As discussed above, in all our computations we implemented the gas models derived in \cite{Tavakoli:2012jx, Pohl:2007dz}, based on the most 
recent HI and CO line surveys available \cite{Kalberla:2005ts,Dame:2000sp}.

\begin{figure}
\centering
	\includegraphics[scale=0.65]{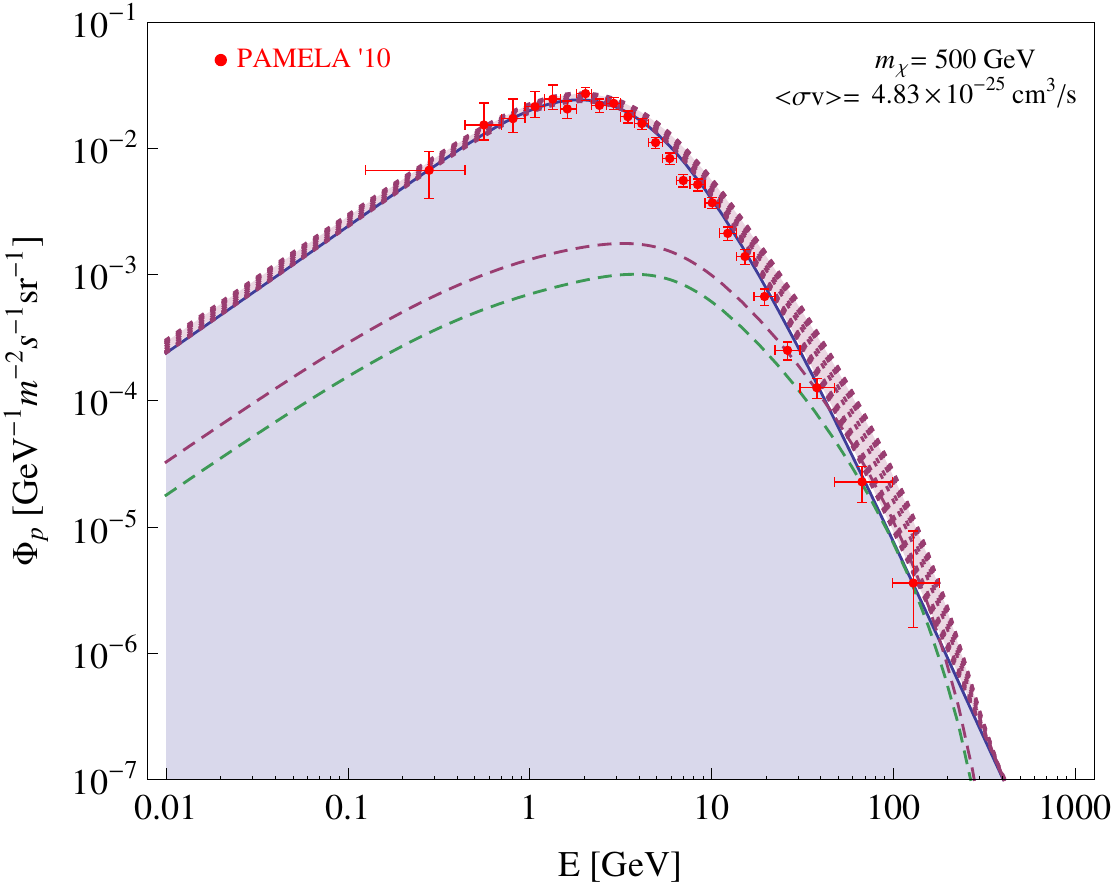}
	\includegraphics[scale=0.65]{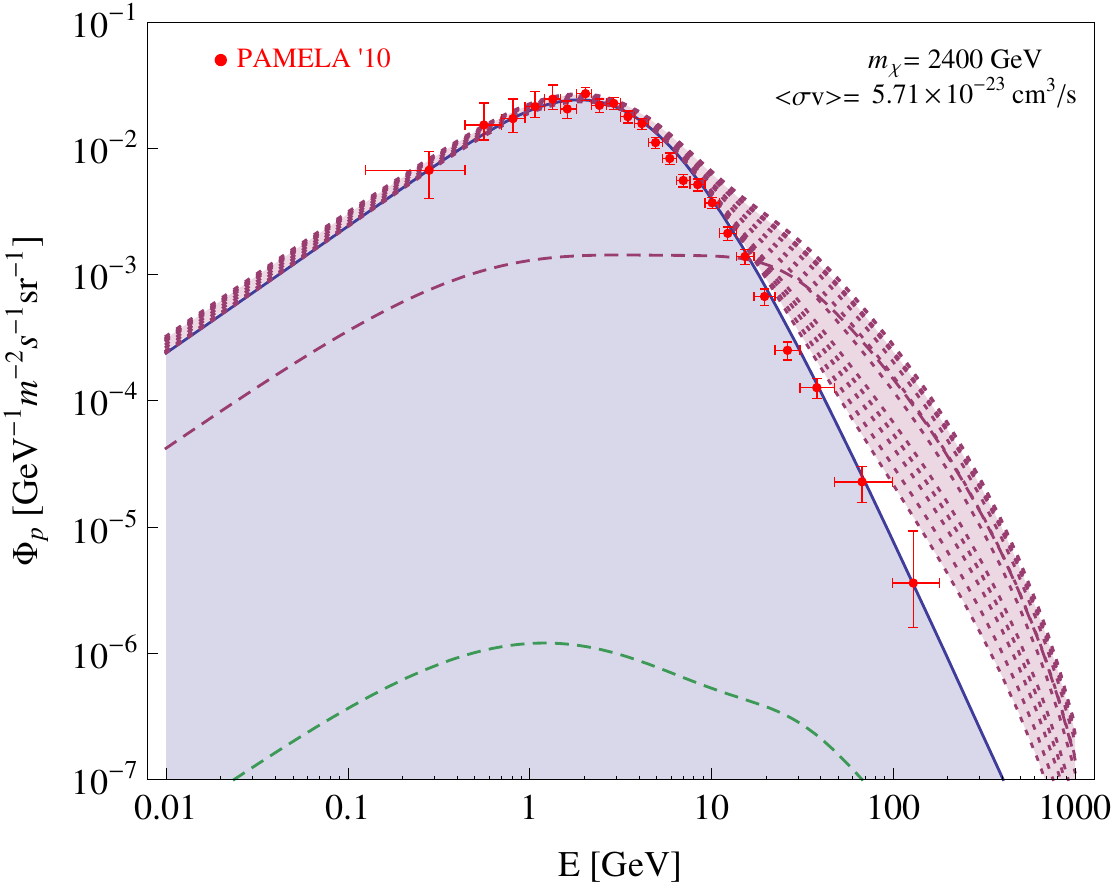}
		\includegraphics[scale=0.65]{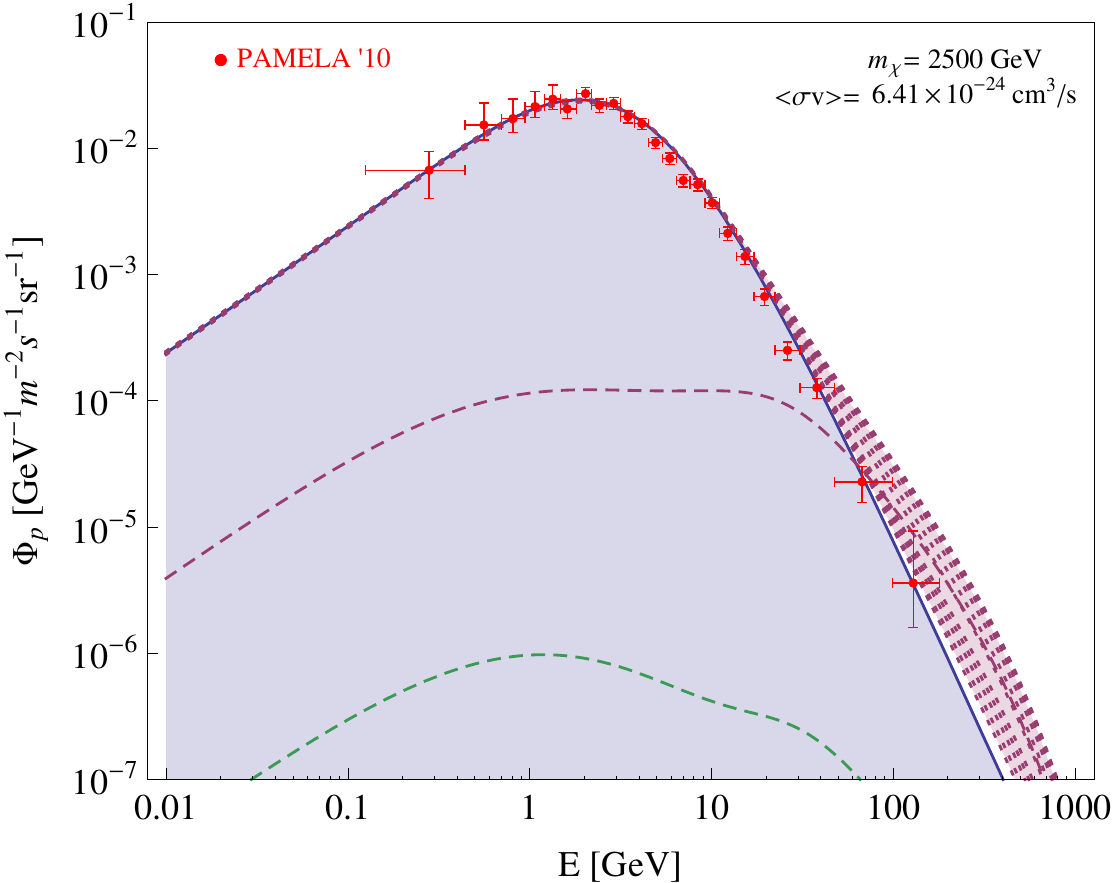}
			\includegraphics[scale=0.65]{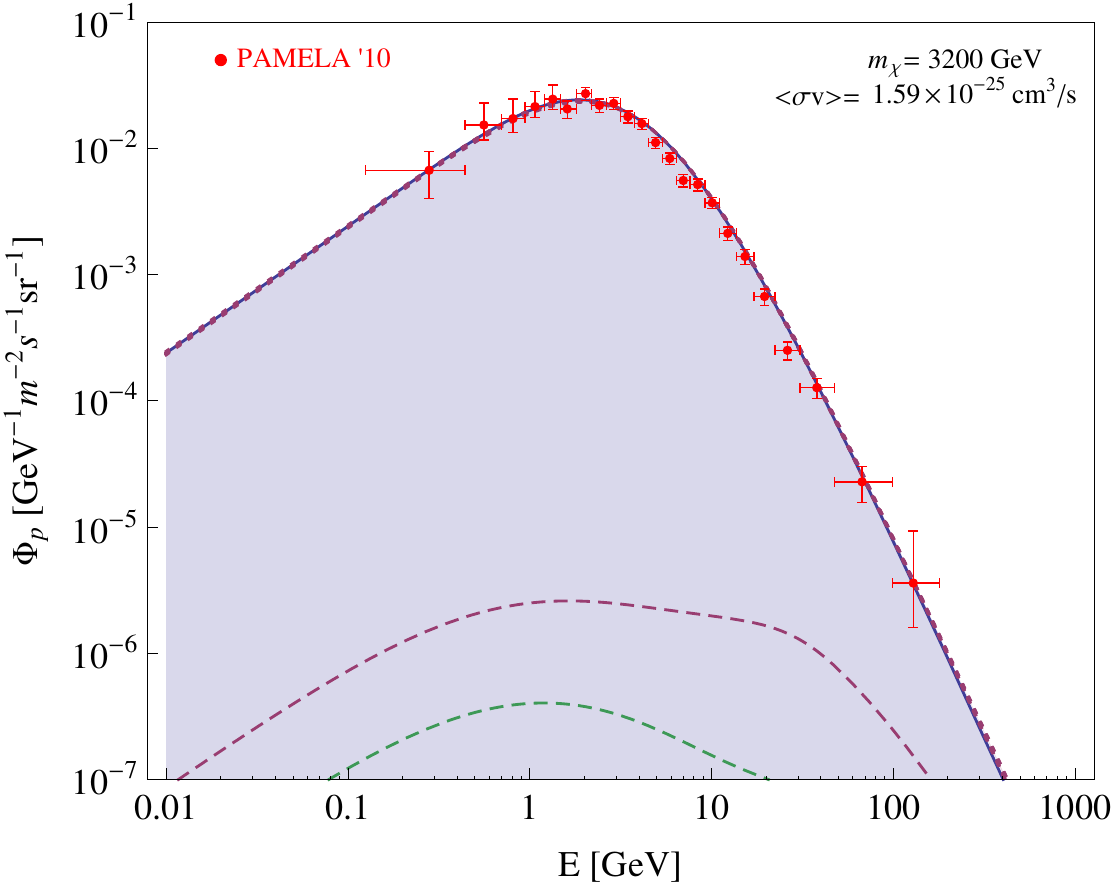}
	\caption{Antiproton fluxes for all the propagation models and $m_\c=0.5,\,2.4,\,2.5,\,3.2$ TeV. The blue shaded region shows the expected background after solar modulation with $\phi=0.51$ GV. The dashed lines represent the dark matter contribution without (green) and with Sommerfeld effect (violet) for the example case of $z_d=4$ kpc model. The total spectrum including background and DM signal is given by the violet shaded regions where the dotted lines correspond to our benchmark models (the $z_d=1$ kpc being the closest to the background and growing with the thickness). The strong boost of the signal for $m_\c= 2.4$ and $2.5$ TeV comes from the resonance in the Sommerfeld effect.}
    \label{fig:antip}
\end{figure}

The propagation of antiprotons is governed mostly by diffusion. The total energy losses timescale is much larger compared to the diffusion timescale in the interstellar medium up to PeV energies; CR energy above which solving the propagation equation at steady state  is not the right approach since PeV CR sources are not smoothly distributed in the Galaxy.
 On the other hand, diffusive reacceleration processes (and possibly convection) can have a strong impact on the low energy tail of the spectrum. For our purposes, however, only the energies larger than roughly 10 GeV matter, since at the TeV scale the Wino DM contribution affects mostly this part of the spectrum.

Indeed, on Fig.~\ref{fig:antip} we plot the predicted antiproton fluxes at Earth coming from the Wino with masses $m_\c=0.5,\;2.4,\;2.5,\;3.2$ TeV. They were chosen in such a way, that the lowest is the generic "lighter" Wino case still in good agreement with the data. For lower masses the annihilation cross section grows and generically (depending however on the propagation model) overshoots the data. The highest one $m_\c=3.2$ TeV gives correct thermal relic density and as can be seen does not produce any noticeable excess over the background. The two middle cases are close to the Sommerfeld resonance, where the total annihilation cross section grows by several orders of magnitude. This can be seen by comparing the violet and green dashed lines on the plots, giving the dark matter contribution with and without Sommerfeld enhancement, respectively. Needless to say, it introduces a huge change in the predicted signal and therefore cannot be ignored.

The violet shaded region corresponds to the total DM plus background flux for all the range of  $z_d$. It clearly shows the importance of the uncertainty of the propagation model in this search channel. The close to resonance case of $m_\c=2.4$ TeV is already excluded (in fact overshooting the data even without background) for thicker diffusion zones, but may be marginally consistent for very thin ones. Moving a bit away from the resonance eases the tension, but still the thickest cases give too much antiprotons.

\begin{figure}
\centering
	\includegraphics[scale=0.67]{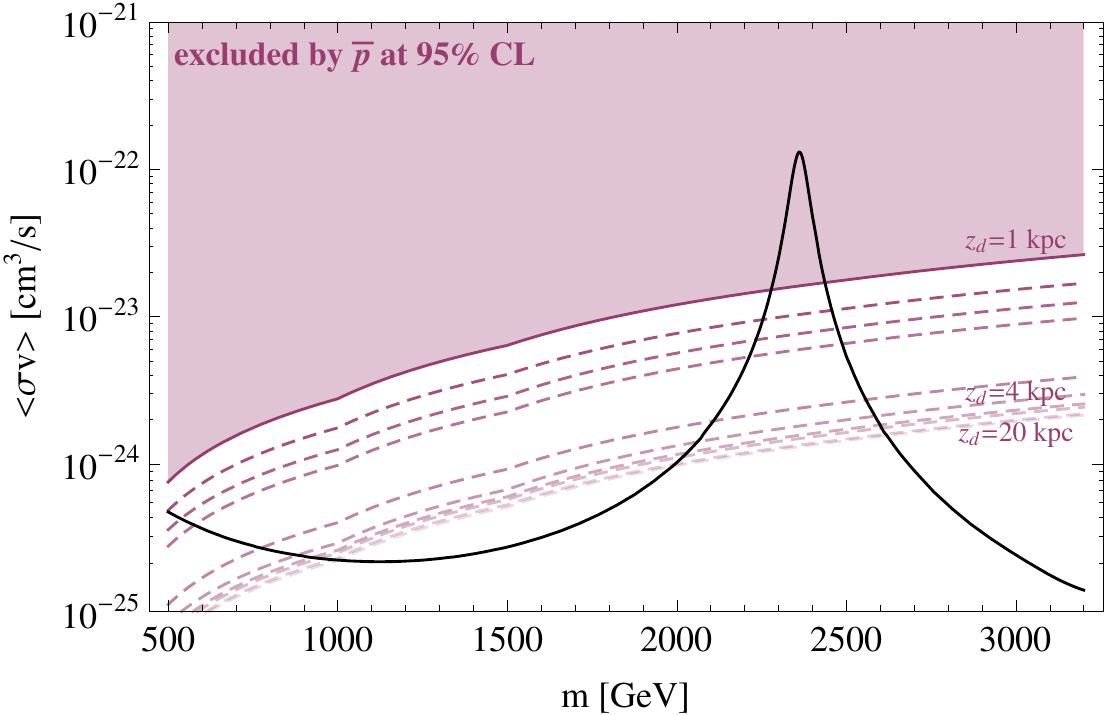}
	\includegraphics[scale=0.665]{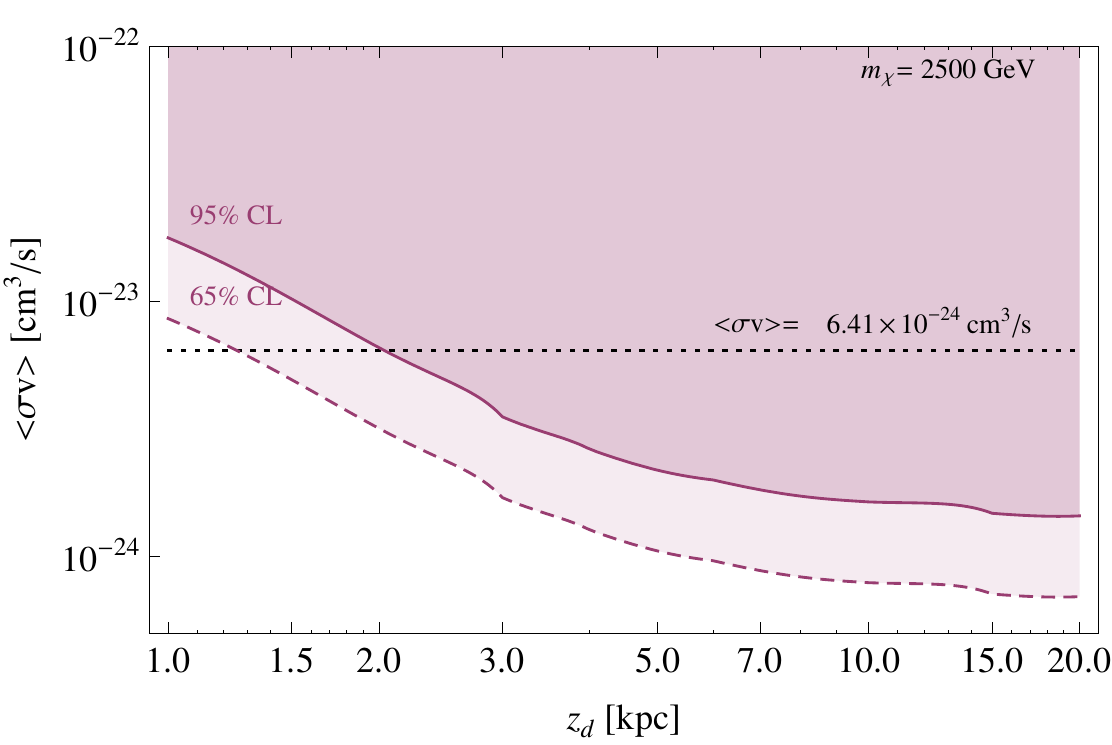}
	\caption{Limits from antiprotons. \textit{Left:} the 95\% CL upper limits on the total annihilation cross section as a function of the Wino mass. The dashed lines correspond to the benchmark propagation models. The black solid line is the actual present-day cross section of the Wino computed at one-loop level with the Sommerfeld effect included. \textit{Right:} the dependence of the upper bound on the thickness of the diffusion zone for an example case of $m_\c=2.5\; \TeV$. The dashed black line shows the value of the one-loop level cross section for a Wino in this case.}
    \label{fig:antip_limits}
\end{figure}

The obtained 95\% CL upper limits on the Wino model are given on Fig.~\ref{fig:antip_limits}. Although, as we advocated at the beginning, the antiproton channel is expected to give one of the most stringent constraint for models with annihilations predominantly to weak gauge bosons, one can see that the obtained limits are not that severe. Indeed, at this stage, when all the benchmark propagation models are allowed, the Wino dark matter can be robustly excluded by the $\bar p$ channel only at low masses (see also \cite{Belanger:2012ta}), i.e.\footnote{The lower limit comes from the fact, that at the Wino masses below $m_W$, the $W^+W^-$ annihilation channel is kinematically not allowed and the cross section is considerably smaller.} $m_W \lesssim  m_\c\lesssim 500$ GeV and in the very proximity of the resonance, i.e. in the range $2.21 \;{\rm TeV} \lesssim m_\c \lesssim 2.46\;{\rm TeV}$. However, as one can see, the thicker the propagation model, the more stringent limits. Therefore, if one can disfavour thin propagation models using other channels, more sever limits apply. We will come back to this point after discussing the searches in the diffuse $\gamma$-rays.

\subsection{Leptons}
\label{sec:leptons}

In the same way as antiprotons, positrons are typically produced as secondaries. They come mostly from the decay of charged mesons ($\pi^+$ and $K^+$) produced in the interactions of the nuclei with the gas. However, CR $e^{\pm}$  can also be produced and accelerated in pulsars \cite{Kobayashi:2003kp,Hooper:2008kg,Profumo:2008ms}. 

Pulsars are fast rotating magnetized neutron stars surrounded by a comoving plasma configuration called âmagnetosphere. Electrons in the magnetosphere loose energy and emit photons, which are energetic enough to produce electron-positron pairs in the intense pulsar magnetic field. This leads to an potentially effective source of primary electrons and positrons, especially coming from the  middle aged pulsars (about $10^5$ years old) \cite{Grasso:2009ma}.

The contribution of pulsars can be effectively described  by an injection spectrum with $E^{-n}$ together with a high energy break related to the cooling time of the electrons and positrons during their propagation. We chose to fit the impact of the 
galactic pulsars population 
following the parametrization of \cite{Malyshev:2009tw}:
\be
Q_{\rm pul}(r,z,E)=N_0 \left(\frac{E}{E_0}\right)^{-n} e^{-E/E_c} f_{\rm pul}(r,z),
\ee
where we will effectively assume that the $e^\pm$ reaching our location are isotropic. 
The injection index $n$, critical energy $E_c$ introducing the cut-off and the normalization $N_0$ will be treated as free parameters of the fit, while the normalization of the energy is fixed to $E_0=5$ GeV;
$f_{\rm pul}(r,z)$ describes their spatial distribution \cite{FaucherGiguere:2005ny}.
It is important to bear in mind, that this is only an effective description, aiming in taking into account contribution of many pulsars. However, as we will show below, the component of this kind added on top of the background (and the dark matter contribution) can give a very good fit to the data.

\begin{figure}
\centering
	\includegraphics[scale=0.65]{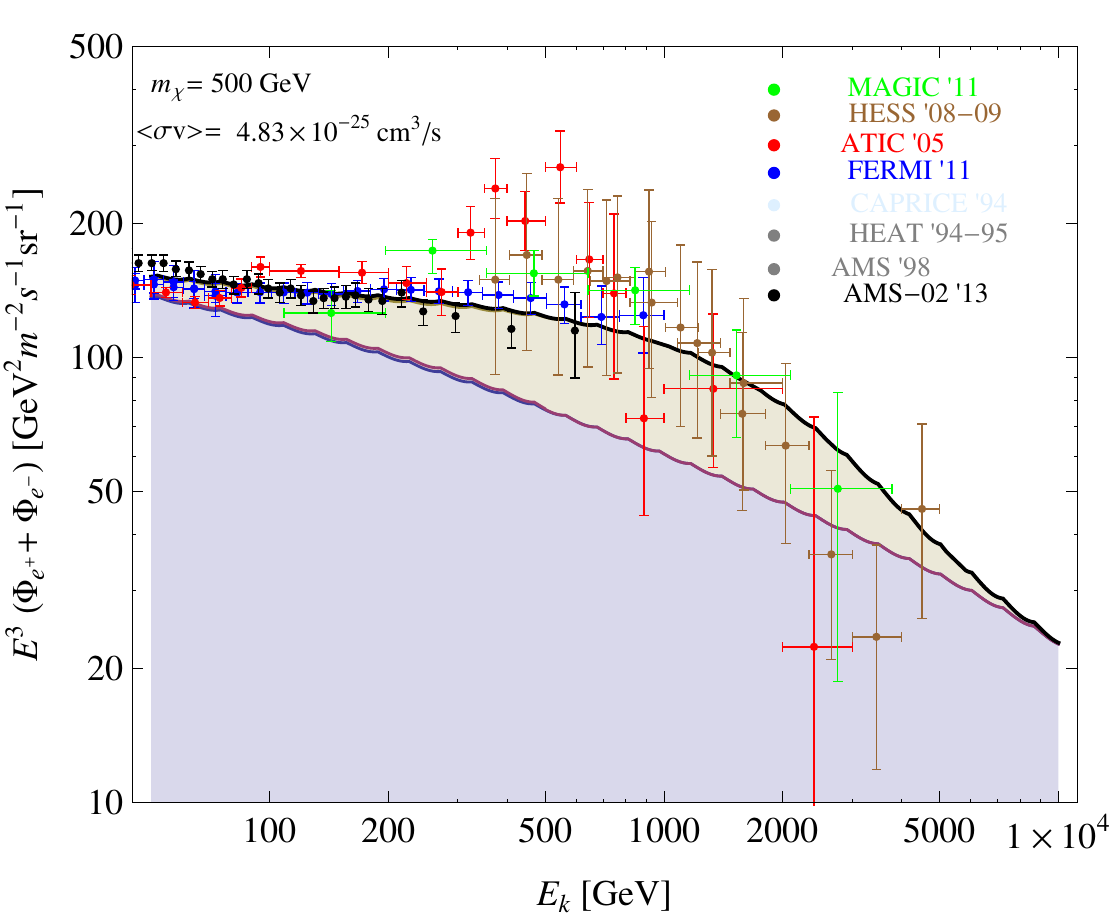}
		\includegraphics[scale=0.65]{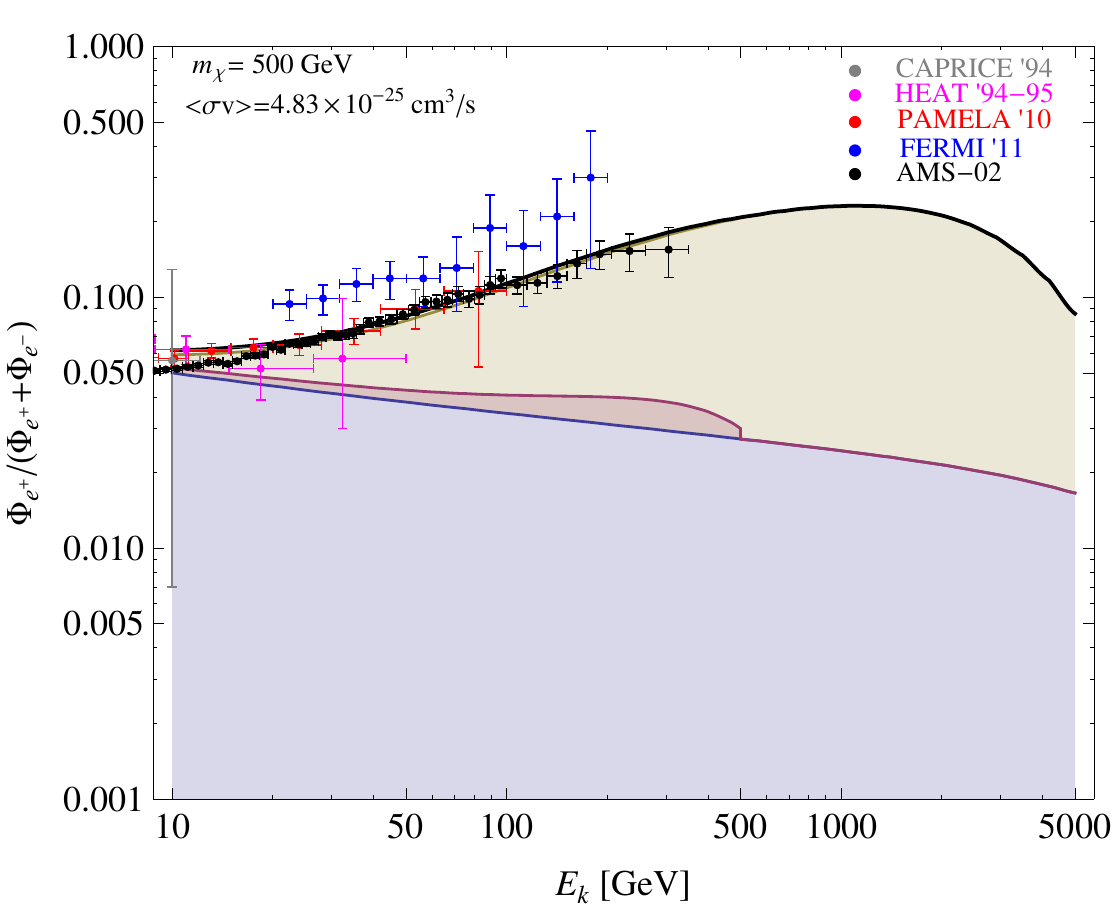}
			\includegraphics[scale=0.65]{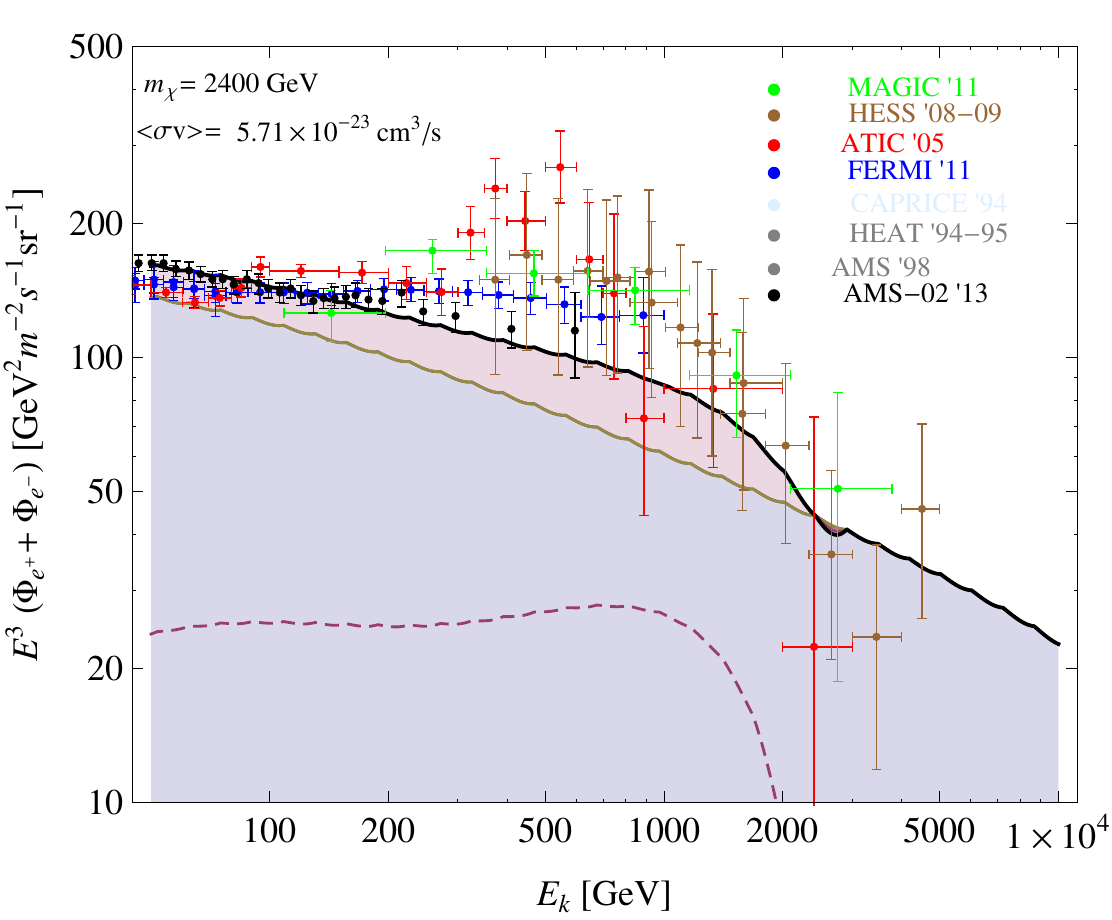}
		\includegraphics[scale=0.65]{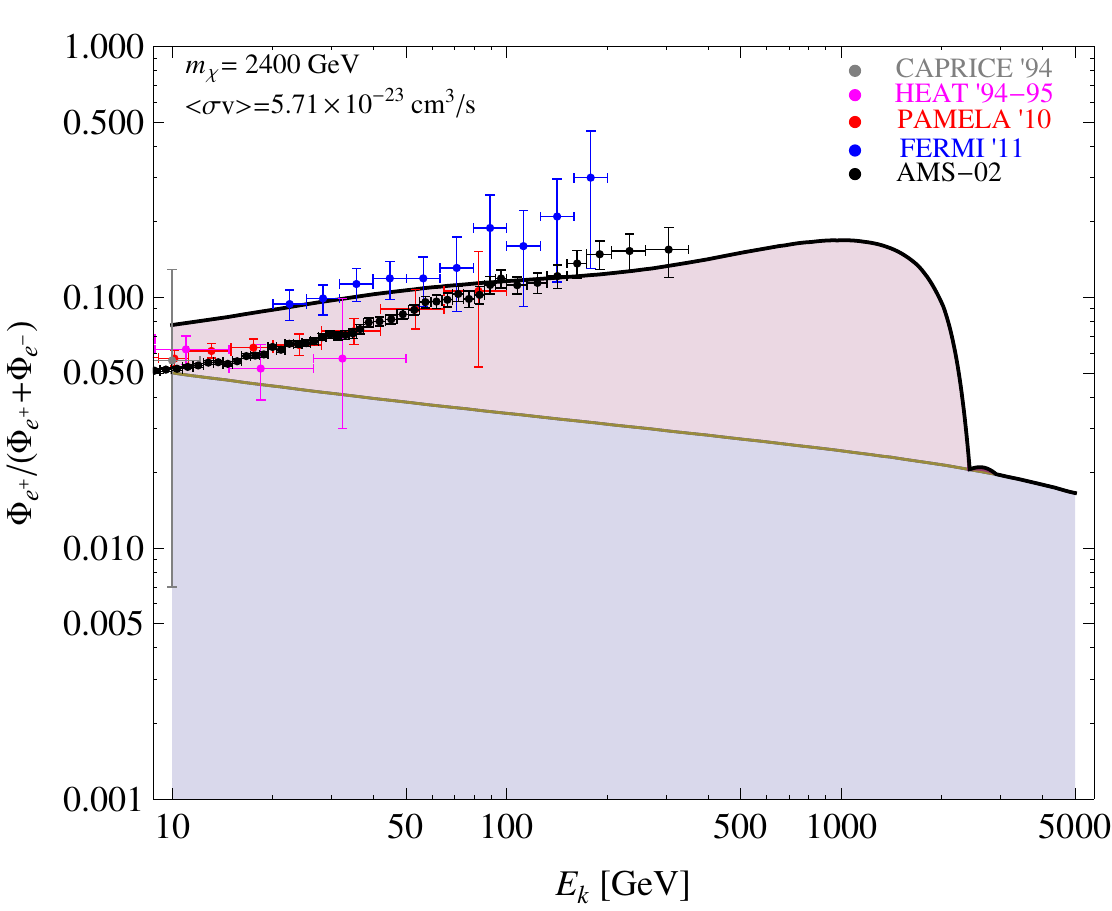}
			\includegraphics[scale=0.65]{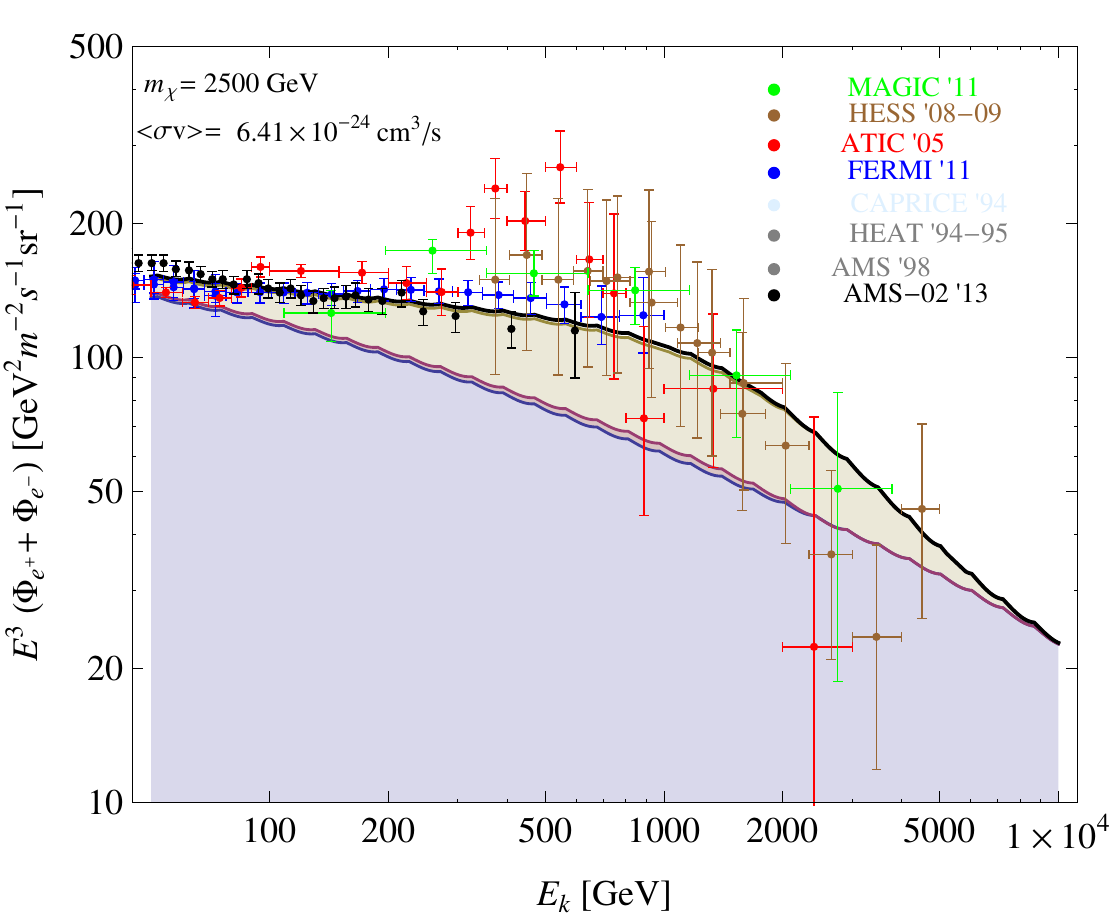}
		\includegraphics[scale=0.65]{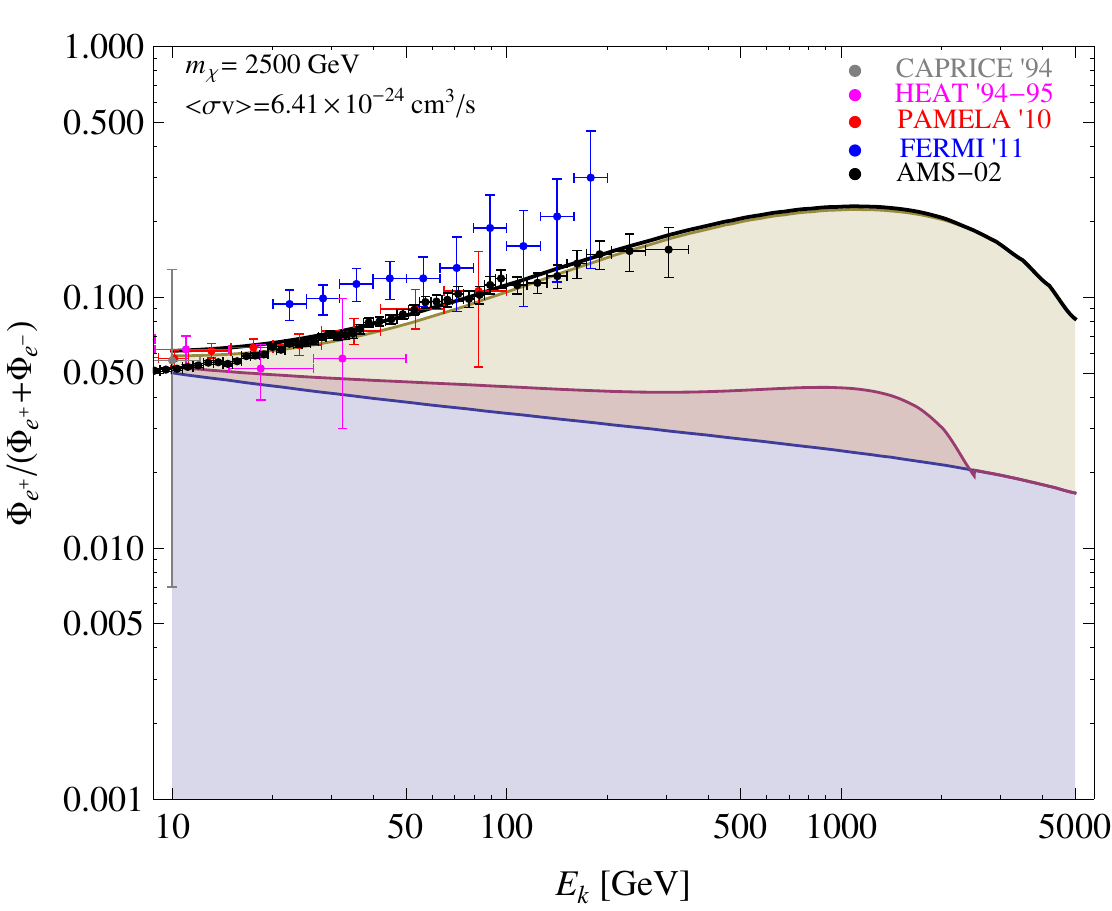}
	\caption{Total $e^- + e^+$ fluxes (\textit{left column}) and positron fraction (\textit{right column}) for the $z_d=4$ kpc propagation model and $m_\c=0.5,\;2.4,\;2.5$ TeV. The blue shaded region represents the expected background with solar modulation. The DM component (violet dashed line) is typically not strong enough to fit the data (together with the background giving violet solid line) and hence additional source of positrons from pulsars (yellow solid) is needed. The solid black line gives total signal including all sources. The cases with larger masses outside of the presented range do not introduce nearly any significant lepton component.}
    \label{fig:posi}
\end{figure}

In the propagation of leptons, the major role is played by the energy losses. Electrons  and positrons loose energy by ICS of CMB photons and infrared or optical galactic starlight. These mechanisms are very effective with increasing energy and therefore the very energetic electrons and positrons measured locally by CR detectors have to come to us from nearby. Diffusion is the dominant process only at low energies, since only then leptons have time to diffuse before loosing most of their energy. It follows, that the main uncertainty in the signals coming from positrons is not attributed to the propagation, but rather the precise knowledge about the energy losses, the interstellar radiation field and the exact values of the primary injection spectra.

We first address the question, of whether the Wino DM model can solve  by itself the CR lepton puzzle. The starting observation is that to fit the positron fraction, one needs a large annihilation cross section, of the order of $\mathcal{O}(10^{-22}\text{--}10^{-23} \;{\rm cm}^3/{\rm s})$ (see e.g. \cite{Cirelli:2008pk}). This is rather difficult to achieve for most models, especially if one insists on the thermal production mechanism, pointing to the cross section of $3\times 10^{-26} \;{\rm cm}^3/{\rm s}$. This is also true for the Wino case. However, the virtue of this particular model is that it posses and efficient mechanism of boosting the present-day cross section to the needed values, i.e. the Sommerfeld effect coming from weak bosons exchange. Far from the resonance, as in the case of $m_\c=500$ GeV (top panel in Fig.~\ref{fig:posi}), the dark matter contribution (given by violet dashed line) added to the expected background (blue region) gives a combined result (violet shaded region) that is much below the data. Therefore, one needs a dominant pulsar contribution to be added to the total flux and positron fraction (background+DM+pulsar, black line).

Being only moderately off the resonance, (bottom panel of Fig.~\ref{fig:posi})
the dark matter contribution is too small. Therefore, only for the masses at the proximity of the resonance, could we potentially fit the \textit{AMS-02} data just from the contribution of the dark matter annihilations (middle panel of Fig.~\ref{fig:posi}). Yet even the $m_\c=2.4$ TeV case, which is close to the resonance, produces typically too few total leptons and, more importantly, too much positrons with energies of around 10 GeV. Even if the former could be adjusted by varying the background, in particular the value of the exponential cut-off in primary electrons, the latter is very hard to evade. 

\begin{figure}
\centering
	\includegraphics[scale=0.67]{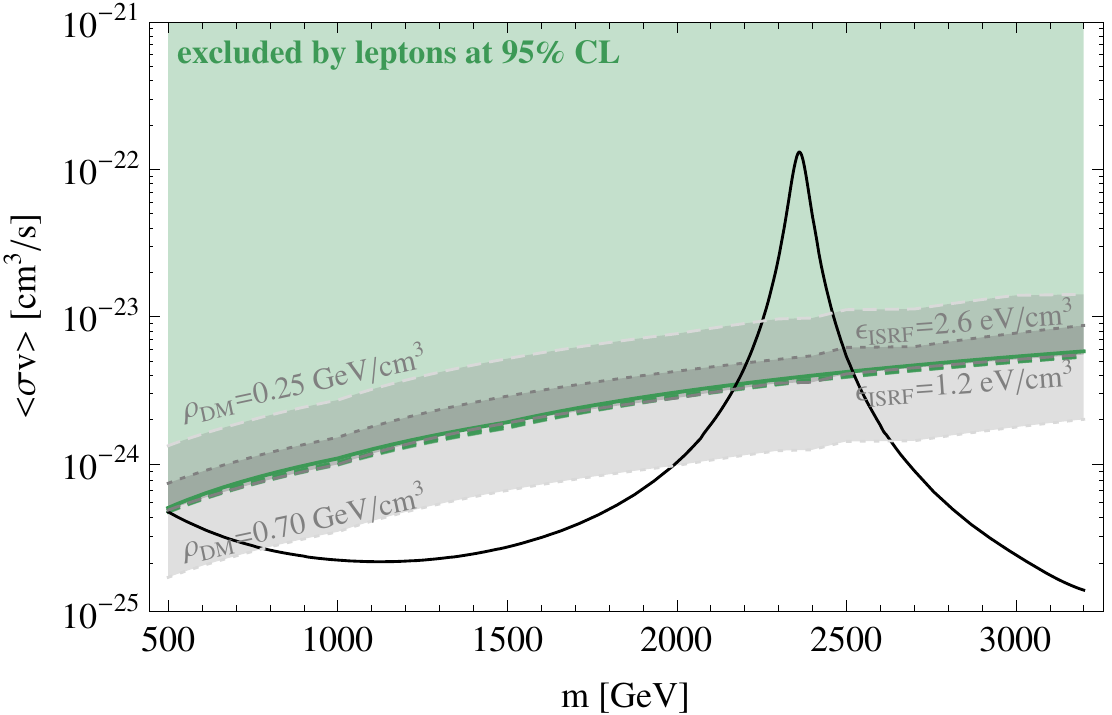}
	\includegraphics[scale=0.665]{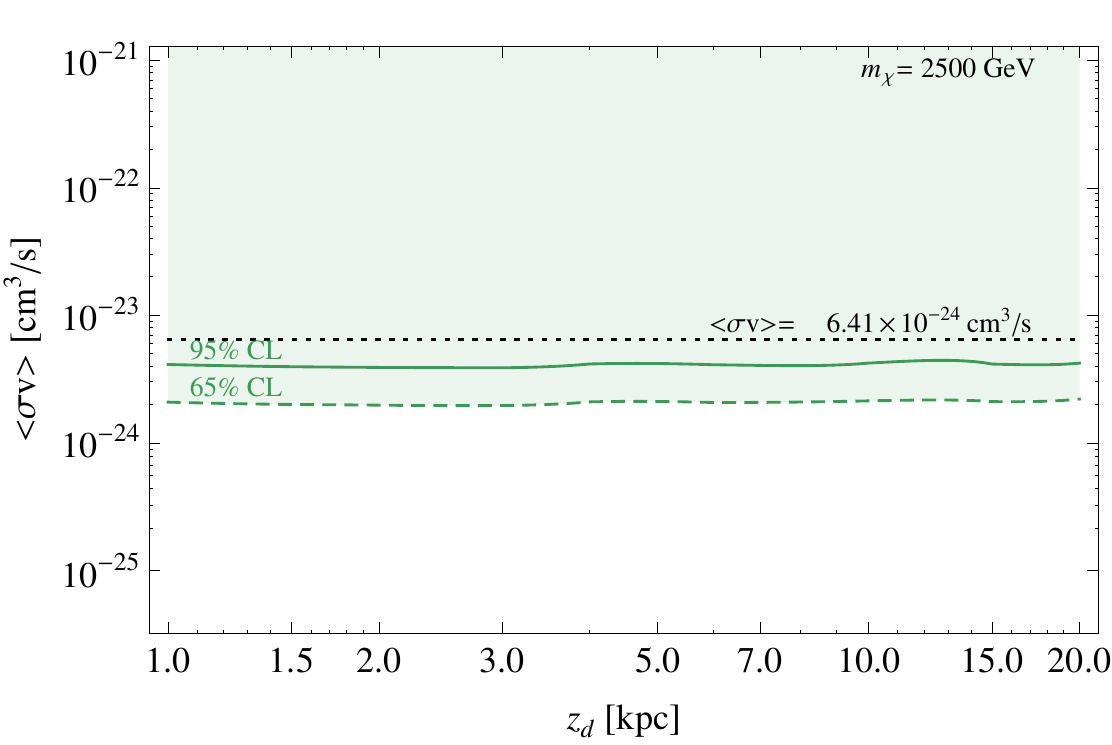}
	\caption{Limits from combination of all lepton datasets. The legend is analogous to Fig.~\ref{fig:antip_limits}. The gray bands show the uncertainty of the limit related to the local DM density and the energy density in the ISRF and the magnetic field.}
    \label{fig:leptons_limits}
\end{figure}

Finally, Fig.~\ref{fig:leptons_limits} shows the 95\% CL upper limits on the cross section obtained from the comparison with combined data sets for electrons, positron fraction and total leptons. The result is analogous to the antiproton case, with two important differences. First, the limits are significantly more stringent. This is partially because of the very high statistics of the positron fraction \textit{AMS-02} data. Secondly, there is nearly no dependence of the limits with the diffusion zone thickness $z_d$. On the other hand, since CR leptons are local, the energy losses and the local DM density uncertainties play an important role. The former can be parametrized by the local energy density of the interstellar radiation field (ISRF) and the magnetic field, which we vary in the range 1.2-2.6 eV/cm$^3$ (dark gray band). The latter we probe by assuming $\rho_{DM}(r_0)$ between 0.25 and 0.7 GeV/cm$^3$ (light gray band).

To summarize, our detailed study shows that the Wino model cannot explain the CR lepton puzzle. In fact, the lepton data can even rule out the very proximity of the resonance, independently of other channels, and not relying much on the propagation model. To fit the data one thus needs to have sources of positrons additional than the DM, at least in this model.

Finally, our results on the CR leptons and protons, have an impact on the calculation of diffuse galactic $\gamma$-rays.
The low energy positron flux is comparable to the electron one and thus introduces a non-negligible contribution of the background positrons to the diffuse $\gamma$-ray emission in the MeV range. Moreover, the high energy leptons also may introduce some (smaller) contribution and for a robust prediction one should take them into account. This means that one needs both the dark matter and the pulsar components in the calculation for the diffuse $\gamma$-rays.

\subsection{Limits on Wino annihilation in the DM halo from diffuse galactic $\gamma$-rays}
\label{sec:GalacticDiffuse}

The \textit{Fermi} $\gamma$-ray background is mainly composed of the galactic diffuse background, the isotropic diffuse background (dominated by unresolved extragalactic sources) and the resolved galactic and extragalactic point sources. In addition extended sources contribute at various latitudes. The galactic diffuse background arises from a combination of astrophysical processes. These are, the inelastic nucleon-nucleon collisions producing mainly $\pi^0$s which subsequently decay to two photons, bremsstrahlung radiation from interactions of CR electrons with the ISM gas and also up-scattering of CMB and galactic radiation field photons. From the point of view of dark matter searches in diffuse $\gamma$-rays, all these processes give rise to a very prominent background. The DM contribution comes from two types of processes: direct emission (prompt $\gamma$-rays) during annihilation or decay process
which includes the hadronization and decay processes that lead to stable SM particles and the secondary contribution coming from
ICS and bremsstrahlung by the produced stable CR electrons and positrons.
 
The dark matter prompt $\gamma$-ray flux is given by: 
\be
\frac{d\Phi_\g}{dE}=\frac{1}{4\pi}\int \langle \sigma v\rangle \frac{\rho^2_{DM}(l,\Omega)}{2m_\c^2}\frac{dN_\g}{dE}\,dl d\Omega\;,
\ee
where $d\Omega$ is the solid angle within which the observation is made, and $l$ the length along the line of sight. In the annihilation spectrum at production $dN_\g/dE$ all the processes of prompt production of decay and radiative emission are taken into account. If the annihilation cross section is homogeneous\footnote{A counterexample would be if $\langle \sigma v\rangle$ does significantly depend on the velocity dispersion and one includes substructures.}, then it simplifies to
\be
\frac{d\Phi_\g}{dE}=\frac{1}{4\pi} \langle \sigma v\rangle \frac{1}{2m_\c^2}\frac{dN_\g}{dE}\, J,
\ee
where all the factors depend on the particle physics properties of the dark matter, except the so-called $J$-factor:
\be
J=\int \rho^2_{DM}(l,\Omega)\,dld\Omega\;,
\ee
depending on the dark matter distribution in the halo. The ICS and bremsstrahlung contributions form DM are evaluated from our codes in the same manner that the equivalent backgrounds are being calculated.

\begin{figure}
\centering
	\includegraphics[scale=0.85]{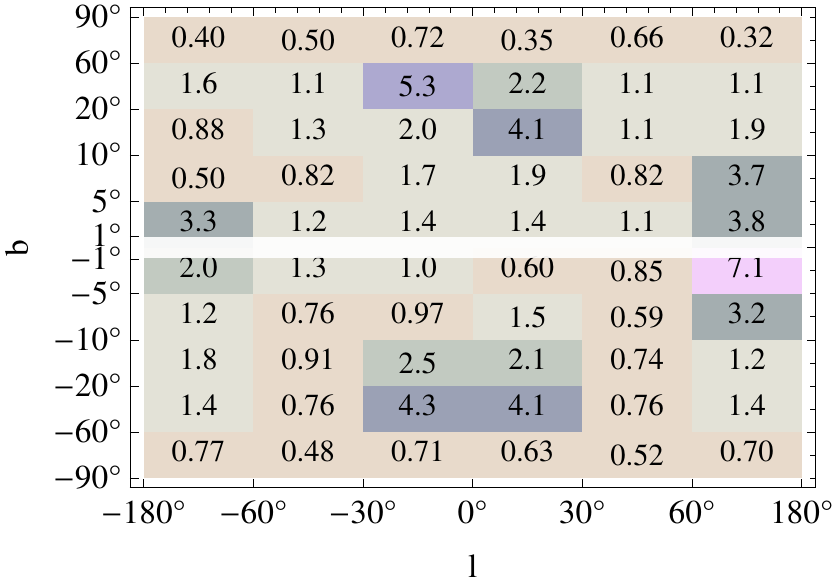}
		\includegraphics[scale=0.85]{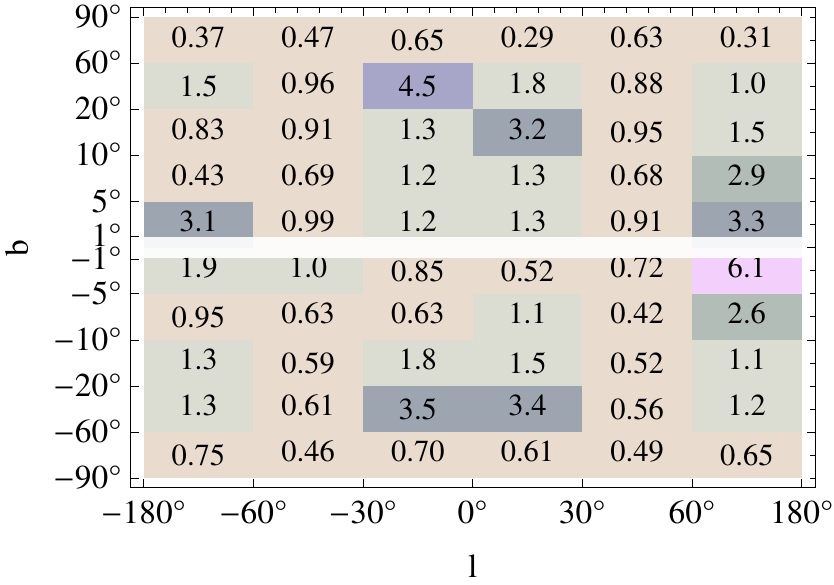}
			\includegraphics[scale=0.85]{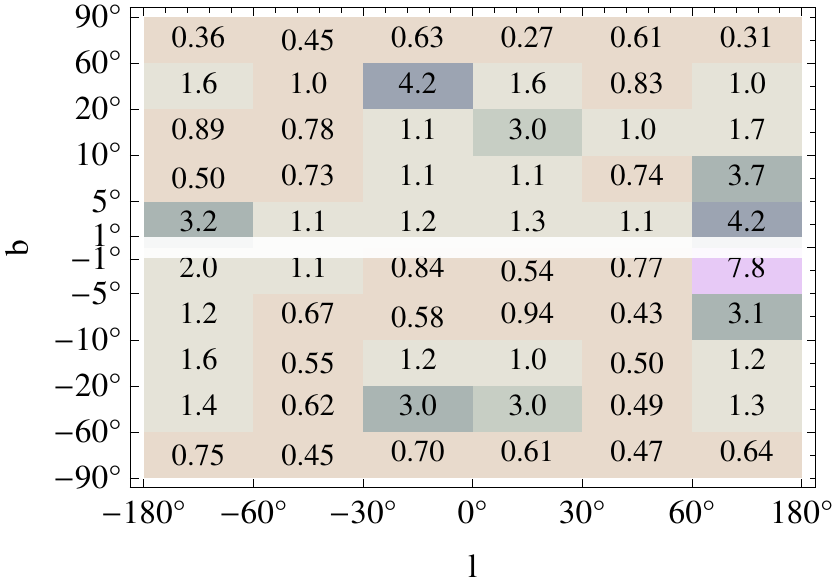}
						\includegraphics[scale=0.85]{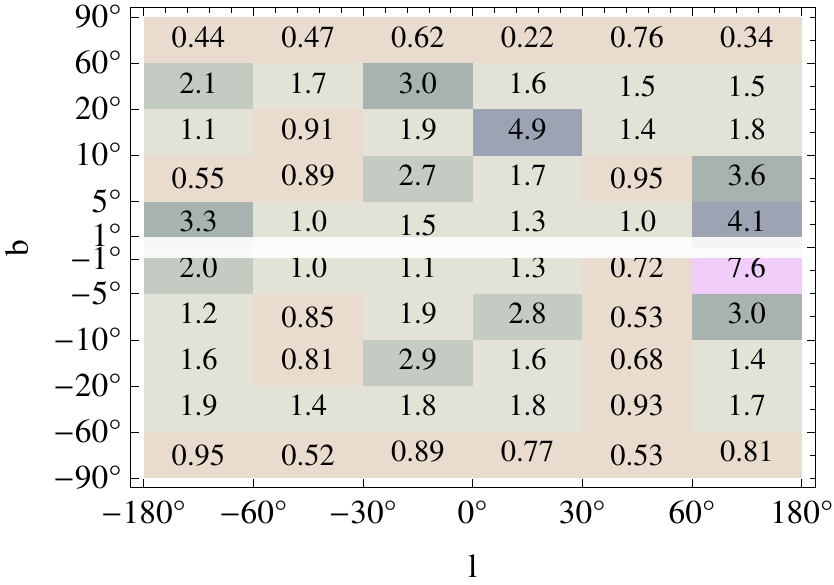}
	\caption{The values of reduced $\chi^2$ of the computed $\gamma$-ray sky-maps confronted with the \textit{Fermi} data for three propagation models $z_d=2$ kpc \textit{(top left)}, $z_d=4$ kpc \textit{(top right)} and $z_d=10$ kpc \textit{(bottom left)}, vs. the latitude $b$ and longitude $l$.  We mask out the region of $\mid b \mid < 1^{\circ}$, where in the interstellar gas has the greatest model dependence.
The predicted background fluxes typically underestimate the \textit{Fermi} data, for thicker models giving however an overall good agreement. Additionally, the \textit{bottom right} panel shows the $z_d=10$ kpc model with included component of the Wino DM with $m_\c=2.5$ TeV. In all cases we allow the gas normalization to be free within a factor of 2 and treat the exposure uncertainty an a nuisance parameter. The impact of the energy resolution on the observed $\gamma$-ray fluxes has been included as well, treated as a nuisance parameter in the fit of the full 4$\pi$ sky and fixed for the remaining analysis.}
    \label{fig:skymaps}
\end{figure}

In order to obtain constraints on the DM component one needs to understand the astrophysical backgrounds first. We follow the approach of \cite{Cholis:2011un} and use DRAGON to compute the galactic diffuse $\gamma$-ray spectra coming from CRs produced and accelerated in astrophysical sources and propagated in the Galaxy, for different assumptions on the galactic properties (using some reference propagation models).
 Next, we confront the results with the \textit{Fermi} data, to see how well the computed background fits the obtained $\gamma$-ray fluxes in all regions of the sky. The obtained $\gamma$-ray sky-maps for three benchmark models $z_d=2,\,4,\,10$ kpc are illustrated on Fig.~\ref{fig:skymaps}.

At very high latitudes, an overall good agreement with the data is seen for all the propagation models shown. 
At these latitudes, the extragalactic and the local galactic diffuse contributions dominate. In these tests of compatibility with the $\gamma$-ray data,  for the former, we use the measurement  of the isotropic $\gamma$-ray flux from \cite{Abdo:2010nz} which has been based on the high latitudes. With regards to the galactic diffuse components, high latitudes probe the interactions of (local) CRs with the (local) interstellar medium gas and the (local) interstellar radiation field; all of which are best understood locally.   Thus in our method, where we first fit the (locally measured) CR spectra, all our calculations are more robust  at high latitudes (where there can be  less unaccounted for d.o.f. due to different possible extrapolations and assumptions). Any model for the CR propagation or interstellar medium properties in the Galaxy that reproduces CR measurements but fails to have agreement with high latitude $\gamma$-ray observations, can be excluded as an actual model for the Galaxy with our method \cite{Cholis:2011un, Tavakoli:2011wz}.   
When decreasing the latitude, one sees that the thicker the propagation model, the better the fit. This is a consequence of the fact, that the observed disagreement with the data comes from underestimating the fluxes of diffuse $\gamma$-rays. We include in these fits the contribution from the Fermi Bubbles/Haze \cite{Dobler:2009xz,Su:2010qj,Dobler:2011mk} and Loop I \cite{Su:2010qj}, but we ignore the contribution from the Northern Arc \cite{Su:2010qj} for simplicity. Also we mask out the region of $|b|<1^{\circ}$ to minimize the impact of the "dark gas" component which is most important at the $|b|<10^{\circ}$ \cite{Tavakoli:2013zva} and which we ignore in the shown windows.

Thicker diffusion zones have a larger diffusion  coefficient $D_0$ (see Tab.~\ref{tab:propagation}), making electrons diffuse faster away from the galactic disk, giving a higher ICS contribution at $|b|>10^\circ$. That results in models with larger $z_d$ having a better agreement with the data at high latitudes. As a consequence of the faster diffusion, towards the inner few degrees in $|b|$, thicker zones give a smaller ICS contribution,  leading in a slightly worse agreement with the data (see  $|b|<5^\circ$ for the $z_d=2$ kpc vs the $z_d=10$ kpc  model).
Yet, on average the thicker zones are in better agreement with the data, with the thinner propagation models being disfavored, by under-predicting the diffuse flux especially at higher latitudes.

On the bottom right plot of Fig.~\ref{fig:skymaps}, we additionally show how the fit changes when adding the dark matter contribution, chosen to be the $m_\c=2.5$ TeV case, with the $z_d=10$ kpc model. Comparing it to the bottom left plot with background only, we observe that in most of the regions adding the DM does not alter the fit much. The biggest change is in lower latitudes (up to $20^{\circ}$), which stems from the fact that the main DM contribution comes from prompt $\gamma$-rays, while the ICS of DM origin is subdominant.\footnote{This is because the portion of energy  from the DM annihilations that goes to high energy electrons is significantly smaller that that going to high energy $\gamma$-rays (by a factor of at least 3 for injected energies above 10 GeV). This effect is seen mostly in lower latitudes, that probe regions where the annihilation rate from the DM halo is larger.} On average, including this particular dark matter model, makes the fits slightly worse. This can be however easily compensated by taking slightly thinner diffusion zone. Yet we clarify that the windows in Fig.~\ref{fig:skymaps} were chosen to study the goodness of fit of different assumptions on the galactic diffuse background and are not optimized for a possible DM signal. 

 \begin{figure}
\centering
\includegraphics[scale=0.35]{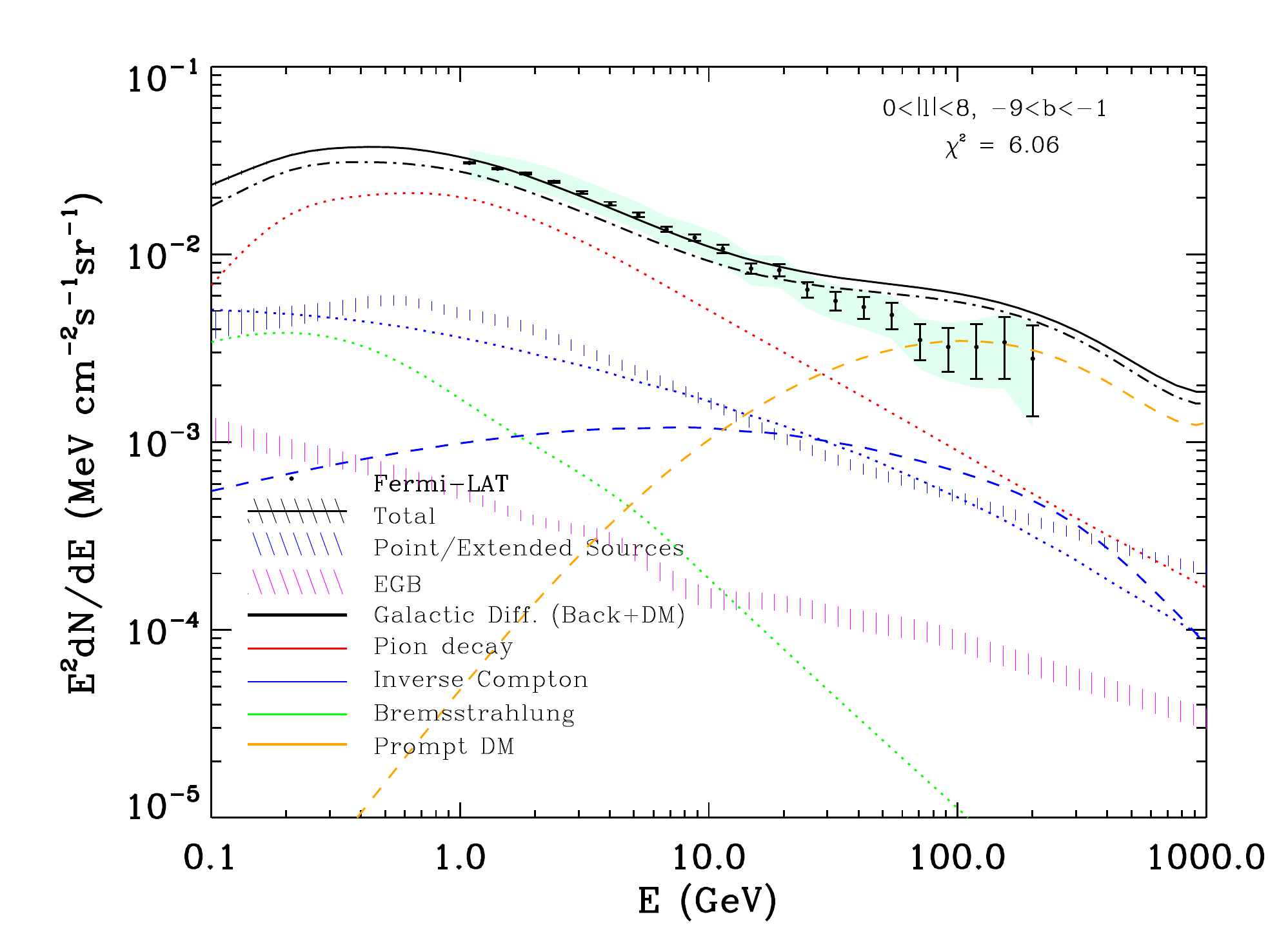}
\includegraphics[scale=0.35]{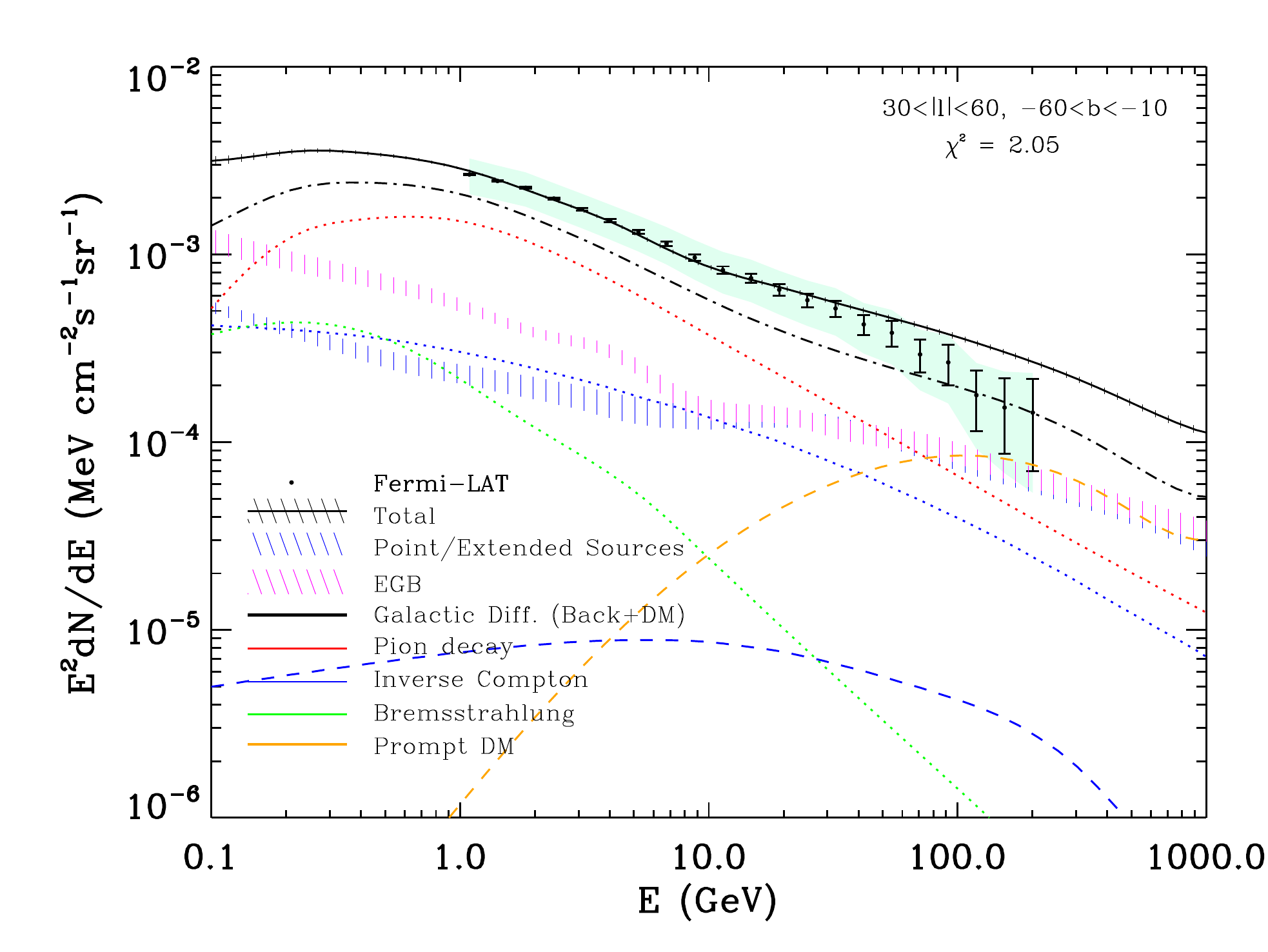}
\includegraphics[scale=0.35]{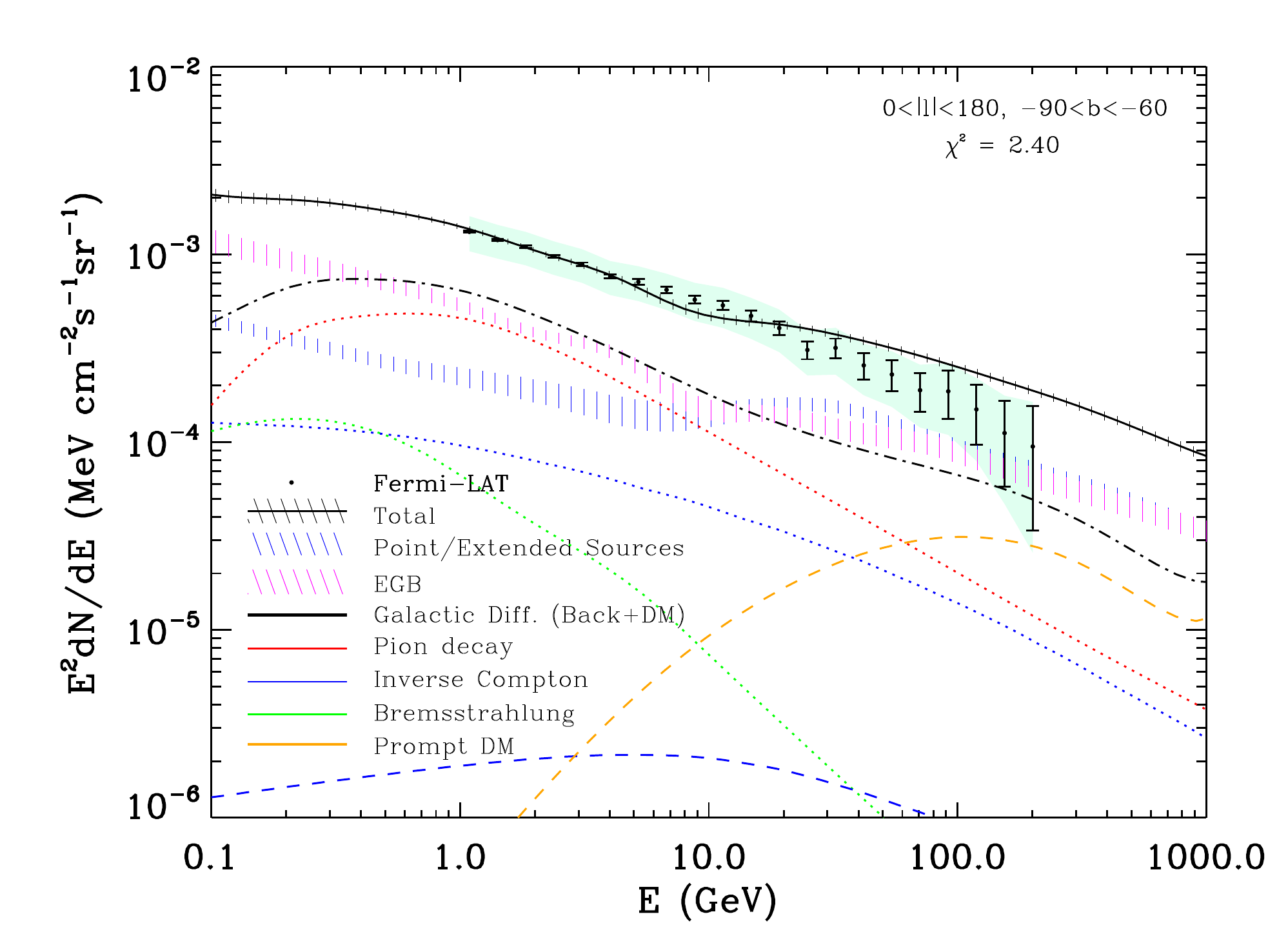}
\includegraphics[scale=0.35]{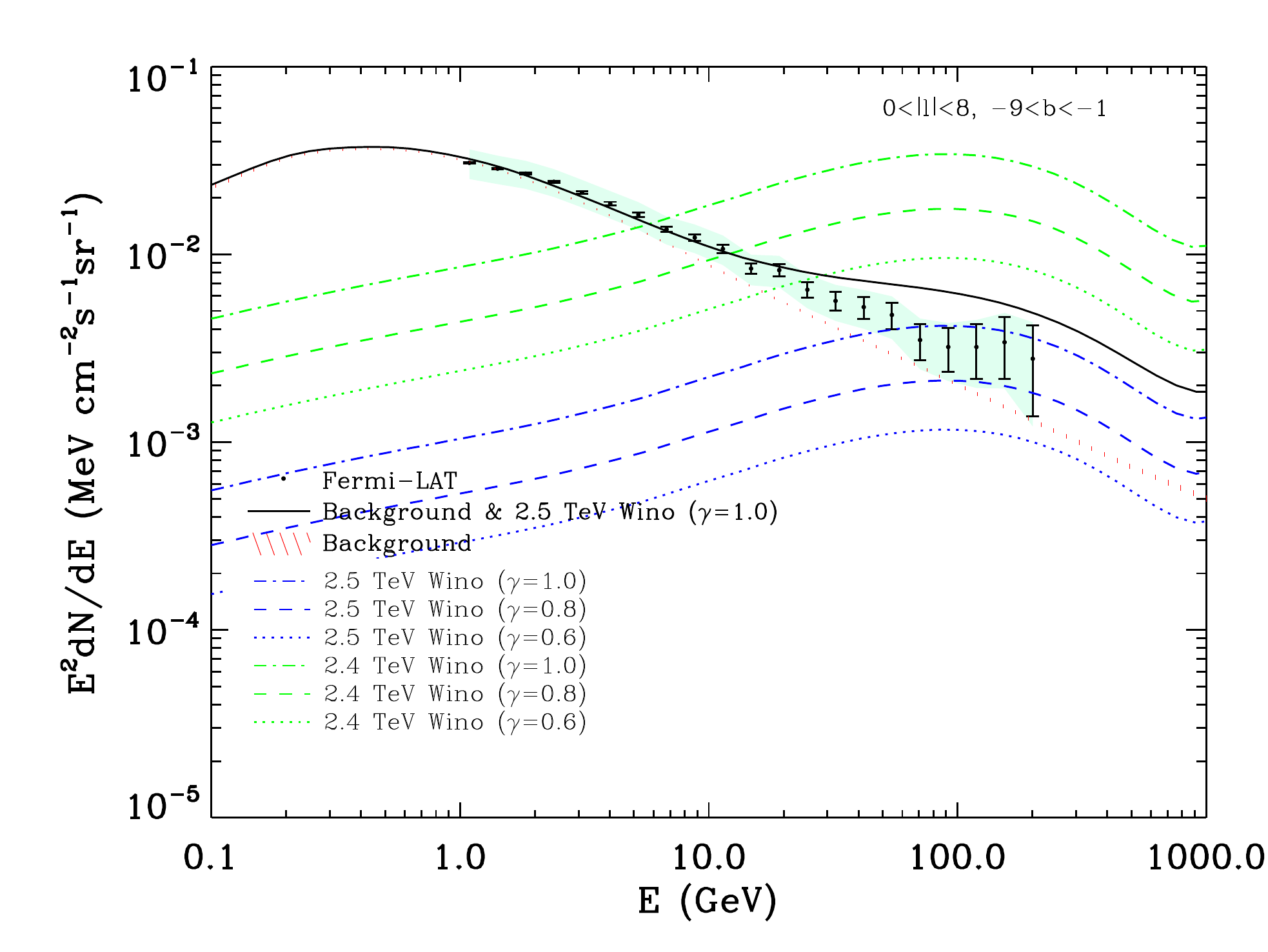} 
	\caption{The $\gamma$-ray fluxes with highlighted all the included contributions, for three chosen windows $0^\circ<|l|<8^\circ$ and $-9^\circ<b<-1^\circ$ \textit{(top left)}, $30^\circ<|l|<60^\circ$ and $-60^\circ<b<-10^\circ$ \textit{(top right)}, $0^\circ<|l|<180^\circ$ and $-90^\circ<b<-60^\circ$ \textit{(bottom left)}.  We use the $z_d=1.4$ kpc propagation model. The astrophysical background (dotted lines for the diffuse) is typically slightly below the data, an effect which is more pronounced for thin diffusion zones. The Wino DM of $m_\c=2.5\;$TeV contribution is given by a dashed orange (prompt) and blue (ICS) lines. Everything, except the point/extended sources and extragalactic background (EGB), is computed with the DRAGON code. The sum of all (background $\&$ DM) diffuse contributions gives the dashed-dotted black line. The total $\gamma$-ray flux, from dark matter and all the backgrounds, is given by the black dashed region, with the black solid line giving the theoretical mean value for reference. The light green band gives the observed $\gamma$-ray flux within a region defined by the statistical and systematic (due to energy resolution and exposure) uncertainties, summed in quadrature. The black error-bars give only the statistical errors, with the systematic uncertainties accounted for in the fit as nuisance parameters, impacting the mean values of the plotted data points. In calculating the $\chi^2$ per d.o.f. of the fit, we added in quadrature the statistical and theoretical errors. The theoretical errors are shown by the black dashed band (sum in quadrature of the EGB and point/extended sources uncertainties). \textit{Bottom right:} the DM total $\gamma$-ray flux close to the GC for $m_\c=2.4\;$TeV (blue) and $2.5\;$TeV (green) with varying $\gamma$ parameter of the generalized NFW profile.}
    \label{fig:gammazd1m2500}
\end{figure}

In order to check systematically the impact of different diffusion zone scale-heights and other galactic propagation assumptions, we concentrate in three windows in the southern sky: $0^\circ<|l|<8^\circ$ and $-9^\circ<b<-1^\circ$, $30^\circ<|l|<60^\circ$ and $-60^\circ<b<-10^\circ$ and also $0^\circ<|l|<180^\circ$  and $-90^\circ<b<-60^\circ$. The window closest to the galactic disk, is chosen to optimize the limits on DM from diffuse $\gamma$-rays while the other two regions probe the DM contribution at intermediate latitudes and close to the south galactic pole. The impact of DM substructures is not included. For these windows we compute diffusive $\gamma$-ray fluxes for all the benchmark propagation models, including both the background and the dark matter contributions. The latter does not depend significantly on the propagation model, since as discussed above, the prompt $\gamma$-ray component is the dominant one. 
The total fluxes for $z_d=1.4\;$kpc and reference assumptions on the ISRF  are presented on Fig.~\ref{fig:gammazd1m2500}, with highlighted all the individual contributions. 
As we can see the case of a 2.5 TeV Wino DM is already excluded by the $\gamma$-ray flux measurement at $0^\circ<|l|<8^\circ$, $-9^\circ<b<-1^\circ$. 
In fact the entire mass range of $m_{\chi}< 0.75$ TeV and 1.9-2.7 TeV is also excluded at 95 $\%$ CL for the  $z_{d} = 1.4$ kpc case. At higher latitudes this mass range can also be excluded (see section~\ref{sec:HighLat}).

In our exclusion result for $z_{d} = 1$ kpc we have already included the possible uncertainty in the interstellar medium gas density along the observed line of sight  (LOS) and angles. These uncertainties can come from the exact CO to H2 relation ($X_{CO}$) assumed to convert CO column density maps to 3-dimensional H2 galactic density models \cite{Timur:2011vv}. An other uncertainty, is the exact assumptions  in converting the Doppler velocity shift to position in the Galaxy . To account for these uncertainties we allow for the total ISM gas column density along our line of sight to be free within a factor of 2, impacting by the same factor our $\pi^0$ and bremsstrahlung diffuse components. 
Allowing for different diffusion zone thickness does not impact much our results (see Table~\ref{tab:gamma}), since in the  $0^\circ<|l|<8^\circ$, $-9^\circ<b<-1^\circ$ window the prompt component (independent of CR diffusion assumptions) is the major component above 10 GeV (Figure~\ref{fig:gammazd1m2500}). 
The only other Wino DM $\gamma$-ray component that contributes significantly to the observed fluxes is the ICS which depends on the assumptions made on the low energy target photons emitted/absorbed/scattered by stars/dust and from the CMB.
We test alternative assumptions on the galactic (non-CMB) component of these photons using the reference models of \cite{Tavakoli:2013zva} (see their Appendix B). Our basic result does not change as shown in Table~\ref{tab:gamma}; the 2.5 TeV Wino is excluded at high significance with the mass range of $m_{\chi}< 0.6$ TeV and  2.0-2.5 TeV excluded at 95 $\%$ CL in all background choices.  

\begin{table}
\centering
\small{
\begin{tabular}{|c|c c c|}
\cline{ 1- 4}
 $z_d$ & $0^\circ<|l|<8^\circ$ & $30^\circ<|l|<60^\circ$ & $0^\circ<|l|<180^\circ$ \\
$\rm \left[kpc\right]$ & $-9^\circ<b<-1^\circ$  & $-60^\circ<b<-10^\circ$ & $-90^\circ<b<-60^\circ$ \\ \hline

 1.0                  &  6.70    &    2.54      &      2.49    \\
 1.4                  &  6.06    &    2.05      &      2.40    \\  
 1.7                  &  6.46    &    2.25      &      2.45    \\ 
 2.0                  &  6.77    &    2.44      &      2.50    \\
 3.0                  &  6.86    &    2.37      &      2.49    \\ 
 4.0                  &  6.80    &    2.28      &      2.47    \\  
 6.0                  &  6.76    &    2.26      &      2.47    \\ 
 8.0                  &  6.76    &    2.25      &      2.47    \\ 
 10                   &  7.00    &    2.37      &      2.49    \\ 
 15                   &  6.97    &    2.35      &      2.49    \\ 
 20                   &  6.74    &    2.23      &      2.46    \\ 
\hline
MaxBulge   ($z_d = 1.4$ kpc)  &   5.65      &   2.05      &      2.40    \\
MinBulge    ($z_d = 1.4$ kpc)  &   6.56      &   2.05      &      2.40   \\            
MaxDisk      ($z_d = 1.4$ kpc)  &   5.94      &   2.07      &      2.40    \\
MinDisk       ($z_d = 1.4$ kpc)  &   6.20      &   2.01      &      2.39    \\
MaxMetalG ($z_d = 1.4$ kpc)  &   5.66      &   2.03      &      2.39    \\
MinMetalG  ($z_d = 1.4$ kpc)  &   6.04      &   2.05      &      2.40    \\\hline
\end{tabular}
}
\caption{The values of the reduced $\chi^2$ for the selected windows. In all propagation models the Wino DM contribution of $m_\c=2.5$ TeV was included. As a reference, we use the case of the $z_d=1.4$ kpc propagation model, and vary the assumptions on the interstellar radiation field, that impacts both the background ICS and that from Wino annihilations.}
\label{tab:gamma}
\end{table}

These results help us put more stringent and robust limits on the Wino dark matter model. As we have shown in section~\ref{sec:antiprotons}, masses close to the resonance were marginally consistent with antiproton measurements, if and only if the diffusion zone was very thin. Including the $\gamma$-ray constraints this possibility is completely ruled out due to disagreement with low latitude $\gamma$-ray fluxes. We have tested that the 2.5 TeV Wino DM combined with a thin diffusion zone is excluded by the $\gamma$-ray data for a wide variation of galactic CR propagation and galactic diffuse background conditions. The excluded region of heavy Wino masses is always $m_{\chi}< 0.65$ TeV and 2.0-2.65 TeV at 95$\%$ CL. 

On the bottom right plot of Fig.~\ref{fig:gammazd1m2500} we also show the results for the GC for the two Wino masses $m_\c=2.4\;$TeV (blue) and $2.5\;$TeV (green) for several different choices of the DM distribution. We have used the generalized NFW profile and varied the value of $\gamma$ parameter from 0.6 to 1. The resulting difference in the $\gamma$-ray fluxes vary up to a factor of 4. As is clear the case of the 2.4 TeV wino is already excluded for all the assumed DM profiles. In fact our limits come from a region that does not include the inner $1^{\circ}$, thus the exact profile assumptions in the inner 150 pc of the Galaxy's DM halo are irrelevant. 
The 2.5 TeV case, is less constrained, and allowed for profiles with inner slope $\gamma <$0.8.

\subsection{Limits from the high latitude $\gamma$-ray spectra}
\label{sec:HighLat}

Up to this point, we have been discussing limits on Wino DM, that can be derived from $\gamma$-rays produced in annihilations 
taking place predominantly in the smooth galactic halo of our Galaxy. Annihilations taking place in distant galaxies, galaxy clusters, 
as well as in small substructures bound in the Milky Way's potential, are expected to give an isotropic signal and can contribute significantly
at  high latitudes. In addition, annihilations in the smooth halo do contribute also at high latitudes and have to be included.

In Table~\ref{tab:gamma}, we present the goodness of fit that specific combinations of CR propagation models for the galactic diffuse 
emission together with the diffuse emission signal from a 2.5 TeV Wino, have for the region of $\mid l \mid <180^{\circ}$, 
$-90^{\circ}<b<-60^{\circ}$. Turning the argument around, the relevant 95$\%$ CL  upper limits get to be at most 8.5$\times 10^{-25}\;$cm$^{3}$s$^{-1}$, 
excluding at least the $m_{\chi}< 0.65$ TeV and  2.04-2.53 TeV mass range. In that case though, only the annihilations taking place in the smooth halo are taken into account.  

The \textit{Fermi}-LAT collaboration, has released measurements of the isotropic $\gamma$-ray background \cite{Abdo:2010nz},
allowing the study of those signals \cite{Abdo:2010dk, Hutsi:2010ai, Fornasa:2012gu, Ando:2013ff}. There is a variety of $\gamma$-ray
sources that have been shown to contribute to the isotropic background, among which, BL Lacertae, Flat Spectrum Radio Quasars, 
starforming and starburst galaxies, radio galaxies and ultra high energy cosmic rays cascading in the intergalactic medium. These 
components have their uncertainties both with respect to the spectral indices and the individual flux contributions. 
 Those uncertainties have to be taken into account when deriving DM limits. Such studies have been carried out in some works as in
 \cite{Donato:2011pe, Calore:2013yia, Cholis:2013ena}. In all cases the non-DM components account for the major part of the isotropic flux (with wide uncertainties though).  
 
 Using \cite{Cholis:2013ena} as a guide, we get that for a 2.5 TeV Wino DM the current 95 $\%$ upper limit on $\sigma$v is 8$\times 10^{-25}$
 cm$^{3}$s$^{-1}$ or that the mass range of $m_{\chi}< 0.75$ TeV and 1.92 to 2.7 TeV is excluded already from the data of \cite{Abdo:2010nz}. Ignoring substructures in the 
 extragalactic signal, the 95$\%$ C.L. upper limit on $\sigma$v is 3$\times 10^{-24}\;$cm$^{3}$s$^{-1}$ (the excluded mass range reduces to 2.12-2.52 TeV).
 In both cases, masses close to the resonance are already excluded just by the current isotropic spectrum. At 10 years of data 
 collection from the \textit{Fermi}-LAT collaboration, the properties of many of the non-DM sources contributing to the isotropic signal, 
 will be better understood and together with higher statistics will provide even tighter limits on DM annihilation \cite{Cholis:2013ena}.
 For a 2.5 TeV Wino, the expected 10 yr limit is 1.5$\times 10^{-25}$ (6$\times10^{-25}$) cm$^{3}$s$^{-1}$ including (ignoring) DM
 substructures. Discussing the Wino mass range probed, the entire mass range (the $m_{\chi}< 0.85$ TeV and 1.85-2.85 TeV mass range) will be probed including 
 (ignoring) extragalactic DM substructures.
    
\subsection{Galaxy clusters}
\label{sec:Clusters}

Galaxy clusters due to their high concentrations of DM, can provide an alternative target for searches of DM annihilation signals in 
$\gamma$-rays \cite{Colafrancesco:2005ji, Baltz:2008wd, Jeltema:2008vu, Pinzke:2009cp}. 
For TeV scale Wino masses, the prompt component has a broad peak at $\sim 100$ GeV (see Fig.~\ref{fig:gammazd1m2500}).
In addition to such a possible DM signal in $\gamma$-rays, galaxy clusters have intergalactic atomic and molecular gas which act as targets for the production of $\gamma$-rays of non DM origin.
These $\gamma$-rays are, like in the case of our own Galaxy,  produced via $\pi^{0}$ decays and bremsstrahlung radiation.
The emissivity of $\pi^{0}$ and bremsstrahlung backgrounds is proportional to the product of the cosmic ray proton or 
electron densities (produced in galaxies), with the atomic and molecular gas densities that exist both inside the galaxies and in 
the intergalactic medium.
There is also, another contribution to $\gamma$-rays from inverse Compton scattering, which may be the most difficult to evaluate 
because of uncertainties in the radiation field in galaxy clusters.
The exact assumptions on the gas distribution at the galaxy cluster and its member galaxies, the radiation field and the CR production and propagation, 
can considerably affect the background contribution between gas rich and gas poor clusters and thus the 
induced limits on DM annihilation. 
Moreover, the $\gamma$-ray radiation from the isotropic and the diffuse Galactic components and from uncorrelated point 
sources, which lay along the same
line of sight, dominate any signal at $\gamma$-rays from these targets.

Finally, the assumptions on distribution of DM substructures and the evolution of the dark matter profile in galaxy 
clusters, can significantly change (factors of up to $10^{3}$) the DM limits from some of these targets, in addition to changing the morphology on the sky of the DM signal (for recent analysis see 
\cite{Pinzke:2011ek, Ando:2012vu, Han:2012au}).

The closest galaxy clusters have a typical extension radius of a few degrees in the sky.
In such small windows very few photons above 100 GeV are observed by the \textit{Fermi}-LAT instrument.
This makes the analysis of the spectra from each one of those targets statistics limited. Stacked analyses of multiple targets, deal with this 
limitation, but introduce systematic limitations associated to the fact that each of the galaxy clusters has their own backgrounds and substructure distribution, that in such analyses are typically not handled as separate degrees of freedom.  

In the analysis of \cite{Han:2012au} and \cite{Hektor:2012kc}  possible signals of DM annihilations have been suggested from 
$\gamma$-ray observations toward nearby galaxy clusters, while \cite{Aharonian:2009bc, Aleksic:2009ir, Ackermann:2010rg, Dugger:2010ys, Zimmer:2011vy, Huang:2011xr} have seen no evident $\gamma$-ray excess and placed only limits on DM annihilations. Yet, even the current limits can not exclude our Wino models at the few TeV mass range. 

\subsection{Dwarf Spheroidals}
\label{sec:dSph}

Dwarf spheroidal galaxies (dSphs) are small dark matter dominated galaxies with typical luminosities $O(10^{7})$ L$_{\odot}$ (see e.g. \cite{Walker:2012td} for a review). They have suppressed star formation rates and low gas densities and due to their smaller sizes the escape timescales of CRs
produced in them are also significantly smaller than in the Milky Way. Therefore, the production of $\gamma$-rays from point sources and interactions between CRs and their local medium is expected to be suppressed. Thanks to these properties, the dSphs provide some of the best targets to look for signals from DM annihilation \cite{Evans:2003sc, Colafrancesco:2006he, Strigari:2006rd, Bovy:2009zs, Scott:2009jn, Perelstein:2010at}. First results from \textit{Fermi} collaboration suggested that there was no clear excess of $\gamma$-rays between 200 MeV and 100 GeV, above the expected background, towards all known dwarf spheroidal galaxies \cite{Abdo:2010ex,Ackermann:2011wa}. 
More recently, some mild excess in $\gamma$-rays at the GeV range has been suggested \cite{Ackermann:2013yva}. 
Therefore, we can ask how strong limits on our DM models, can we obtain from the current data.

Here we will closely follow the approach of Ref. \cite{Cholis:2012ve} and provide limits coming from the two of the most 
background$+$foreground emission
clean targets: the Ursa Minor and Sextans. 
Following  a conservative approach we choose these two dwarf galaxies for two reasons; first, a relatively large amount of kinematical data to extract the $J$-factors, is available for both of them \cite{Salucci:2011ee}, and second we can properly account for the galactic foreground and extragalactic background $\gamma$-ray contributions  \cite{Cholis:2012ve}. Thus we have under control, the main sources of uncertainty in this channel of indirect DM searches.
Before presenting our limits, we summarize here some of the main features for Ursa Minor that provides the tightest limits among the two dSphs.  The wide  
field photometry analysis in \cite{carrera2002} finds that Ursa Minor hosts a predominantly old  
stellar population, with virtually all the stars formed before 10 Gyr ago, and 90$\%$ of them  formed before 13 Gyr ago,
making it the only dSph Milky Way satellite hosting a pure old stellar population.
Using the magnitude of the horizontal branch stars and comparing with Hipparcos data on globular clusters, \cite{carrera2002} 
determine the distance of Ursa Minor from the Sun to be  $76 \pm 4$~kpc, which, given its galactic coordinates  
$(l,b)$=($104.95^{\circ}$,$44,80^{\circ}$), translates into a distance from the Galactic center of about 78~kpc.\footnote{The quoted value for the heliocentric distance, which is in agreement with the determination from~\cite{Bellazzini2002},  
is larger than the mid 1980's value of $66 \pm 3$~kpc from \cite{Olszewski1985,Cudworth1986} often adopted even in the recent
literature; as pointed out in~\cite{Piatek2005}, the difference is mainly due to the absolute magnitude calibration of the horizontal  
branch, so \cite{Piatek2005} suggests to the more modern determination; standing in between are the values of $70 \pm 9$~kpc~\cite{Nemec1988} and $69 \pm 4$~kpc~\cite{Mighell1999}.} Moreover, Ursa Minor is
the "classical" dwarf with the shortest distance to the Galactic center  except for Sagittarius, and is currently close to the 
apocenter  on a rather eccentric orbit, with its pericenter estimated in~\cite{lux2010} to be at $40\pm20$~kpc or $30\pm10$~kpc 
depending on the model for the Milky Way potential well.  It is then likely, that Ursa Minor has been affected by tides. 
Using N-body simulation,~\cite{munoz2008} suggests that, analogously to Carina, 
Ursa Minor may be an object  that, starting out as embedded in a much more extended dark matter halo, has then suffered a 
mass loss until the luminous component is exposed to tidal effects, and, while still having very  
large mass to light ratio, can be modeled as  a "mass-follow-light" object.
Finally we note that the stellar surface brightness of Ursa Minor has the largest  ellipticity among all classical dwarfs (excluding again Sagittarius  
that is suffering heavy tidal disruption), with mean value of $\epsilon \equiv 1-b/a$ (where $b/a$ is the minor  over major axis ratio)
estimated in~\cite{Irwin1995} to be $0.56\pm0.05$. Nevertheless, most analyses (including ours) treat Ursa Minor as a spherically symmetric system. This is a simplification on the DM profile, for which the accurate triaxial treatment, we leave for future work. 

The method for obtaining the limits is the same as in \cite{Cholis:2012ve}, from where we adopted also the values for the $J$-
factors, definitions of regions of interest which are used for estimating the signal and the treatment of subtracting the galactic 
foreground and extragalactic background. Referring for all the details to this paper, here we will just give the results for the Wino 
model. 

\begin{table}
\centering
\begin{tabular}{|c|c c c|c c c|}
\hline
\multicolumn{1}{|l|}{} & \multicolumn{ 3}{c|}{Ursa Minor} & \multicolumn{ 3}{c|}{Sextans} \\ \hline
\multicolumn{1}{|l|}{Mass [GeV]} & \multicolumn{1}{l}{90\% CL} & \multicolumn{1}{l}{ 95\% CL} & \multicolumn{1}{l|}{99.9\% CL} & \multicolumn{1}{l}{90\% CL} & \multicolumn{1}{l}{ 95\% CL} & \multicolumn{1}{l|}{99.9\% CL} \\ \hline
500 & 0.66 & 0.94 & 2.13 & 1.52 & 2.02 & 4.02 \\ 
1000 & 3.19 & 4.77 & 11.8 & 7.17 & 9.67 & 19.7 \\
2360 & 0.020 & 0.031 & 0.079 & 0.043 & 0.059 & 0.123 \\
2400 & 0.033 & 0.051 & 0.132 & 0.071 & 0.097 & 0.203 \\ 
2500 & 0.55 & 0.83 & 2.14 & 1.16 & 1.59 & 3.31 \\ 
2700 & 3.69 & 5.61 & 14.46 & 7.84 & 10.7 & 22.4 \\
3200 & 14.7 & 22.4 & 58.0 & 30.4 & 41.9 & 87.9 \\ \hline
\end{tabular}
\caption{The upper limits on the boost factors coming from two dSphs: Ursa Minor and Sextans. Values of BFs smaller than one suggest the model is excluded.}
\label{tab:dsph}
\end{table}

\begin{figure}
\centering
\includegraphics[width=3.10in,angle=0]{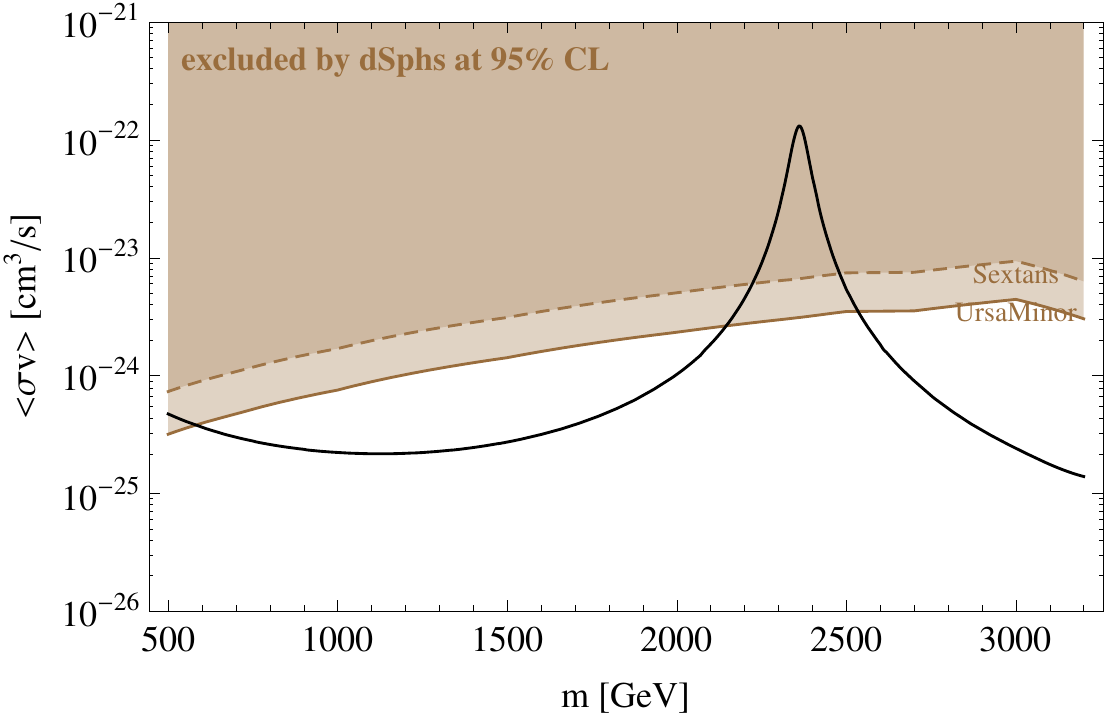}
\caption{ 95\% CL upper limits on the total annihilation cross-section. The limits are derived individually from Ursa Minor and from Sextans dwarf spheroidal galaxies, using the \textit{Fermi}-LAT data (see text for details).}
\label{fig:dsphs}
\end{figure}

After computing the $\gamma$-ray flux for a given mass of the Wino, we confront it with the data collected by \textit{Fermi}-LAT between August 2008 and December 2013, with energies between 1 and 100 GeV. From this we compute what is the additional boost factor (BF) i.e. the multiplication factor on the cross section for each model, that is allowed by the observed residual $\gamma$-ray spectrum at the given confidence levels (CL). Our results are summarized in Tab.~\ref{tab:dsph}, where three different confidence levels are presented; and in Fig.~\ref{fig:dsphs} where we show the 95
$\%$ CL from Ursa Minor and Sextans. Again the results show that only masses very close to the resonance, $m_\c=2.4$ TeV, can be ruled out. When moving a bit further from the resonance the allowed boost factors are getting close to $1$, and effectively all other masses are not constrained in any way by this search channel. 

Note, that the biggest uncertainty here comes from the evaluation of the $J$-factor. The \textit{Fermi} collaboration itself following different assumptions on the uncertainties of the $J$-factors and the modeling of the background$+$foreground emission  has provided limits on several "standard" channels of annihilation by doing a stacked and a source by source analysis \cite{Ackermann:2011wa, Ackermann:2013yva}.
For such an analysis our limits would be for the individual Ursa Minor dSph a factor of $\sim$3 weaker than  those of Tab.~\ref{tab:dsph}. With a joint likelihood analysis though the limits are a factor of 2 stronger, thus confining also the cases of 0.5 and 2.5 TeV mass. Also the authors of \cite{GeringerSameth:2011iw} using a different joint analysis strategy for the Milky Way dSphs, have shown the significance in the uncertainties of the $J$-factors with their weaker limits being a factor of 5 weaker than those of 
Tab.~\ref{tab:dsph}.

\subsection{Limits from a $\gamma$-ray line feature}
\label{sec:GammaLine}

\begin{figure}
\centering
\includegraphics[width=3.10in,angle=0]{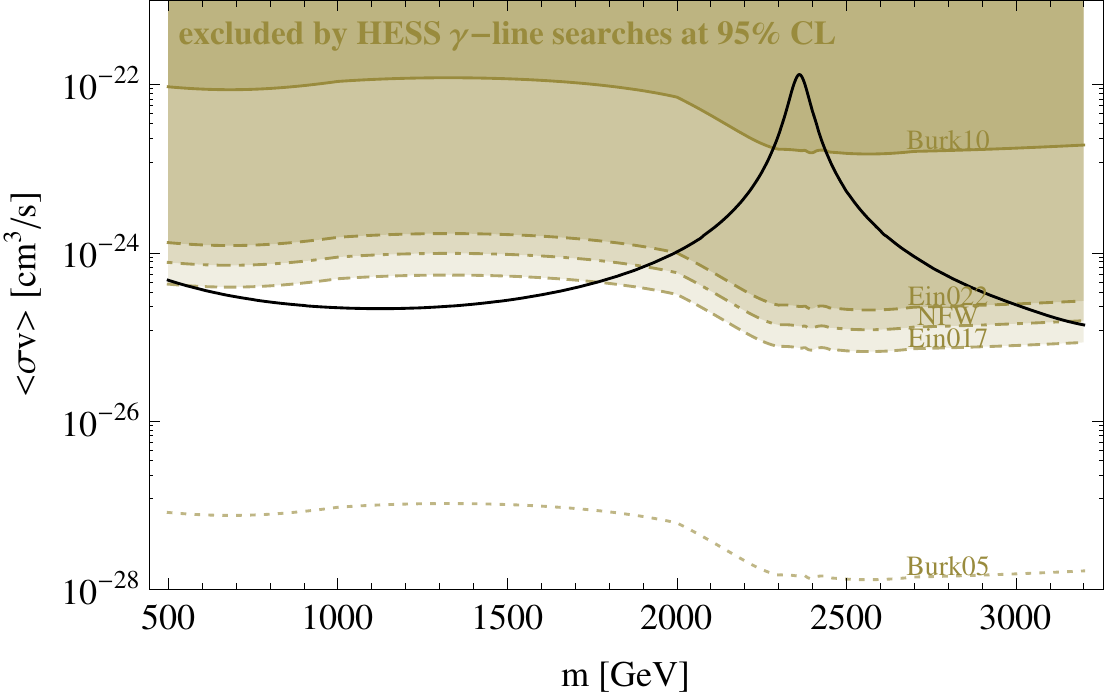}
\caption{95 $\%$ CL upper limits on the total annihilation cross-section. The limits are based on Wino annihilations giving monochromatic $\gamma$-rays: $\chi \chi \longrightarrow \gamma \gamma$, $\gamma Z$ and $\gamma$-rays from internal bremsstrahlung. We use the data from HESS \cite{Abramowski:2013ax} and show five different DM profiles; from top to bottom: Burkert with $R_{B} = 10$ kpc, Einasto with $\alpha = 0.22$, NFW, Einasto with $\alpha = 0.17$ and Burkert with $R_{B} = 0.5$ kpc (see text for details).}
\label{fig:GammaLine}
\end{figure}

A smocking gun signature in $\gamma$-rays, from DM annihilations in the Galaxy is the presence of a line feature toward the galactic center (see \cite{Bringmann:2012ez} for a recent review).
Both \textit{Fermi} and HESS have recently searched for such an excess and published upper limits on monochromatic $\gamma$-ray fluxes \cite{Fermi-LAT:2013uma, Abramowski:2013ax}.
As can be indicated from Fig.~\ref{fig:spectfin}, all the Wino mass range has a branching ratio to monochromatic $\gamma$-rays at the high energy end of the spectrum, allowing for such a search mode. For the case at hand, since we are mostly interested in the masses at the TeV scale, the data of interest are those from HESS.
\cite{Cohen:2013ama, Fan:2013faa}, have recently suggested that the absence of any evident $\gamma$-ray line feature in the HESS data,
can place very strong limits on the Wino DM models in most of the mass range. In Fig.~\ref{fig:GammaLine}, using five different DM profiles for the smooth halo, we show the $95\%$ CL upper limits on the total annihilation cross-section coming from the $\chi \chi \longrightarrow \gamma \gamma$ and  $\chi \chi \longrightarrow \gamma Z$ partial cross-sections and also the internal bremsstrahlung. Given that the HESS energy resolution in the energy range of interest, is $\simeq 15 \%$, the three components can not be discriminated. \footnote{We consider a 15$\%$ energy resolution for the entire mass range of 0.5 to 3.2 TeV.}

The five DM halo profiles that we use, are the standard NFW profile (\ref{eq:NFWprofile}) with $R_{c} = 20$ kpc, the Einasto DM profile \cite{Graham:2006ae}:
\be
\rho(r)=\rho_{0} \exp\left(-\frac{2}{\alpha}\frac{r^{\alpha}-R_{\odot}^{\alpha}}{R_{c}^{\alpha}} \right),
\ee
with a slope $\alpha$ of 0.22 and 0.17, and $R_{c} = 10$ kpc ($R_{\odot} = 8.5$ kpc is the Sun's galacto-centric distance)
and the Burkert DM profile \cite{Burkert:1995yz}:
\be
\rho(r)=\rho_{B} \frac{1}{(1+r/R_{B})}\frac{1}{(1+ (r/R_{B})^{2})},
\ee
with $R_{B} = 10$ and 0.5 kpc. The normalizations $\rho_{0}$, $\rho_{NFW}$ and $\rho_{B}$ are taken such that in all cases the local DM density is 0.4 GeV/cm$^{3}$. 
The two Burkert cases are shown only for direct comparison with the work of \cite{Cohen:2013ama}. Both of them either over-predict or under-predict the included DM mass in the inner 8 kpc of the Milky Way. Especially the case of $R_{B}=0.5$ kpc,  grossly violates bounds from local dynamical measurements, as well as of the inner and outer rotation curve. Moreover, the Burkert  case with $R_{B}=0.5$ kpc core radius is not a cored profile, but in fact a profile with $\rho ( r ) \propto r^{-3}$; thus the term "Burkert" is slightly deceiving. For these reasons we only present it in Fig.~\ref{fig:GammaLine} for reference and we disregard it in our final conclusions. Instead, the NFW and the two Einasto profiles are representative of the DM density in the Galaxy.

As can be seen from Fig.~\ref{fig:GammaLine}, Winos with masses heavier than 2 TeV can be excluded at 95 $\%$ CL from the HESS data. 
Being very conservative, and using the Burkert profile with $R_{B} = 10$ kpc, we can only exclude the mass range around the resonance: 2.25-2.45 TeV.  
Our results are slightly more conservative than those of \cite{Cohen:2013ama}, since we include one-loop contributions to the annihilation cross-section which reduce the partial annihilation cross-sections to the monochromatic $\gamma$s. Yet, we also include the internal bremsstrahlung, which at the HESS energy resolution can not be discriminated from the nearly monochromatic photons from $\chi \chi \longrightarrow \gamma \gamma$, $\gamma Z$.

\subsection{Neutrinos from the Galaxy halo}
\label{sec:neutrinos}

\begin{figure}
\centering
\includegraphics[width=3.10in,angle=0]{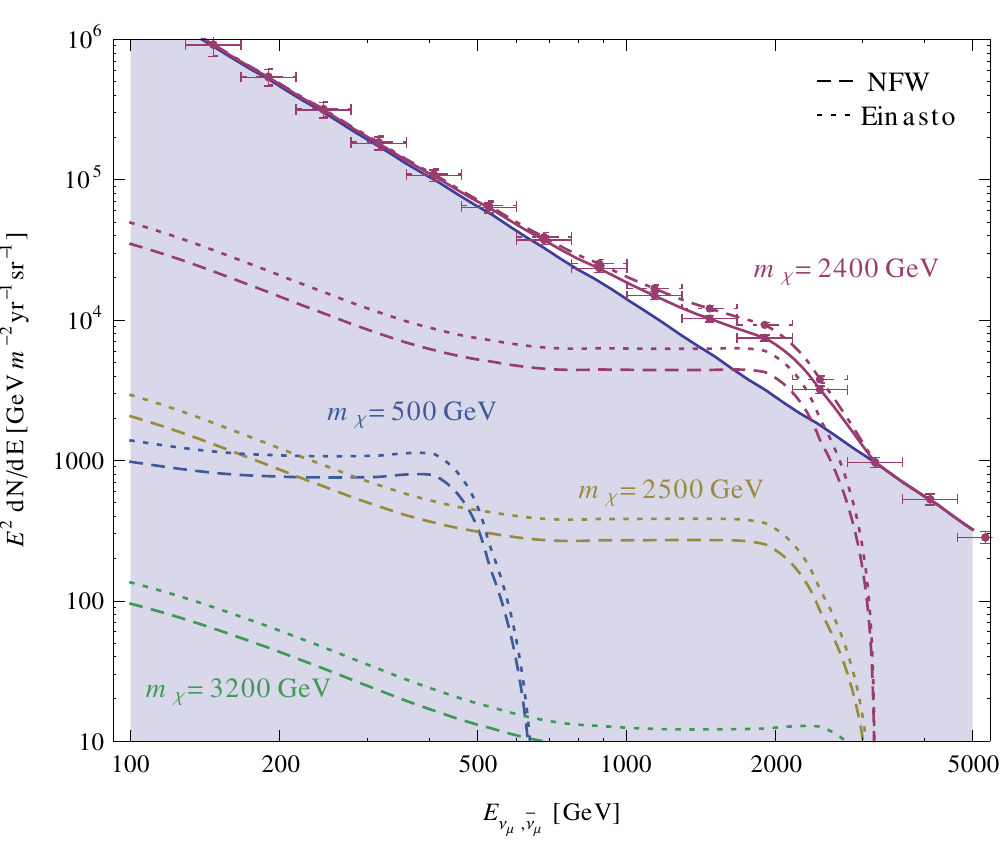}
\caption{KM3NeT simulated $\nu_{\mu}$ upwards going/moving fluxes, from our Wino models. We consider 
3 yrs of data taking and show the fluxes within the window of $\mid l \mid < 5^{\circ}$,
$5^{\circ} < \mid b \mid < 15^{\circ}$. \textit{Continuous power-law line}: atmospheric background. 
\textit{Dashed lines}: NFW DM profile, \textit{dotted lines}: Einasto DM profile. The highest flux comes from 
the 2.4 TeV case while the lowest flux from the 3.2 TeV case.}
\label{fig:NeutrinoSignals}
\end{figure}
An alternative probe to search for indirect signals of DM annihilation is 
high energy neutrinos towards the galactic center (see also \cite{Dasgupta:2012bd} for neutrino signals from galaxy clusters). A signal to look for, is a hardening 
or a "bump" in the spectrum of upward moving neutrino events in km$^{3}$ telescopes.
The background for such events is dominated by the isotropically distributed (over 
long observation time periods) atmospheric $\nu_{\mu}, \bar{\nu}_{\mu}$ flux.
The background spectrum is known to be described by an almost featureless 
power-law with index of $dN_{\nu_{\mu}, \bar{\nu}_{\mu}}/dE \propto E^{-3.7}$.
These events have a much better angular resolution than the shower events, allowing us to 
optimize the analysis by choosing the region of interest with the best signal to background (see \cite{Cholis:2012fr}).
We study only the upwards $\nu_{\mu}, \bar{\nu}_{\mu}$ from DM and backgrounds
 and use for the atmospheric one the parametrization of \cite{Honda:2006qj}. 
Additionally TeV neutrino point sources and diffuse neutrino flux from collisions of 
CR protons with the ISM contribute. Since these components peak at the disk, we 
avoid having our results depending on predictions towards the inner few degrees in latitude.  
Following \cite{Cholis:2012fr} we choose to search for a signal in the region of 
$\mid l \mid < 5^{\circ}$, $5^{\circ} < \mid b \mid < 15^{\circ}$.\footnote{For more on searches of 
DM signals from the GC with km$^{3}$ telescopes and the expected distribution of the 
$\nu_{\mu}, \bar{\nu}_{\mu}$ events on the sky see \cite{Cholis:2012fr}.}

Recently IceCube detected neutrinos at energies above 50 TeV and up to a PeV, at a rate, that is 4.3$\sigma$ 
above the expected atmospheric background \cite{Aartsen:2013jdh}; suggesting detection for the first time, of 
astrophysical neutrinos of galactic or extragalactic origin (see for a review \cite{Anchordoqui:2013dnh}).   
Yet, IceCube does not have the sensitivity towards the GC to distinguish a signal from DM annihilation,
while a km$^{3}$ telescope in the northern hemisphere as KM3NeT will be able to probe such signals.
Using the HOURS simulation \cite{Tsirigotis:2012nr}, for the reconstruction upward going 
$\nu_{\mu}, \bar{\nu}_{\mu}$ we show in Fig.~\ref{fig:NeutrinoSignals} the expected fluxes 
from DM annihilation in the window of $\mid l \mid < 5^{\circ}$, $5^{\circ} < \mid b \mid < 15^{\circ}$.
Error-bars refer to 3 years of collecting data. Given the uncertainties on the DM halo distribution 
we show results for both the NFW and the Einasto profiles (locally normalized to the value of 0.4 GeV cm$^{-3}$
\cite{Catena:2009mf, Salucci:2010qr}).

Among our four reference Wino models only the 2.4 TeV case where the cross section is close to its resonance 
can be observed. We calculate after 3 yrs 1025 atmospheric background events between 
600 GeV and 2.8 TeV, 926 from the Einasto DM profile and 627 from the NFW profile (more than 20$\sigma$ detection).\footnote{Currently, here is a 20-30 $\%$ uncertainty in the normalization of the atmospheric background \cite{Honda:2006qj}. Since this background is also being measured by IceCude, at that energy range, these uncertainties are expected to decrease.}

\subsection{CMB constraints}
\label{sec:Synch}

The CMB temperature and polarization power spectra have been used to place constrains on DM  annihilation models  \cite{Galli:2009zc, Slatyer:2009yq, Lopez-Honorez:2013cua, Galli:2013dna}.
Based on the \textit{WMAP}-5 \cite{Komatsu:2008hk} data, and using \cite{Slatyer:2009yq} as a guide, we can exclude at the 95$\%$ CL only the 
mass range around the resonance, between 2.30 and 2.42 TeV. 
With the \textit{WMAP}-9 \cite{Hinshaw:2012aka} data, the constraints based on \cite{Lopez-Honorez:2013cua} 
exclude at 95$\%$ CL a slightly larger mass range, around the resonance 2.25-2.46 TeV.
While combining \textit{WMAP}-9 with ACT \cite{Sievers:2013ica} or SPT \cite{Story:2012wx} the limits of \cite{Lopez-Honorez:2013cua}, exclude the mass range of 2.18-2.5 TeV. 

Finally the \text{Planck} data are expected to place constraints by up to a factor of 10 tighter than those from the \textit{WMAP}-5 data \cite{Galli:2009zc, Slatyer:2009yq}, with the current constraints \cite{Ade:2013zuv}, being still weak, since they do not include the polarization information at intermediate and high multipoles. 

\subsection{Antideuterons}
\label{sec:IDantid}

Antideutrons have been proposed as a prospective, clean channel for DM searches already many years ago in \cite{Donato:1999gy}. They are produced in the high energy collision of a $p$, $\bar p$ or He impinging onto the interstellar gas (mainly H and He). The production cross section is very low and has a relatively high threshold. For the anitdeuteron to be formed, an impinging particle needs to have an energy (in the rest frame of the gas) $E \geq 17m_p$. For the possible dark matter detection what is even more important is its low binding energy, $B_d\approx 2.2$ MeV. It means that they are easily destroyed and do not propagate long enough to loose most of their energy. This leads to very low background of astrophysical antideuterons with $E_k/n<1$ GeV.

The downside of its rareness is that its cross sections for production, elastic and inelastic scattering are not well known. It is an important source of uncertainty in the predictions of its signals coming from dark matter annihilation. In our work we adopted all these cross sections from the work \cite{Duperray:2005si}, based on fitting the experimental data under some reasonable assumptions.\footnote{For all the details see the original work and also \cite{Donato:2008uq}. We would also like to thank David Maurin for sharing these cross sections.} We then implemented them into DRAGON code \cite{Evoli:2008dv} for the production of secondary and tertiary antideuterons, as well as their propagation in the Galaxy.

The standard treatment of how $\bar d$ is produced, the one which is also implemented in \ds, follows the "coalescence model", see e.g. \cite{Donato:1999gy,Donato:2008uq,Braeuninger:2009bh}.
This approach is based on an assumption that $\bar p$ and $\bar n$ will combine (coalesce) to an $\bar d$, if and only if:
\be
| \vec{k}_{\bar p}-\vec{k}_{\bar n} | \leq p_0,
\ee
where $p_0$ is called \textit{coalescence momentum}. As a rough estimate which gives some intuition, one can obtain this value from $p_0\sim \sqrt{m_d B_d}\sim 60$ MeV, where the mass of deuteron is $m_d=1.8756$ GeV.

The more precise values are derived from experimental data (e.g. formation of $\bar d$ from $e^+e^-$ collisions or hadronic $Z$ decays). Different works used a bit different values: in \cite{Donato:1999gy} $p_0=58$ MeV, which is also the default value \ds, then with the new data it was updated to a value $p_0=79$ MeV \cite{Donato:2008uq}, while independently \cite{Kadastik:2009fk,Cui:2010ud} have found value $p_0=80$ MeV from their own fit to the ALEPH data.\footnote{In fact value used in the two latter works is 160 MeV, but they use a different definition of the normalization of the antideuteron yield by a factor of $2^3$ coming from the change in $p_0$ by factor 2.}

In the spherical approximation, in which the final $\bar p$ and $\bar n$ are distributed uniformly over all $4\pi$, the dependence of the final result on $p_0$ is through an overall normalization factor, being the volume of 3-dim sphere in momentum space with radius $p_0$.
The $\bar d$ yield is given by
\be
\g_d\frac{dN_d}{d^3k_d}=\frac{4\pi p_0^3}{3}\g_n\frac{dN_n}{d^3k_n}\g_p\frac{dN_p}{d^3k_p},
\ee
where the Lorenz factors are approximately equal $\g_d\approx \g_n\approx \g_p$. We have also
\be
d^3k_X=4\pi k_X^2 dk_X=4\pi k_X^2\frac{E_X}{\sqrt{E_X^2-m_X^2}}dE_X,
\ee
where $k_X=\sqrt{E_X^2-m_X^2}$. In order for $\bar d$ to form, the difference in momentum of $\bar p$ and $\bar n$ has to be less then $p_0\ll k_n,k_p$. Thus, the momenta approximately satisfy the relation:
\be
\frac{k_d}{2}\approx k_p\approx k_n \equiv k\;,
\ee
and thus also $E_p\approx E_n\equiv E$. Finally, the antideuteron spectrum can be computed from the proton one via:\footnote{This is based on the assumption that the spectra of antiprotons and antineutrons in a single annihilation are approximately equal. For a more refined approach see \cite{Ibarra:2013qt}.}
\be
\frac{dN_d}{dE_d}=
\frac{4 p_0^3}{3}\frac{\gamma_p}{E\sqrt{E^2-m_p^2}}
\left(\frac{dN_p}{dE} \right)^2\;.
\ee

Apart from the update in value of $p_0$ there is another, much more important difference in the newer works \cite{Kadastik:2009fk,Cui:2010ud,Ibarra:2013qt,Fornengo:2013osa} from the older ones: although they rely on the coalescence model they do not make the spherical approximation, but run a dedicated Monte Carlo codes to compute the $\bar d$ yields. The authors argue, that this is correct way to proceed, in contrary to previous works of \cite{Donato:2008uq,Donato:1999gy,Braeuninger:2009bh}, 
which they claim to be oversimplified. The important phenomenological difference, is the behavior of the flux for higher dark matter masses, where the spherical approximation gives $m_\c^{-2}$ dependence, while Monte Carlo gives a more flat distribution.

The physical reason why the spherical coalescence model is not sufficient is that after dark matter annihilates, final states are very energetic and go in back-to-back jets, rather than distributed over all $4\pi$. Therefore, the $\bar p$ and $\bar n$ are typically produced with much smaller separation angle. This is especially pronounced for high $m_\c$ and explains why the in this regime the difference is the largest.

\begin{figure}[t]
\centering
	\includegraphics[scale=0.7]{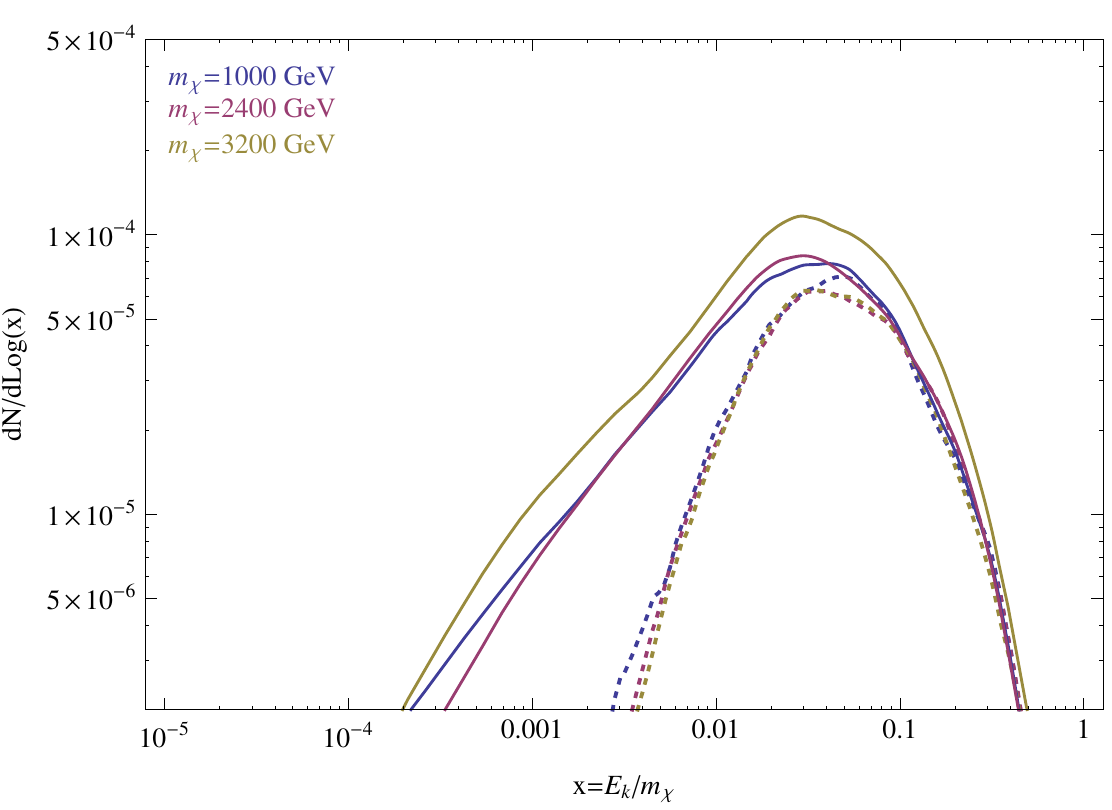}
	\caption{Antideuteron spectra from Wino annihilation for masses $m_\c=1,\,2.4,\,3.2$ TeV. Dotted lines show the tree level results, while solid the Sommerfeld enhanced ones.}
    \label{fig:spectantiD}
\end{figure}

In our work we adopted the results from the Monte Carlo approach of Ref.~\cite{Kadastik:2009fk}, which is publicly  available via the PPPC 4 DM ID code. However, as before, for the electroweak corrections we used our computation and also incorporated the Sommerfeld effect.

We show the antideuterons fluxes at production on Fig.~\ref{fig:spectantiD} for the same set of masses as before. It is clearly visible that the spectra are very similar to each other. As discussed above, this is indeed on contrary to $m_\c^{-2}$ scaling as found in pioneering works on this topic. The main effect of electroweak and Sommerfeld corrections is to increase the soft part of the spectrum, which seems promising for the detection prospects (see Sec.~\ref{sec:IDantid}). Note also, that including the Sommerfeld effect makes the fluxes even larger for larger $m_\c$.

Although the production mechanism is rather well understood, the precise computations are not that well under control. Indeed, very recently authors of \cite{Dal:2012my} showed, that the result is very sensitive to the fragmentation model used in the Monte Carlo codes for the computation of the fluxes at production. In particular, \pyt which uses string fragmentation model gives results generically different by a factor of 2-3  than HERWIG, based on cluster hadronization model. Furthermore, close to the kinematical thresholds this discrepancy grows rapidly. Therefore, one still needs better understanding of the particle physics underlying the antideuteron production to make very robust claims.

\begin{figure}
\centering
	\includegraphics[scale=0.65]{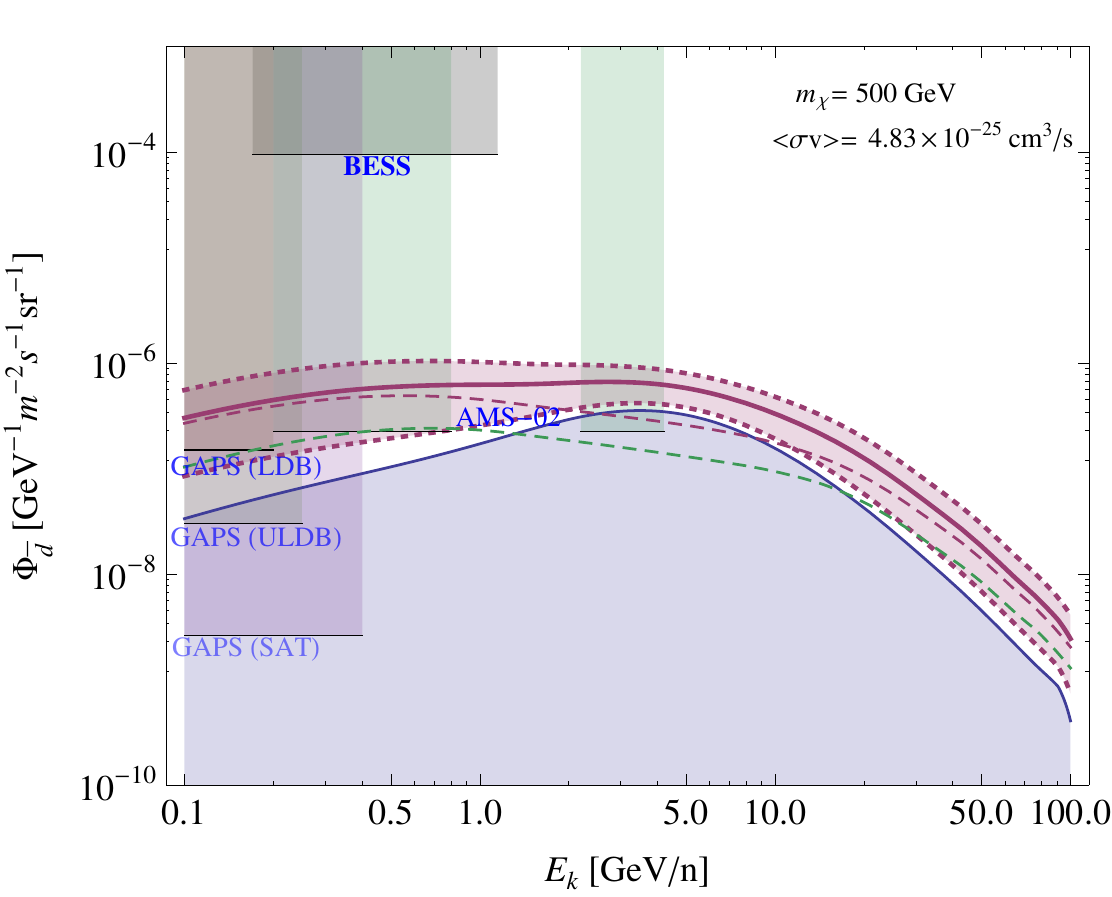}
	\includegraphics[scale=0.65]{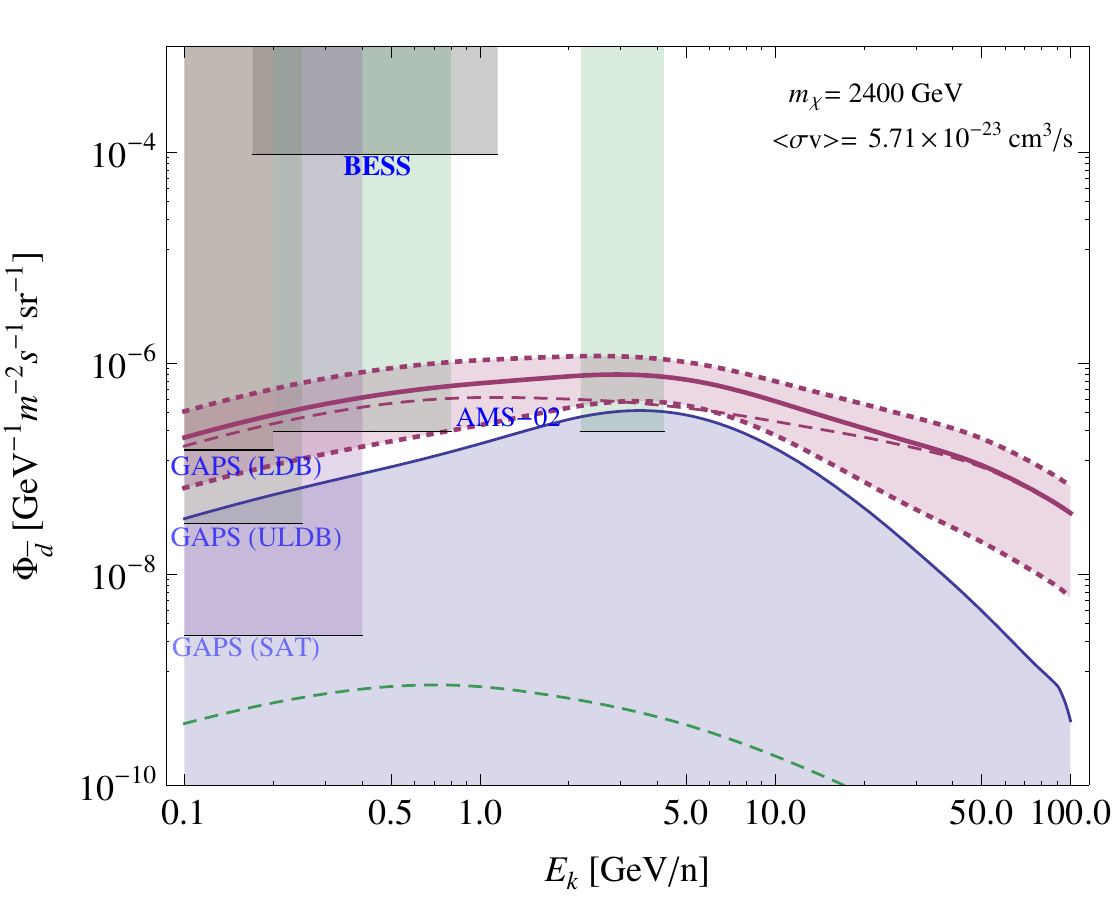}
		\includegraphics[scale=0.65]{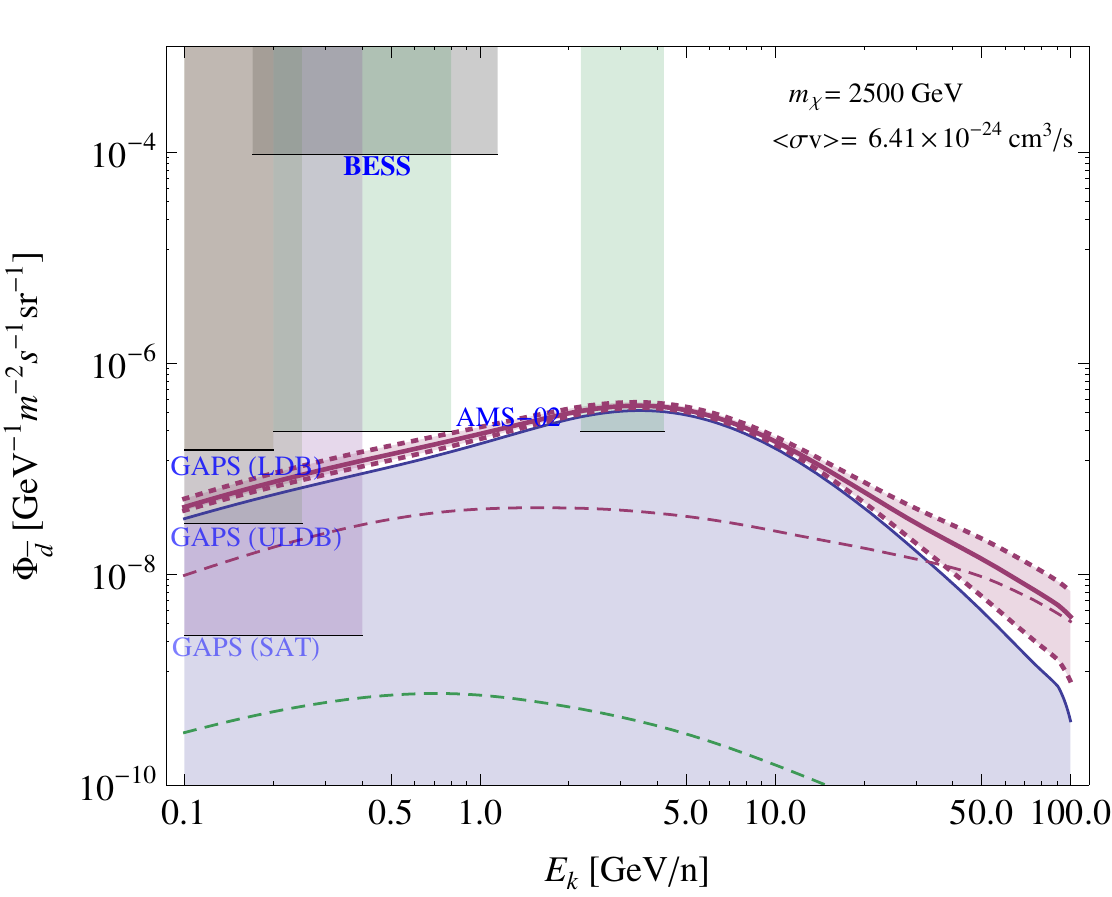}
			\includegraphics[scale=0.65]{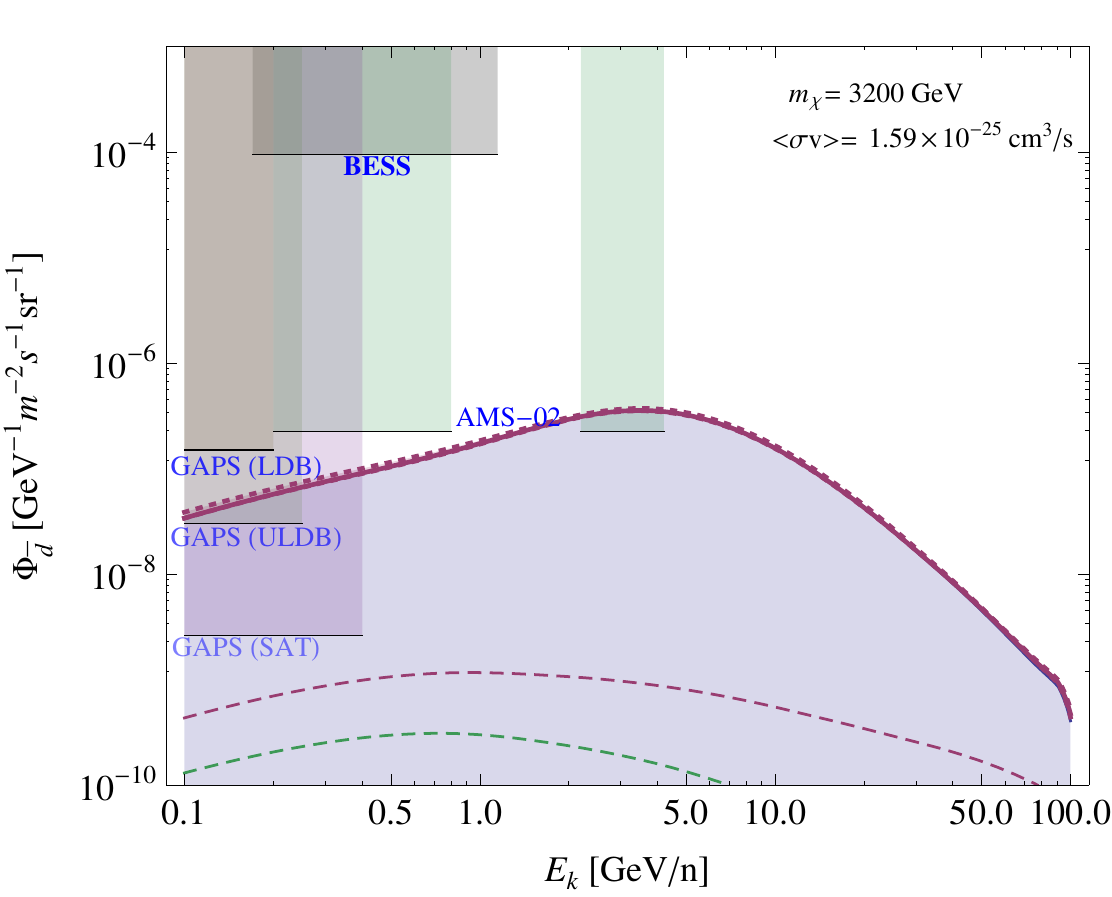}
	\caption{Antideutron fluxes for $m_\c=0.5,\,2.4,\,2.5,\,3.2$ TeV. The blue shaded region shows the expected background after solar modulation with the modulation potentail assumed to be the same as for the antiprotons. Violet dashed line gives the DM component for $z_d=4$ kpc model, while the solid line gives the total flux. The shaded violet region spans different propagation models, for the minimal and maximal denoted with dotted lines. For comparison, the dashed green line shows the DM contribution without the Sommerfeld effect. The shaded regions in the upper left give the exclusion by BESS experiment and projected sensitivity of GAPS and AMS-02.}
    \label{fig:antid}
\end{figure}

Having all that said, it is nevertheless interesting to check what could be the potential signatures of our Wino model in this channel. The results for our four benchmark masses are plotted on Fig.~\ref{fig:antid}. The overall values of the flux are very low, and that indeed the dark matter component can be dominant in some cases. This happens especially in low mass and resonance regions, i.e. whenever the annihilation cross section is large enough. In addition, the expected antidetueron flux can be of reach of not only the future planned experiment GAPS \cite{GAPS}, but possibly also AMS-02 \cite{AMSsite}, which is already collecting data at the International Space Station. The bound set by BESS \cite{Fuke:2005it} and the predicted sensitivities of AMS-02 and GAPS, plotted as a shaded regions, are summarized in Tab.~\ref{tab:antidlimits}. Unfortunately, the most clear signatures are expected in models with cross sections too large to be allowed by previous, more robust channels. For the case $m_\c=2.5$ TeV the signal to background ratio at low energies is only about 2, which is way too small to be giving a clear signature, given the large uncertainties.

\begin{table}
\centering
\begin{tabular}{|c|c|c|}
\hline
Experiment & Energy/nucleon [GeV/n] & Upper bound/sensitivity [${\rm m}^{-2} {\rm s}^{-1} {\rm sr}^{-1} {\rm GeV}^{-1}$] \\ \hline
BESS & $0.17\leq E_k/n \leq 1.15$ & $0.95\times 10^{-4}$ \\ \hline
\multicolumn{1}{|c|}{\multirow{2}{*}{AMS-02}} & $0.2\leq E_k/n \leq 0.8$ & $2.25\times 10^{-7}$ \\ 
& $2.2\leq E_k/n \leq 4.2$ & $2.25\times 10^{-7}$ \\ \hline
GAPS (LDB) & $0.1\leq E_k/n \leq 0.2$ & $1.5\times 10^{-7}$ \\ \hline
GAPS (ULDB) & $0.05\leq E_k/n \leq 0.25$ & $3.0\times 10^{-8}$ \\ \hline
GAPS (SAT) & $0.1\leq E_k/n \leq 0.4$ & $\sim 2.6\times 10^{-9}$ \\ \hline
\end{tabular}
\caption{Limits on antideuteron flux. The BESS limit is an only actual upper bound. The AMS-02 predicted sensitivity is given following \cite{amslimit} and refers to 3 years of data taking. For GAPS the three proposals are the Long Duration Balloon (LDB), the Ultra-Long Duration Balloon (ULDB), and a Satellite  (SAT) mission. Limits taken from \cite{Cui:2010ud}.}
\label{tab:antidlimits}
\end{table}

Indeed, in this case uncertainties are very large. Note, that on the plot the scale has ticks every two orders of magnitude, so that the propagation uncertainty introduces more than an order of magnitude effect, especially in the experimentally interesting low energy window. Additionally, since the expected signature is at energies not exceeding around 1 GeV, the solar modulation plays a very important role and a more sophisticated approach than the force field approximation would be needed (see e.g. \cite{Maccione:2012cu,Fornengo:2013osa}). 
Moreover, as we mentioned in the beginning it is hard to quantify our lack of full understanding of the production mechanisms. In fact, we also do not know very well the background, which comes from impinging of the cosmic ray protons and He on the interstellar gas. The reason is that not only there are no measurements of the CR antideuterons, but also the cross sections of their secondary (and tertiary) production mechanisms are based on fitting to small sample of data and rely on some (reasonable) theoretical assumptions.\footnote{For more details on the uncertainties and cross sections see \cite{Donato:2008uq}.} 
Nevertheless, at low energies the background is expected to be suppressed much more than in the case of antiprotons, what gives hope for exploiting this channel for dark matter detection.

In conclusion, the antidueteron channel may be very promising in the future, when some progress will be made on both experiential and theoretical sides. Then, it is conceivable that the Wino dark matter can be constrained (or maybe detected) by this search channel.

\section{Discussion and Conclusions}
\label{sec:concl}

In this work we perform the multichannel analysis of the indirect detection limits for the Wino DM model. We put an emphasis on the consistent determination of all the background and DM signals (in CRs and galactic $\gamma$-rays) within the same galactic propagation framework. This is an important, though often unappreciated point, as it allows to cross correlate the obtained results from many different search channels. The aim of this study is threefold: \textit{i)} examine what can we learn from cross correlating different channels, \textit{ii)} discuss the relevance of various of uncertainties and \textit{iii)} taking all that into account and inferring more robust limits on the Wino DM model. 

The analysis is carried in several consecutive steps. First we compute the annihilation spectra at production at one-loop order and including the electroweak Sommerfeld effect (see Sec.~\ref{sec:fluxes}). 
The former significantly alters the low energy part of the CR and photon spectra from DM annihilation, while the latter changes the value of the total cross section even by few orders of magnitude.

Next we determine a set of benchmark propagation models, whose parameters are fitted to the available CR data (see Sec.~\ref{sec:methodology}). This allows for predictions of the CR background. Using this propagation models, we compute the expected signal of the Wino in antiprotons, antiproton to proton ratio, positrons and antideuterons, from which we derive the limits on the cross section for a given Wino mass. Then we study the diffuse $\gamma$-rays in the galactic halo, whose predicted backgrounds are determined based on the previously obtained CR lepton fluxes. We compare with the high and low latitude \textit{Fermi} data to rule out the very thin propagation models, which in result make the constraints from antiprotons stronger, and in addition place strong limits themselves on the DM component. Additionally, we calculate the limits from dwarf spheroidal galaxies and a monochromatic $\gamma$-ray line feature from the Galactic Center. We also discuss the diffuse emission close to the GC, galaxy clusters and CMB constraints. Finally, we derive the prospects for neutrino and antideuteron searches.

\begin{figure}[t]
\centering
	\includegraphics[scale=0.8]{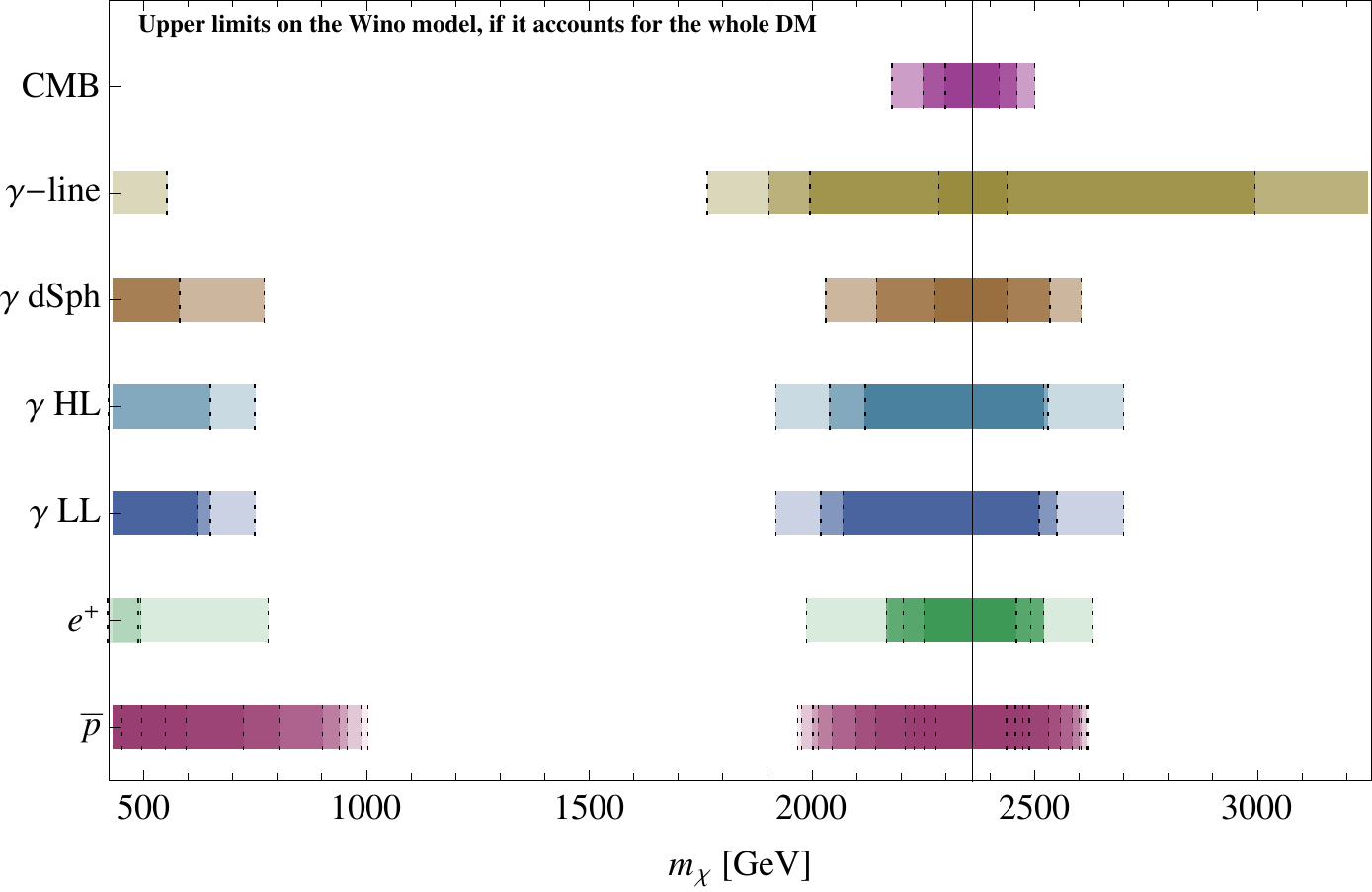}
	\caption{Combination of the 95\% CL upper limits for the Wino DM mass. The black vertical line shows the position of the peak of the resonance. For a given channel different shaded regions correspond to limits derived using different assumptions. For antiprotons ($\bar{p}$) this is related to the diffusion zone thickness, leptons ($e^+$) the local DM density and energy density in the ISRF and magnetic field, low latitude $\gamma$-rays ($\gamma$ LL) the radiation field in the inner galaxy and interstellar gas, high latitude $\gamma$-rays ($\gamma$ HL) the extragalactic DM substructures, for dwarf spheroidal galaxies ($\gamma$ dSph) the $J$-factor and foreground emission, for $\gamma$-line the DM profile in the inner 1$^\circ$ and in the case of the CMB constraints different combinations of data sets. See text and sections corresponding to its given channel for details.
	}
    \label{fig:limits}
\end{figure}

The final set of limits for the Wino DM model is presented on Fig.~\ref{fig:limits}. All the channels give limits around the resonance and most of them also in the low mass range, which allow to exclude these parameter regions with high level of confidence. The precise constraints depend on various assumptions, which we have analysed in detail. For every search channel one can single out one physical quantity that has the strongest impact on the derived limits. To show  the impact in a given channel, of the uncertainty on the leading quantity, we calculate the 95\% CL upper limits under different choices/assumptions on that leading quantity. This is shown on Fig..~\ref{fig:limits} with different color shading.

For the case of antiprotons, the largest impact comes from the diffusion zone scale height assumption and therefore we parametrize its effect by varying the diffusion zone height $z_d$ from 1 to 20 kpc (see Sec.~\ref{sec:antiprotons}). Cosmic ray leptons, are very local and as we have shown, propagation has a very minor influence. On the other hand, the energy losses and even more importantly the local DM density can significantly change the predicted fluxes. The former can be parametrized by the local energy density of the ISRF and the magnetic field, which we vary in the range 1.2-2.6 eV/cm$^{3}$, while the uncertainty on the latter, we probe by assuming $\rho$ between 0.25 and 0.7 GeV$/$cm$^3$ (see Sec.~\ref{sec:leptons}). From $\gamma$-ray observations we derive limits from the GC, from low and high latitudes and from observations toward known dwarf spheroidal galaxies. The low latitude $\gamma$-ray limits, mainly depend on assumptions on the radiation field in the inner galaxy and on assumptions on the interstellar gas (see Sec.~\ref{sec:GalacticDiffuse}). The high latitude $\gamma$-rays contain mostly the extragalactic component, for which the biggest uncertainty comes from the knowledge of the DM substructures on distant galaxies and galaxy clusters (see Sec.~\ref{sec:HighLat}). For the dSphs, the main limiting factor is the uncertainty related to the $J$-factor and the foreground emission, which varies between targets (see Sec.~\ref{sec:dSph}). The constraints from the $\gamma$-line toward the GC are limited by the knowledge of the DM profile in the inner 1$^\circ$. Depending whether a cored or a cuspy profile is assumed the limits change by a few orders of magnitude, nevertheless for quite a wide range of DM profiles, the Wino DM is severely constrained (see Sec.~\ref{sec:GammaLine}). Finally, the CMB constraints depend on the different combinations of data sets used (see Sec.~\ref{sec:Synch}). On the plot we do not include galaxy clusters because the uncertainties are too large to infer any unchallenged limits (see discussion in Sec.~\ref{sec:Clusters}).

\begin{figure}[t]
\centering
	\includegraphics[scale=0.8]{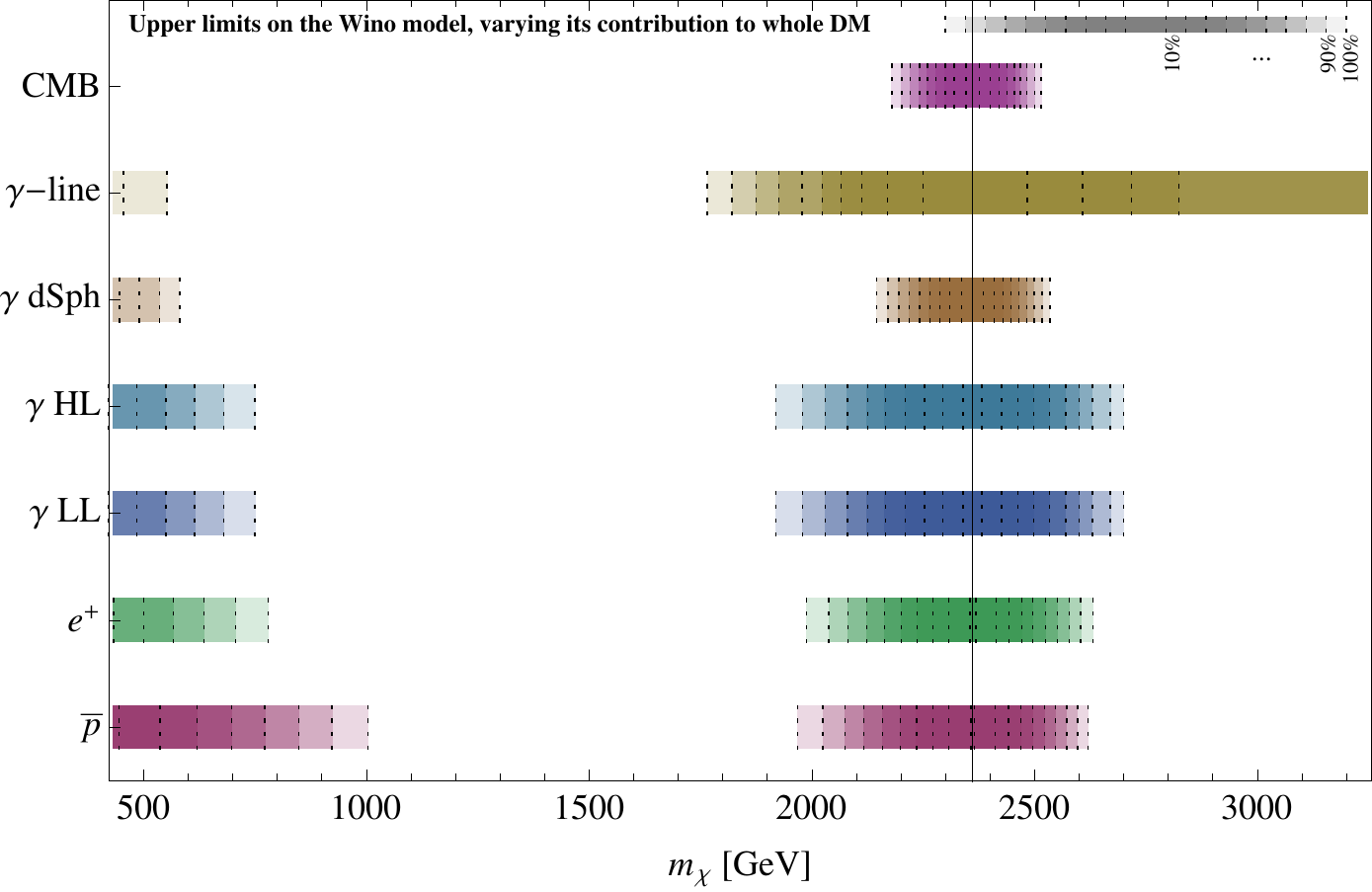}
	\caption{Combination of the most stringent 95\% CL upper limits for the Wino DM mass, relaxing the condition of the Wino accounting for the whole dark matter. The shaded regions correspond to reducing the Wino contribution from 100\% (lighter regions) to 10\% (darker regions) in steps of 10\%. Specifically for the case of dwarf spheroidal limits, we use those derived by Ursa Minor which we derive ourselves (see section~\ref{sec:dSph} and following \cite{Cholis:2012ve}) rather than those from a joint likelihood analysis as those from \cite{Ackermann:2011wa} which can exclude an additional $\simeq$100 GeV in mass on each side of the maximum excluded mass range. The black vertical line shows the position of the peak of the resonance.}
    \label{fig:limitsfreerho}
\end{figure}

Additionally, on Fig.~\ref{fig:limitsfreerho} we show how the limits change when Wino is constituting only a fraction of the whole dark matter. For every channel we take the most stringent limits and reduce the Wino abundance from 100\% (lighter shaded regions) to 10\% (darker shaded regions) in steps of 10\%.

Apart from the above limits, we also determined future prospects for the neutrinos and antideuteron searches. In the former, only the KM3NeT experiment could be able to probe the vicinity of the resonance, a region excluded by all the above discussed channels. The antideuterons on the other hand constitute also a quite promising channel for the low mass region, however masses below around 400 GeV (500 GeV) are already ruled out (constrained) by antiprotons, dSphs and leptons. Moreover, this channel has very large uncertainties related to both propagation and antideuterons' production. More promising prospects, can be expected in the high latitude $\gamma$-rays, where the 10 yr \textit{Fermi} data will have the potential to probe the entire (cosmologically relevant) Wino mass range discussed, if the extragalactic DM substructures are included (see \cite{Cholis:2013ena}). Finally the CMB limits from \textit{Planck} are expected to be a factor $\sim$5 tighter than the current ones after the polarization data are released.

\acknowledgments
We would like to thank Mirko Boezio, Timothy Cohen, Carmelo Evoli, Dan Hooper and Luca Maccione for valuable discussions. 
The work of A.H. is supported by the Gottfried
Wilhelm Leibniz programme of the Deutsche Forschungsgemeinschaft (DFG)
and the DFG cluster of excellence "Origin and Structure of the
Universe". This work has been supported by the US Department of Energy (I.C.). I.C. would like to thank also the Aspen Center for Physics and the NSF Grant \#1066293 for hospitality during the latter stages of this project. P.U. acknowledges partial support from the European Union FP7 ITN INVISIBLES (Marie Curie Actions, PITN-GA-2011-289442).

\appendix

\section{The Sommerfeld enhanced spectrum}
\label{app:spectrum}

In this appendix we give the explicit formulas used to compute the primary annihilation spectrum. The differential cross section including the SE is given by:
\begin{equation}
\frac{d\sigma_{SE}}{dx}=\sigma_2^{tot} \delta(1-x) + |s_0|^2\frac{d\sigma^{00}_{WW\gamma,Z}}{dx}
+ |s_\pm|^2\frac{d\sigma^{+-}_{WW\gamma,Z}}{dx}
+ 2\Rea s_0 s_\pm^*\frac{d\sigma^{mix}_{WW\gamma,Z}}{dx}\;, 
\end{equation} 
where we used a short notation $d\sigma^{00}_{WW\gamma,Z}/dx=d\sigma^{00}_{WW\gamma}/dx
+d\sigma^{00}_{WWZ}/dx$. The appearance of the last term is the result of computing the SE on an amplitude level and $\sigma^{mix}$ denotes the "cross section" obtained from integrating over phase space the mixed term with both $A_{\c^0\c^0}$ and $A_{\c^+\c^-}$.

The two-body enhanced cross section is equal to:
\begin{eqnarray}
\sigma_2^{tot} &=& |s_0|^2 \sigma_{tree}^{00\to WW} + |s_\pm|^2 \sigma_{tree}^{+-\to WW} + 2\Rea s_0 s_\pm^* \sigma_{tree}^{mix}  + |s_\pm|^2  \left( \sigma_{tree}^{+-\to ZZ} + \sigma_{tree}^{+-\to Z\gamma} + \sigma_{tree}^{+-\to \gamma\gamma} \right) \nonumber\\
&+&  |s_0|^2 \sigma_{loop}^{00\to WW}  + |s_\pm|^2 \sigma_{loop}^{+-\to WW} + 2\Rea s_0 s_\pm^* \sigma_{loop}^{mix} + |s_\pm|^2  \left( \sigma_{loop}^{+-\to ZZ} + \sigma_{loop}^{+-\to Z\gamma} + \sigma_{loop}^{+-\to \gamma\gamma} \right) \nonumber\\
&+& 2\Rea s_0 s_\pm^*  \left( \sigma_{loop}^{mix\to ZZ} + \sigma_{loop}^{mix\to Z\gamma} + \sigma_{loop}^{mix\to \gamma\gamma} \right).
\end{eqnarray}
First of the two lines of the above expression give a contribution of the order $\mathcal{O}(g^4)$, while the rest $\mathcal{O}(g^6)$. Note, that the are no terms like $ |s_0|^2  \sigma_{loop}^{00\to \gamma\gamma}$, since this is of higher order.\footnote{Here, by writing only one delta function for all kind of final states we make an assumption that in the annihilation to $Z\gamma$ we can treat the resulting particles as having the same mass. It is a good approximation for the DM mass of several TeV. However, even for lower masses this approximation affects only the slight shift of the $Z\gamma$ line component.}	

The splitting functions for gauge boson emission we obtain from our three-body initial spectra and the total annihilation cross section, via:
\begin{eqnarray}
D_\gamma(z)&=&\frac{\sigma_{WW\gamma}}{\sigma_{tot}}\frac{dN^\gamma_{WW\gamma}}{dz} ,
\\
D_Z(z)&=&\frac{\sigma_{WWZ}}{\sigma_{tot}}\frac{dN^Z_{WWZ}}{dz} ,
\\
D_W^{tree}(z)&=&\frac{\sigma^{WW}_{tree}}{\sigma_{tot}} \delta(1-z) ,
\\
D_W^{loop+rp}(z)&=& \frac{\sigma^{WW}_{loop}}{\sigma_{tot}} \delta(1-z) + \frac{\sigma_{WW\gamma}}{\sigma_{tot}}\frac{dN^W_{WW\gamma}}{dz} + \frac{\sigma_{WWZ}}{\sigma_{tot}}\frac{dN^W_{WWZ}}{dz} .
\end{eqnarray}

The final spectra we obtain then from:
\begin{eqnarray*}
   \frac{dN^f_{tot}}{dx} (M,x) &=& \delta_{f\gamma} D_\gamma(x) \\
      &+& \int_x^1 dz \,S_W\, \left(D_{W}^{tree}(z)+D_{W}^{loop+rp}(z)\right)\,
   \frac{dN^{\rm DS}_{W\to f}}{dx}\left(zM,\frac{x}{z}\right) \\
   &+& \int_x^1 dz \,S_Z\, D_{Z}(z)\,
   \frac{dN^{\rm DS}_{Z\to f}}{dx}\left(zM,\frac{x}{z}\right).
\end{eqnarray*}

In this way we include properly annihilations to gauge bosons with the radiative corrections. However, \pyt also does not include electroweak splitting of gauge bosons into fermions (it includes only the decay). We incorporate them by adding all the additional splitting functions, taken from \cite{Ciafaloni:2010ti}, computed at the leading double log level.

The complete \textit{Sommerfeld corrected} $D(z)$ functions, separated with respect to the perturbation theory order, are then:
\begin{eqnarray}
\sigma_{SE}  D^{(4)}_W(x)&=& \left(|s_0|^2  \sigma_{tree}^{00\to WW} + |s_\pm|^2  \sigma_{tree}^{+-\to WW}  + 2\Rea s_0 s_\pm^* \sigma_{tree}^{mix\to WW}  \right) \delta(1-x)
\\
\sigma_{SE}  D^{(6)}_W(x)&=& \left(|s_0|^2  \sigma_{loop}^{00\to WW} + |s_\pm|^2  \sigma_{loop}^{+-\to WW}  + 2\Rea s_0 s_\pm^* \sigma_{loop}^{mix\to WW}  \right) \delta(1-x) \\
&+&  |s_0|^2 \sigma^{00}_{WWZ}\frac{dN^W_{WWZ}}{dx} + |s_\pm|^2 \sigma^{+-}_{WWZ}\frac{dN^W_{WWZ}}{dx} 
+ 2\Rea s_0 s_\pm^* \sigma^{mix}_{WWZ}\frac{dN^W_{WWZ}}{dx} \nonumber\\
&+&  |s_0|^2 \sigma^{00}_{WW\gamma}\frac{dN^W_{WW\gamma}}{dx} + |s_\pm|^2 \sigma^{+-}_{WW\gamma}\frac{dN^W_{WW\gamma}}{dx} 
+ 2\Rea s_0 s_\pm^* \sigma^{mix}_{WW\gamma}\frac{dN^W_{WW\gamma}}{dx} \nonumber
\end{eqnarray}

\begin{eqnarray}
\sigma_{SE}  D^{(4)}_\gamma(x)&=& |s_\pm|^2 \left( 2\sigma_{tree}^{+-\to \gamma\gamma} + \sigma_{tree}^{+-\to Z\gamma}\right) \delta(1-x)
\\
\sigma_{SE}  D^{(6)}_\gamma(x)&=& |s_0|^2 \sigma^{00}_{WW\gamma}\frac{dN^\gamma_{WW\gamma}}{dx} + |s_\pm|^2 \sigma^{+-}_{WW\gamma}\frac{dN^\gamma_{WW\gamma}}{dx} 
+ 2\Rea s_0 s_\pm^* \sigma^{mix}_{WW\gamma}\frac{dN^\gamma_{WW\gamma}}{dx} \\
&+& |s_\pm|^2 \left( 2\sigma_{loop}^{+-\to \gamma\gamma} + \sigma_{loop}^{+-\to Z\gamma}\right) \delta(1-x) +  2\Rea s_0 s_\pm^* \left(2 \sigma_{loop}^{mix\to \gamma\gamma} + \sigma_{loop}^{mix\to Z\gamma}\right) \delta(1-x) \nonumber
\end{eqnarray}
and analogously for $D_Z$. Note the multiplicity factor of $2$ in front of the terms with production of two identical gauge bosons.

In the same way as before, the final spectra we obtain then from:
\begin{eqnarray*}
   \frac{dN^f_{SE}}{dx} (M,x) &=& \delta_{f\gamma} D^{(4)}_\gamma(x) +  \delta_{f\gamma} D^{(6)}_\gamma(x) \\
      &+& \int_x^1 dz \,S_W\, \left(D_{W}^{(4)}(z)+D_{W}^{(6)}(z)\right)\,
   \frac{dN^{\rm DS}_{W\to f}}{dx}\left(zM,\frac{x}{z}\right) \\
    &+& \int_x^1 dz \,S_Z\, \left(D_{Z}^{(4)}(z)+D_{Z}^{(6)}(z)\right)\,
   \frac{dN^{\rm DS}_{Z\to f}}{dx}\left(zM,\frac{x}{z}\right) \,.
\end{eqnarray*}

\bibliography{paper}

\begin{thebibliography}{150}
\expandafter\ifx\csname natexlab\endcsname\relax\def\natexlab#1{#1}\fi
\expandafter\ifx\csname bibnamefont\endcsname\relax
  \def\bibnamefont#1{#1}\fi
\expandafter\ifx\csname bibfnamefont\endcsname\relax
  \def\bibfnamefont#1{#1}\fi
\expandafter\ifx\csname citenamefont\endcsname\relax
  \def\citenamefont#1{#1}\fi
\expandafter\ifx\csname url\endcsname\relax
  \def\url#1{\texttt{#1}}\fi
\expandafter\ifx\csname urlprefix\endcsname\relax\def\urlprefix{URL }\fi
\providecommand{\bibinfo}[2]{#2}
\providecommand{\eprint}[2][]{\url{#2}}

\bibitem[{\citenamefont{Hinshaw et~al.}(2013)}]{Hinshaw:2012aka}
\bibinfo{author}{\bibfnamefont{G.}~\bibnamefont{Hinshaw}} \bibnamefont{et~al.}
  (\bibinfo{collaboration}{WMAP}), \bibinfo{journal}{Astrophys.J.Suppl.}
  \textbf{\bibinfo{volume}{208}}, \bibinfo{pages}{19} (\bibinfo{year}{2013}),
  \eprint{1212.5226}.

\bibitem[{\citenamefont{Ade et~al.}(2013)}]{Ade:2013zuv}
\bibinfo{author}{\bibfnamefont{P.}~\bibnamefont{Ade}} \bibnamefont{et~al.}
  (\bibinfo{collaboration}{Planck Collaboration}) (\bibinfo{year}{2013}),
  \eprint{1303.5076}.

\bibitem[{\citenamefont{Hisano et~al.}(2007)\citenamefont{Hisano, Matsumoto,
  Nagai, Saito, and Senami}}]{Hisano:2006nn}
\bibinfo{author}{\bibfnamefont{J.}~\bibnamefont{Hisano}},
  \bibinfo{author}{\bibfnamefont{S.}~\bibnamefont{Matsumoto}},
  \bibinfo{author}{\bibfnamefont{M.}~\bibnamefont{Nagai}},
  \bibinfo{author}{\bibfnamefont{O.}~\bibnamefont{Saito}}, \bibnamefont{and}
  \bibinfo{author}{\bibfnamefont{M.}~\bibnamefont{Senami}},
  \bibinfo{journal}{Phys.Lett.} \textbf{\bibinfo{volume}{B646}},
  \bibinfo{pages}{34} (\bibinfo{year}{2007}), \eprint{hep-ph/0610249}.

\bibitem[{\citenamefont{Hryczuk}(2012)}]{HryczukPHD}
\bibinfo{author}{\bibfnamefont{A.}~\bibnamefont{Hryczuk}}, Ph.D. thesis,
  \bibinfo{school}{SISSA} (\bibinfo{year}{2012}).

\bibitem[{\citenamefont{Baudis}(2012)}]{Baudis:2012bc}
\bibinfo{author}{\bibfnamefont{L.}~\bibnamefont{Baudis}}
  (\bibinfo{collaboration}{DARWIN Consortium}) (\bibinfo{year}{2012}),
  \eprint{1201.2402}.

\bibitem[{\citenamefont{Chattopadhyay et~al.}(2007)\citenamefont{Chattopadhyay,
  Das, Konar, and Roy}}]{Chattopadhyay:2006xb}
\bibinfo{author}{\bibfnamefont{U.}~\bibnamefont{Chattopadhyay}},
  \bibinfo{author}{\bibfnamefont{D.}~\bibnamefont{Das}},
  \bibinfo{author}{\bibfnamefont{P.}~\bibnamefont{Konar}}, \bibnamefont{and}
  \bibinfo{author}{\bibfnamefont{D.}~\bibnamefont{Roy}},
  \bibinfo{journal}{Phys.Rev.} \textbf{\bibinfo{volume}{D75}},
  \bibinfo{pages}{073014} (\bibinfo{year}{2007}), \eprint{hep-ph/0610077}.

\bibitem[{\citenamefont{Ullio}(2001)}]{Ullio:2001uq}
\bibinfo{author}{\bibfnamefont{P.}~\bibnamefont{Ullio}},
  \bibinfo{journal}{JHEP} \textbf{\bibinfo{volume}{0106}}, \bibinfo{pages}{053}
  (\bibinfo{year}{2001}), \eprint{hep-ph/0105052},
  \urlprefix\url{http://arxiv.org/abs/hep-ph/0105052}.

\bibitem[{\citenamefont{Baer et~al.}(2005)\citenamefont{Baer, Mustafayev, Park,
  and Profumo}}]{Baer:2005zc}
\bibinfo{author}{\bibfnamefont{H.}~\bibnamefont{Baer}},
  \bibinfo{author}{\bibfnamefont{A.}~\bibnamefont{Mustafayev}},
  \bibinfo{author}{\bibfnamefont{E.-K.} \bibnamefont{Park}}, \bibnamefont{and}
  \bibinfo{author}{\bibfnamefont{S.}~\bibnamefont{Profumo}},
  \bibinfo{journal}{JHEP} \textbf{\bibinfo{volume}{0507}}, \bibinfo{pages}{046}
  (\bibinfo{year}{2005}), \eprint{hep-ph/0505227}.

\bibitem[{\citenamefont{Grajek et~al.}(2008)\citenamefont{Grajek, Kane, Phalen,
  Pierce, and Watson}}]{Grajek:2008jb}
\bibinfo{author}{\bibfnamefont{P.}~\bibnamefont{Grajek}},
  \bibinfo{author}{\bibfnamefont{G.}~\bibnamefont{Kane}},
  \bibinfo{author}{\bibfnamefont{D.~J.} \bibnamefont{Phalen}},
  \bibinfo{author}{\bibfnamefont{A.}~\bibnamefont{Pierce}}, \bibnamefont{and}
  \bibinfo{author}{\bibfnamefont{S.}~\bibnamefont{Watson}}
  (\bibinfo{year}{2008}), \eprint{0807.1508}.

\bibitem[{\citenamefont{Belanger et~al.}(2012)\citenamefont{Belanger, Boehm,
  Cirelli, Da~Silva, and Pukhov}}]{Belanger:2012ta}
\bibinfo{author}{\bibfnamefont{G.}~\bibnamefont{Belanger}},
  \bibinfo{author}{\bibfnamefont{C.}~\bibnamefont{Boehm}},
  \bibinfo{author}{\bibfnamefont{M.}~\bibnamefont{Cirelli}},
  \bibinfo{author}{\bibfnamefont{J.}~\bibnamefont{Da~Silva}}, \bibnamefont{and}
  \bibinfo{author}{\bibfnamefont{A.}~\bibnamefont{Pukhov}},
  \bibinfo{journal}{JCAP} \textbf{\bibinfo{volume}{1211}}, \bibinfo{pages}{028}
  (\bibinfo{year}{2012}), \eprint{1208.5009}.

\bibitem[{\citenamefont{Hisano et~al.}(2006)\citenamefont{Hisano, Matsumoto,
  Saito, and Senami}}]{Hisano:2005ec}
\bibinfo{author}{\bibfnamefont{J.}~\bibnamefont{Hisano}},
  \bibinfo{author}{\bibfnamefont{S.}~\bibnamefont{Matsumoto}},
  \bibinfo{author}{\bibfnamefont{O.}~\bibnamefont{Saito}}, \bibnamefont{and}
  \bibinfo{author}{\bibfnamefont{M.}~\bibnamefont{Senami}},
  \bibinfo{journal}{Phys.Rev.} \textbf{\bibinfo{volume}{D73}},
  \bibinfo{pages}{055004} (\bibinfo{year}{2006}), \eprint{hep-ph/0511118}.

\bibitem[{\citenamefont{Grajek et~al.}(2009)\citenamefont{Grajek, Kane, Phalen,
  Pierce, and Watson}}]{Grajek:2008pg}
\bibinfo{author}{\bibfnamefont{P.}~\bibnamefont{Grajek}},
  \bibinfo{author}{\bibfnamefont{G.}~\bibnamefont{Kane}},
  \bibinfo{author}{\bibfnamefont{D.}~\bibnamefont{Phalen}},
  \bibinfo{author}{\bibfnamefont{A.}~\bibnamefont{Pierce}}, \bibnamefont{and}
  \bibinfo{author}{\bibfnamefont{S.}~\bibnamefont{Watson}},
  \bibinfo{journal}{Phys.Rev.} \textbf{\bibinfo{volume}{D79}},
  \bibinfo{pages}{043506} (\bibinfo{year}{2009}), \eprint{0812.4555}.

\bibitem[{\citenamefont{Kane et~al.}(2009)\citenamefont{Kane, Lu, and
  Watson}}]{Kane:2009if}
\bibinfo{author}{\bibfnamefont{G.}~\bibnamefont{Kane}},
  \bibinfo{author}{\bibfnamefont{R.}~\bibnamefont{Lu}}, \bibnamefont{and}
  \bibinfo{author}{\bibfnamefont{S.}~\bibnamefont{Watson}},
  \bibinfo{journal}{Phys.Lett.} \textbf{\bibinfo{volume}{B681}},
  \bibinfo{pages}{151} (\bibinfo{year}{2009}), \eprint{0906.4765}.

\bibitem[{\citenamefont{Finkbeiner et~al.}(2011)\citenamefont{Finkbeiner,
  Goodenough, Slatyer, Vogelsberger, and Weiner}}]{Finkbeiner:2010sm}
\bibinfo{author}{\bibfnamefont{D.~P.} \bibnamefont{Finkbeiner}},
  \bibinfo{author}{\bibfnamefont{L.}~\bibnamefont{Goodenough}},
  \bibinfo{author}{\bibfnamefont{T.~R.} \bibnamefont{Slatyer}},
  \bibinfo{author}{\bibfnamefont{M.}~\bibnamefont{Vogelsberger}},
  \bibnamefont{and} \bibinfo{author}{\bibfnamefont{N.}~\bibnamefont{Weiner}},
  \bibinfo{journal}{JCAP} \textbf{\bibinfo{volume}{1105}}, \bibinfo{pages}{002}
  (\bibinfo{year}{2011}), \eprint{1011.3082}.

\bibitem[{\citenamefont{Hryczuk et~al.}(2011)\citenamefont{Hryczuk, Iengo, and
  Ullio}}]{Hryczuk:2010zi}
\bibinfo{author}{\bibfnamefont{A.}~\bibnamefont{Hryczuk}},
  \bibinfo{author}{\bibfnamefont{R.}~\bibnamefont{Iengo}}, \bibnamefont{and}
  \bibinfo{author}{\bibfnamefont{P.}~\bibnamefont{Ullio}},
  \bibinfo{journal}{JHEP} \textbf{\bibinfo{volume}{1103}}, \bibinfo{pages}{069}
  (\bibinfo{year}{2011}), \eprint{1010.2172}.

\bibitem[{\citenamefont{Zavala et~al.}(2010)\citenamefont{Zavala, Vogelsberger,
  and White}}]{Zavala:2009mi}
\bibinfo{author}{\bibfnamefont{J.}~\bibnamefont{Zavala}},
  \bibinfo{author}{\bibfnamefont{M.}~\bibnamefont{Vogelsberger}},
  \bibnamefont{and} \bibinfo{author}{\bibfnamefont{S.~D.} \bibnamefont{White}},
  \bibinfo{journal}{Phys.Rev.} \textbf{\bibinfo{volume}{D81}},
  \bibinfo{pages}{083502} (\bibinfo{year}{2010}), \eprint{0910.5221}.

\bibitem[{\citenamefont{Beneke et~al.}(2013)\citenamefont{Beneke, Hellmann, and
  Ruiz-Femenia}}]{Beneke:2012tg}
\bibinfo{author}{\bibfnamefont{M.}~\bibnamefont{Beneke}},
  \bibinfo{author}{\bibfnamefont{C.}~\bibnamefont{Hellmann}}, \bibnamefont{and}
  \bibinfo{author}{\bibfnamefont{P.}~\bibnamefont{Ruiz-Femenia}},
  \bibinfo{journal}{JHEP} \textbf{\bibinfo{volume}{1303}}, \bibinfo{pages}{148}
  (\bibinfo{year}{2013}), \eprint{1210.7928}.

\bibitem[{\citenamefont{Hellmann and
  Ruiz-Femen{\'\i}a}(2013)}]{Hellmann:2013jxa}
\bibinfo{author}{\bibfnamefont{C.}~\bibnamefont{Hellmann}} \bibnamefont{and}
  \bibinfo{author}{\bibfnamefont{P.}~\bibnamefont{Ruiz-Femen{\'\i}a}},
  \bibinfo{journal}{JHEP} \textbf{\bibinfo{volume}{1308}}, \bibinfo{pages}{084}
  (\bibinfo{year}{2013}), \eprint{1303.0200}.

\bibitem[{\citenamefont{Hryczuk and Iengo}(2012)}]{Hryczuk:2011vi}
\bibinfo{author}{\bibfnamefont{A.}~\bibnamefont{Hryczuk}} \bibnamefont{and}
  \bibinfo{author}{\bibfnamefont{R.}~\bibnamefont{Iengo}},
  \bibinfo{journal}{JHEP} \textbf{\bibinfo{volume}{1201}}, \bibinfo{pages}{163}
  (\bibinfo{year}{2012}), \eprint{1111.2916}.

\bibitem[{\citenamefont{Ciafaloni and Urbano}(2010)}]{Ciafaloni:2010qr}
\bibinfo{author}{\bibfnamefont{P.}~\bibnamefont{Ciafaloni}} \bibnamefont{and}
  \bibinfo{author}{\bibfnamefont{A.}~\bibnamefont{Urbano}},
  \bibinfo{journal}{Phys.Rev.} \textbf{\bibinfo{volume}{D82}},
  \bibinfo{pages}{043512} (\bibinfo{year}{2010}), \eprint{1001.3950}.

\bibitem[{\citenamefont{Gribov and Lipatov}(1972)}]{Gribov:1972ri}
\bibinfo{author}{\bibfnamefont{V.}~\bibnamefont{Gribov}} \bibnamefont{and}
  \bibinfo{author}{\bibfnamefont{L.}~\bibnamefont{Lipatov}},
  \bibinfo{journal}{Sov.J.Nucl.Phys.} \textbf{\bibinfo{volume}{15}},
  \bibinfo{pages}{438} (\bibinfo{year}{1972}).

\bibitem[{\citenamefont{Altarelli and Parisi}(1977)}]{Altarelli:1977zs}
\bibinfo{author}{\bibfnamefont{G.}~\bibnamefont{Altarelli}} \bibnamefont{and}
  \bibinfo{author}{\bibfnamefont{G.}~\bibnamefont{Parisi}},
  \bibinfo{journal}{Nucl.Phys.} \textbf{\bibinfo{volume}{B126}},
  \bibinfo{pages}{298} (\bibinfo{year}{1977}).

\bibitem[{\citenamefont{Dokshitzer}(1977)}]{Dokshitzer:1977sg}
\bibinfo{author}{\bibfnamefont{Y.~L.} \bibnamefont{Dokshitzer}},
  \bibinfo{journal}{Sov.Phys.JETP} \textbf{\bibinfo{volume}{46}},
  \bibinfo{pages}{641} (\bibinfo{year}{1977}).

\bibitem[{\citenamefont{Sjostrand et~al.}(2006)\citenamefont{Sjostrand, Mrenna,
  and Skands}}]{Sjostrand:2006za}
\bibinfo{author}{\bibfnamefont{T.}~\bibnamefont{Sjostrand}},
  \bibinfo{author}{\bibfnamefont{S.}~\bibnamefont{Mrenna}}, \bibnamefont{and}
  \bibinfo{author}{\bibfnamefont{P.~Z.} \bibnamefont{Skands}},
  \bibinfo{journal}{JHEP} \textbf{\bibinfo{volume}{05}}, \bibinfo{pages}{026}
  (\bibinfo{year}{2006}), \eprint{hep-ph/0603175}.

\bibitem[{\citenamefont{Corcella et~al.}(2002)\citenamefont{Corcella, Knowles,
  Marchesini, Moretti, Odagiri et~al.}}]{Corcella:2002jc}
\bibinfo{author}{\bibfnamefont{G.}~\bibnamefont{Corcella}},
  \bibinfo{author}{\bibfnamefont{I.}~\bibnamefont{Knowles}},
  \bibinfo{author}{\bibfnamefont{G.}~\bibnamefont{Marchesini}},
  \bibinfo{author}{\bibfnamefont{S.}~\bibnamefont{Moretti}},
  \bibinfo{author}{\bibfnamefont{K.}~\bibnamefont{Odagiri}},
  \bibnamefont{et~al.} (\bibinfo{year}{2002}), \eprint{hep-ph/0210213}.

\bibitem[{\citenamefont{Gondolo et~al.}(2004)\citenamefont{Gondolo, Edsjo,
  Ullio, Bergstrom, Schelke et~al.}}]{Gondolo:2004sc}
\bibinfo{author}{\bibfnamefont{P.}~\bibnamefont{Gondolo}},
  \bibinfo{author}{\bibfnamefont{J.}~\bibnamefont{Edsjo}},
  \bibinfo{author}{\bibfnamefont{P.}~\bibnamefont{Ullio}},
  \bibinfo{author}{\bibfnamefont{L.}~\bibnamefont{Bergstrom}},
  \bibinfo{author}{\bibfnamefont{M.}~\bibnamefont{Schelke}},
  \bibnamefont{et~al.}, \bibinfo{journal}{JCAP}
  \textbf{\bibinfo{volume}{0407}}, \bibinfo{pages}{008} (\bibinfo{year}{2004}),
  \eprint{astro-ph/0406204}.

\bibitem[{\citenamefont{Ciafaloni et~al.}(2011)\citenamefont{Ciafaloni,
  Comelli, Riotto, Sala, Strumia et~al.}}]{Ciafaloni:2010ti}
\bibinfo{author}{\bibfnamefont{P.}~\bibnamefont{Ciafaloni}},
  \bibinfo{author}{\bibfnamefont{D.}~\bibnamefont{Comelli}},
  \bibinfo{author}{\bibfnamefont{A.}~\bibnamefont{Riotto}},
  \bibinfo{author}{\bibfnamefont{F.}~\bibnamefont{Sala}},
  \bibinfo{author}{\bibfnamefont{A.}~\bibnamefont{Strumia}},
  \bibnamefont{et~al.}, \bibinfo{journal}{JCAP}
  \textbf{\bibinfo{volume}{1103}}, \bibinfo{pages}{019} (\bibinfo{year}{2011}),
  \eprint{1009.0224}.

\bibitem[{\citenamefont{Cirelli et~al.}(2011)\citenamefont{Cirelli, Corcella,
  Hektor, H{\"u}tsi, Kadastik, Panci, Raidal, Sala, and
  Strumia}}]{Cirelli:2011fk}
\bibinfo{author}{\bibfnamefont{M.}~\bibnamefont{Cirelli}},
  \bibinfo{author}{\bibfnamefont{G.}~\bibnamefont{Corcella}},
  \bibinfo{author}{\bibfnamefont{A.}~\bibnamefont{Hektor}},
  \bibinfo{author}{\bibfnamefont{G.}~\bibnamefont{H{\"u}tsi}},
  \bibinfo{author}{\bibfnamefont{M.}~\bibnamefont{Kadastik}},
  \bibinfo{author}{\bibfnamefont{P.}~\bibnamefont{Panci}},
  \bibinfo{author}{\bibfnamefont{M.}~\bibnamefont{Raidal}},
  \bibinfo{author}{\bibfnamefont{F.}~\bibnamefont{Sala}}, \bibnamefont{and}
  \bibinfo{author}{\bibfnamefont{A.}~\bibnamefont{Strumia}},
  \bibinfo{journal}{JCAP} \textbf{\bibinfo{volume}{1103}}, \bibinfo{pages}{051}
  (\bibinfo{year}{2011}), \eprint{1012.4515}.

\bibitem[{\citenamefont{Ginzburg~(ed.)
  et~al.}(1990)\citenamefont{Ginzburg~(ed.), Dogiel, Berezinsky, Bulanov, and
  Ptuskin}}]{Ginzburg:1990sk}
\bibinfo{author}{\bibfnamefont{V.}~\bibnamefont{Ginzburg~(ed.)}},
  \bibinfo{author}{\bibfnamefont{V.}~\bibnamefont{Dogiel}},
  \bibinfo{author}{\bibfnamefont{V.}~\bibnamefont{Berezinsky}},
  \bibinfo{author}{\bibfnamefont{S.}~\bibnamefont{Bulanov}}, \bibnamefont{and}
  \bibinfo{author}{\bibfnamefont{V.}~\bibnamefont{Ptuskin}},
  \emph{\bibinfo{title}{{Astrophysics of cosmic rays}}} (\bibinfo{year}{1990}).

\bibitem[{\citenamefont{\url{http://www.dragonproject.org}}()}]{DRAGONweb}
\bibinfo{author}{\bibnamefont{\url{http://www.dragonproject.org}}}.

\bibitem[{\citenamefont{Evoli et~al.}(2008)\citenamefont{Evoli, Gaggero,
  Grasso, and Maccione}}]{Evoli:2008dv}
\bibinfo{author}{\bibfnamefont{C.}~\bibnamefont{Evoli}},
  \bibinfo{author}{\bibfnamefont{D.}~\bibnamefont{Gaggero}},
  \bibinfo{author}{\bibfnamefont{D.}~\bibnamefont{Grasso}}, \bibnamefont{and}
  \bibinfo{author}{\bibfnamefont{L.}~\bibnamefont{Maccione}},
  \bibinfo{journal}{JCAP} \textbf{\bibinfo{volume}{0810}}, \bibinfo{pages}{018}
  (\bibinfo{year}{2008}), \eprint{0807.4730}.

\bibitem[{\citenamefont{Tavakoli}(2012)}]{Tavakoli:2012jx}
\bibinfo{author}{\bibfnamefont{M.}~\bibnamefont{Tavakoli}}
  (\bibinfo{year}{2012}), \eprint{1207.6150}.

\bibitem[{\citenamefont{Pohl et~al.}(2008)\citenamefont{Pohl, Englmaier, and
  Bissantz}}]{Pohl:2007dz}
\bibinfo{author}{\bibfnamefont{M.}~\bibnamefont{Pohl}},
  \bibinfo{author}{\bibfnamefont{P.}~\bibnamefont{Englmaier}},
  \bibnamefont{and} \bibinfo{author}{\bibfnamefont{N.}~\bibnamefont{Bissantz}},
  \bibinfo{journal}{Astrophys.J.} \textbf{\bibinfo{volume}{677}},
  \bibinfo{pages}{283} (\bibinfo{year}{2008}), \eprint{0712.4264}.

\bibitem[{\citenamefont{Kalberla et~al.}(2005)\citenamefont{Kalberla, Burton,
  Hartmann, Arnal, Bajaja et~al.}}]{Kalberla:2005ts}
\bibinfo{author}{\bibfnamefont{P.~M.} \bibnamefont{Kalberla}},
  \bibinfo{author}{\bibfnamefont{W.}~\bibnamefont{Burton}},
  \bibinfo{author}{\bibfnamefont{D.}~\bibnamefont{Hartmann}},
  \bibinfo{author}{\bibfnamefont{E.}~\bibnamefont{Arnal}},
  \bibinfo{author}{\bibfnamefont{E.}~\bibnamefont{Bajaja}},
  \bibnamefont{et~al.}, \bibinfo{journal}{Astron.Astrophys.}
  \textbf{\bibinfo{volume}{440}}, \bibinfo{pages}{775} (\bibinfo{year}{2005}),
  \eprint{astro-ph/0504140}.

\bibitem[{\citenamefont{Gleeson and Axford}(1968)}]{Gleeson:1968zza}
\bibinfo{author}{\bibfnamefont{L.}~\bibnamefont{Gleeson}} \bibnamefont{and}
  \bibinfo{author}{\bibfnamefont{W.}~\bibnamefont{Axford}},
  \bibinfo{journal}{Astrophys.J.} \textbf{\bibinfo{volume}{154}},
  \bibinfo{pages}{1011} (\bibinfo{year}{1968}).

\bibitem[{\citenamefont{Putze et~al.}(2010)\citenamefont{Putze, Maurin, and
  Donato}}]{Putze:2010fr}
\bibinfo{author}{\bibfnamefont{A.}~\bibnamefont{Putze}},
  \bibinfo{author}{\bibfnamefont{D.}~\bibnamefont{Maurin}}, \bibnamefont{and}
  \bibinfo{author}{\bibfnamefont{F.}~\bibnamefont{Donato}}
  (\bibinfo{year}{2010}), \eprint{1011.0989}.

\bibitem[{\citenamefont{AMS-02}(2013)}]{AMSsite}
\bibinfo{author}{\bibnamefont{AMS-02}},
  \bibinfo{journal}{http://www.ams02.org/}  (\bibinfo{year}{2013}).

\bibitem[{\citenamefont{Carbone}(2013)}]{PAMELABtoC}
\bibinfo{author}{\bibfnamefont{R.}~\bibnamefont{Carbone}},
  \emph{\bibinfo{title}{{Galactic Boron and Carbon fluxes by the PAMELA
  experiment}}}, \bibinfo{howpublished}{Talk at the ICRC 2013, Rio de Janeiro}
  (\bibinfo{year}{2013}).

\bibitem[{\citenamefont{Adriani et~al.}(2011{\natexlab{a}})}]{Adriani:2011cu}
\bibinfo{author}{\bibfnamefont{O.}~\bibnamefont{Adriani}} \bibnamefont{et~al.}
  (\bibinfo{year}{2011}{\natexlab{a}}), \eprint{1103.4055}.

\bibitem[{\citenamefont{{PAMELA Collaboration}
  et~al.}(2010)\citenamefont{{PAMELA Collaboration}, {Adriani}, {Barbarino},
  {Bazilevskaya}, {Bellotti}, {Boezio}, {Bogomolov}, {Bonechi}, {Bongi},
  {Bonvicini} et~al.}}]{2010arXiv1007.0821P}
\bibinfo{author}{\bibnamefont{{PAMELA Collaboration}}},
  \bibinfo{author}{\bibfnamefont{O.}~\bibnamefont{{Adriani}}},
  \bibinfo{author}{\bibfnamefont{G.~C.} \bibnamefont{{Barbarino}}},
  \bibinfo{author}{\bibfnamefont{G.~A.} \bibnamefont{{Bazilevskaya}}},
  \bibinfo{author}{\bibfnamefont{R.}~\bibnamefont{{Bellotti}}},
  \bibinfo{author}{\bibfnamefont{M.}~\bibnamefont{{Boezio}}},
  \bibinfo{author}{\bibfnamefont{E.~A.} \bibnamefont{{Bogomolov}}},
  \bibinfo{author}{\bibfnamefont{L.}~\bibnamefont{{Bonechi}}},
  \bibinfo{author}{\bibfnamefont{M.}~\bibnamefont{{Bongi}}},
  \bibinfo{author}{\bibfnamefont{V.}~\bibnamefont{{Bonvicini}}},
  \bibnamefont{et~al.}, \bibinfo{journal}{ArXiv e-prints}
  (\bibinfo{year}{2010}), \eprint{1007.0821}.

\bibitem[{\citenamefont{Cirelli et~al.}(2009)\citenamefont{Cirelli, Kadastik,
  Raidal, and Strumia}}]{Cirelli:2008pk}
\bibinfo{author}{\bibfnamefont{M.}~\bibnamefont{Cirelli}},
  \bibinfo{author}{\bibfnamefont{M.}~\bibnamefont{Kadastik}},
  \bibinfo{author}{\bibfnamefont{M.}~\bibnamefont{Raidal}}, \bibnamefont{and}
  \bibinfo{author}{\bibfnamefont{A.}~\bibnamefont{Strumia}},
  \bibinfo{journal}{Nucl.Phys.} \textbf{\bibinfo{volume}{B813}},
  \bibinfo{pages}{1} (\bibinfo{year}{2009}), \eprint{0809.2409}.

\bibitem[{\citenamefont{Donato et~al.}(2009)\citenamefont{Donato, Maurin, Brun,
  Delahaye, and Salati}}]{Donato:2008jk}
\bibinfo{author}{\bibfnamefont{F.}~\bibnamefont{Donato}},
  \bibinfo{author}{\bibfnamefont{D.}~\bibnamefont{Maurin}},
  \bibinfo{author}{\bibfnamefont{P.}~\bibnamefont{Brun}},
  \bibinfo{author}{\bibfnamefont{T.}~\bibnamefont{Delahaye}}, \bibnamefont{and}
  \bibinfo{author}{\bibfnamefont{P.}~\bibnamefont{Salati}},
  \bibinfo{journal}{Phys.Rev.Lett.} \textbf{\bibinfo{volume}{102}},
  \bibinfo{pages}{071301} (\bibinfo{year}{2009}), \eprint{0810.5292}.

\bibitem[{\citenamefont{Cholis}(2011)}]{Cholis:2010xb}
\bibinfo{author}{\bibfnamefont{I.}~\bibnamefont{Cholis}},
  \bibinfo{journal}{JCAP} \textbf{\bibinfo{volume}{1109}}, \bibinfo{pages}{007}
  (\bibinfo{year}{2011}), \eprint{1007.1160}.

\bibitem[{\citenamefont{Evoli et~al.}(2012)\citenamefont{Evoli, Cholis, Grasso,
  Maccione, and Ullio}}]{Evoli:2011id}
\bibinfo{author}{\bibfnamefont{C.}~\bibnamefont{Evoli}},
  \bibinfo{author}{\bibfnamefont{I.}~\bibnamefont{Cholis}},
  \bibinfo{author}{\bibfnamefont{D.}~\bibnamefont{Grasso}},
  \bibinfo{author}{\bibfnamefont{L.}~\bibnamefont{Maccione}}, \bibnamefont{and}
  \bibinfo{author}{\bibfnamefont{P.}~\bibnamefont{Ullio}},
  \bibinfo{journal}{Phys.Rev.} \textbf{\bibinfo{volume}{D85}},
  \bibinfo{pages}{123511} (\bibinfo{year}{2012}), \eprint{1108.0664}.

\bibitem[{\citenamefont{Adriani et~al.}(2011{\natexlab{b}})}]{Adriani:2011xv}
\bibinfo{author}{\bibfnamefont{O.}~\bibnamefont{Adriani}} \bibnamefont{et~al.}
  (\bibinfo{collaboration}{PAMELA Collaboration}),
  \bibinfo{journal}{Phys.Rev.Lett.} \textbf{\bibinfo{volume}{106}},
  \bibinfo{pages}{201101} (\bibinfo{year}{2011}{\natexlab{b}}),
  \eprint{1103.2880}.

\bibitem[{\citenamefont{Ackermann
  et~al.}(2010{\natexlab{a}})}]{Ackermann:2010ij}
\bibinfo{author}{\bibfnamefont{M.}~\bibnamefont{Ackermann}}
  \bibnamefont{et~al.} (\bibinfo{collaboration}{Fermi LAT}),
  \bibinfo{journal}{Phys. Rev.} \textbf{\bibinfo{volume}{D82}},
  \bibinfo{pages}{092004} (\bibinfo{year}{2010}{\natexlab{a}}),
  \eprint{1008.3999}.

\bibitem[{\citenamefont{Aharonian et~al.}(2008)}]{Aharonian:2008aa}
\bibinfo{author}{\bibfnamefont{F.}~\bibnamefont{Aharonian}}
  \bibnamefont{et~al.} (\bibinfo{collaboration}{H.E.S.S. Collaboration}),
  \bibinfo{journal}{Phys.Rev.Lett.} \textbf{\bibinfo{volume}{101}},
  \bibinfo{pages}{261104} (\bibinfo{year}{2008}), \eprint{0811.3894}.

\bibitem[{\citenamefont{Aharonian et~al.}(2009)}]{Aharonian:2009ah}
\bibinfo{author}{\bibfnamefont{F.}~\bibnamefont{Aharonian}}
  \bibnamefont{et~al.} (\bibinfo{collaboration}{H.E.S.S. Collaboration}),
  \bibinfo{journal}{Astron.Astrophys.} \textbf{\bibinfo{volume}{508}},
  \bibinfo{pages}{561} (\bibinfo{year}{2009}), \eprint{0905.0105}.

\bibitem[{\citenamefont{Borla~Tridon et~al.}(2011)\citenamefont{Borla~Tridon,
  Colin, Cossio, Doro, and Scalzotto}}]{BorlaTridon:2011dk}
\bibinfo{author}{\bibfnamefont{D.}~\bibnamefont{Borla~Tridon}},
  \bibinfo{author}{\bibfnamefont{P.}~\bibnamefont{Colin}},
  \bibinfo{author}{\bibfnamefont{L.}~\bibnamefont{Cossio}},
  \bibinfo{author}{\bibfnamefont{M.}~\bibnamefont{Doro}}, \bibnamefont{and}
  \bibinfo{author}{\bibfnamefont{V.}~\bibnamefont{Scalzotto}}
  (\bibinfo{collaboration}{MAGIC Collaboration}) (\bibinfo{year}{2011}),
  \eprint{1110.4008}.

\bibitem[{\citenamefont{Adriani et~al.}(2009)}]{Adriani:2008zr}
\bibinfo{author}{\bibfnamefont{O.}~\bibnamefont{Adriani}} \bibnamefont{et~al.}
  (\bibinfo{collaboration}{PAMELA}), \bibinfo{journal}{Nature}
  \textbf{\bibinfo{volume}{458}}, \bibinfo{pages}{607} (\bibinfo{year}{2009}),
  \eprint{0810.4995}.

\bibitem[{\citenamefont{Aguilar et~al.}(2013)}]{Aguilar:2013qda}
\bibinfo{author}{\bibfnamefont{M.}~\bibnamefont{Aguilar}} \bibnamefont{et~al.}
  (\bibinfo{collaboration}{AMS Collaboration}),
  \bibinfo{journal}{Phys.Rev.Lett.} \textbf{\bibinfo{volume}{110}},
  \bibinfo{pages}{141102} (\bibinfo{year}{2013}).

\bibitem[{\citenamefont{Moore and Diemand}(2010)}]{Moore:1900zz}
\bibinfo{author}{\bibfnamefont{B.}~\bibnamefont{Moore}} \bibnamefont{and}
  \bibinfo{author}{\bibfnamefont{J.}~\bibnamefont{Diemand}}, in
  \emph{\bibinfo{booktitle}{Particle dark matter}}, edited by
  \bibinfo{editor}{\bibfnamefont{G.~e.} \bibnamefont{Bertone}}
  (\bibinfo{publisher}{Cambridge University Press}, \bibinfo{year}{2010}), pp.
  \bibinfo{pages}{14--37}.

\bibitem[{\citenamefont{{Gallagher} et~al.}(1984)\citenamefont{{Gallagher},
  {Hunter}, and {Tutukov}}}]{1984ApJ...284..544G}
\bibinfo{author}{\bibfnamefont{J.~S.} \bibnamefont{{Gallagher}},
  \bibfnamefont{III}}, \bibinfo{author}{\bibfnamefont{D.~A.}
  \bibnamefont{{Hunter}}}, \bibnamefont{and}
  \bibinfo{author}{\bibfnamefont{A.~V.} \bibnamefont{{Tutukov}}},
  \bibinfo{journal}{\apj} \textbf{\bibinfo{volume}{284}}, \bibinfo{pages}{544}
  (\bibinfo{year}{1984}).

\bibitem[{\citenamefont{Strigari et~al.}(2007)\citenamefont{Strigari,
  Koushiappas, Bullock, and Kaplinghat}}]{Strigari:2006rd}
\bibinfo{author}{\bibfnamefont{L.~E.} \bibnamefont{Strigari}},
  \bibinfo{author}{\bibfnamefont{S.~M.} \bibnamefont{Koushiappas}},
  \bibinfo{author}{\bibfnamefont{J.~S.} \bibnamefont{Bullock}},
  \bibnamefont{and}
  \bibinfo{author}{\bibfnamefont{M.}~\bibnamefont{Kaplinghat}},
  \bibinfo{journal}{Phys.Rev.} \textbf{\bibinfo{volume}{D75}},
  \bibinfo{pages}{083526} (\bibinfo{year}{2007}), \eprint{astro-ph/0611925}.

\bibitem[{\citenamefont{Navarro et~al.}(1996)\citenamefont{Navarro, Frenk, and
  White}}]{Navarro:1995iw}
\bibinfo{author}{\bibfnamefont{J.~F.} \bibnamefont{Navarro}},
  \bibinfo{author}{\bibfnamefont{C.~S.} \bibnamefont{Frenk}}, \bibnamefont{and}
  \bibinfo{author}{\bibfnamefont{S.~D.} \bibnamefont{White}},
  \bibinfo{journal}{Astrophys.J.} \textbf{\bibinfo{volume}{462}},
  \bibinfo{pages}{563} (\bibinfo{year}{1996}), \eprint{astro-ph/9508025}.

\bibitem[{\citenamefont{Catena and Ullio}(2010)}]{Catena:2009mf}
\bibinfo{author}{\bibfnamefont{R.}~\bibnamefont{Catena}} \bibnamefont{and}
  \bibinfo{author}{\bibfnamefont{P.}~\bibnamefont{Ullio}},
  \bibinfo{journal}{JCAP} \textbf{\bibinfo{volume}{1008}}, \bibinfo{pages}{004}
  (\bibinfo{year}{2010}), \eprint{0907.0018}.

\bibitem[{\citenamefont{Salucci et~al.}(2010)\citenamefont{Salucci, Nesti,
  Gentile, and Martins}}]{Salucci:2010qr}
\bibinfo{author}{\bibfnamefont{P.}~\bibnamefont{Salucci}},
  \bibinfo{author}{\bibfnamefont{F.}~\bibnamefont{Nesti}},
  \bibinfo{author}{\bibfnamefont{G.}~\bibnamefont{Gentile}}, \bibnamefont{and}
  \bibinfo{author}{\bibfnamefont{C.}~\bibnamefont{Martins}},
  \bibinfo{journal}{Astron.Astrophys.} \textbf{\bibinfo{volume}{523}},
  \bibinfo{pages}{A83} (\bibinfo{year}{2010}), \eprint{1003.3101}.

\bibitem[{\citenamefont{Moroi and Randall}(2000)}]{Moroi:1999zb}
\bibinfo{author}{\bibfnamefont{T.}~\bibnamefont{Moroi}} \bibnamefont{and}
  \bibinfo{author}{\bibfnamefont{L.}~\bibnamefont{Randall}},
  \bibinfo{journal}{Nucl.Phys.} \textbf{\bibinfo{volume}{B570}},
  \bibinfo{pages}{455} (\bibinfo{year}{2000}), \eprint{hep-ph/9906527}.

\bibitem[{\citenamefont{Acharya et~al.}(2007)\citenamefont{Acharya, Bobkov,
  Kane, Kumar, and Shao}}]{Acharya:2007rc}
\bibinfo{author}{\bibfnamefont{B.~S.} \bibnamefont{Acharya}},
  \bibinfo{author}{\bibfnamefont{K.}~\bibnamefont{Bobkov}},
  \bibinfo{author}{\bibfnamefont{G.~L.} \bibnamefont{Kane}},
  \bibinfo{author}{\bibfnamefont{P.}~\bibnamefont{Kumar}}, \bibnamefont{and}
  \bibinfo{author}{\bibfnamefont{J.}~\bibnamefont{Shao}},
  \bibinfo{journal}{Phys.Rev.} \textbf{\bibinfo{volume}{D76}},
  \bibinfo{pages}{126010} (\bibinfo{year}{2007}), \eprint{hep-th/0701034}.

\bibitem[{\citenamefont{Dame et~al.}(2001)\citenamefont{Dame, Hartmann, and
  Thaddeus}}]{Dame:2000sp}
\bibinfo{author}{\bibfnamefont{T.}~\bibnamefont{Dame}},
  \bibinfo{author}{\bibfnamefont{D.}~\bibnamefont{Hartmann}}, \bibnamefont{and}
  \bibinfo{author}{\bibfnamefont{P.}~\bibnamefont{Thaddeus}},
  \bibinfo{journal}{Astrophys.J.} \textbf{\bibinfo{volume}{547}},
  \bibinfo{pages}{792} (\bibinfo{year}{2001}), \eprint{astro-ph/0009217}.

\bibitem[{\citenamefont{Kobayashi et~al.}(2004)\citenamefont{Kobayashi, Komori,
  Yoshida, and Nishimura}}]{Kobayashi:2003kp}
\bibinfo{author}{\bibfnamefont{T.}~\bibnamefont{Kobayashi}},
  \bibinfo{author}{\bibfnamefont{Y.}~\bibnamefont{Komori}},
  \bibinfo{author}{\bibfnamefont{K.}~\bibnamefont{Yoshida}}, \bibnamefont{and}
  \bibinfo{author}{\bibfnamefont{J.}~\bibnamefont{Nishimura}},
  \bibinfo{journal}{Astrophys. J.} \textbf{\bibinfo{volume}{601}},
  \bibinfo{pages}{340} (\bibinfo{year}{2004}), \eprint{astro-ph/0308470}.

\bibitem[{\citenamefont{Hooper et~al.}(2009)\citenamefont{Hooper, Blasi, and
  Serpico}}]{Hooper:2008kg}
\bibinfo{author}{\bibfnamefont{D.}~\bibnamefont{Hooper}},
  \bibinfo{author}{\bibfnamefont{P.}~\bibnamefont{Blasi}}, \bibnamefont{and}
  \bibinfo{author}{\bibfnamefont{P.~D.} \bibnamefont{Serpico}},
  \bibinfo{journal}{JCAP} \textbf{\bibinfo{volume}{0901}}, \bibinfo{pages}{025}
  (\bibinfo{year}{2009}), \eprint{0810.1527}.

\bibitem[{\citenamefont{Profumo}(2011)}]{Profumo:2008ms}
\bibinfo{author}{\bibfnamefont{S.}~\bibnamefont{Profumo}},
  \bibinfo{journal}{Central Eur.J.Phys.} \textbf{\bibinfo{volume}{10}},
  \bibinfo{pages}{1} (\bibinfo{year}{2011}), \eprint{0812.4457}.

\bibitem[{\citenamefont{Grasso et~al.}(2009)}]{Grasso:2009ma}
\bibinfo{author}{\bibfnamefont{D.}~\bibnamefont{Grasso}} \bibnamefont{et~al.}
  (\bibinfo{collaboration}{FERMI-LAT Collaboration}),
  \bibinfo{journal}{Astropart.Phys.} \textbf{\bibinfo{volume}{32}},
  \bibinfo{pages}{140} (\bibinfo{year}{2009}), \eprint{0905.0636}.

\bibitem[{\citenamefont{Malyshev et~al.}(2009)\citenamefont{Malyshev, Cholis,
  and Gelfand}}]{Malyshev:2009tw}
\bibinfo{author}{\bibfnamefont{D.}~\bibnamefont{Malyshev}},
  \bibinfo{author}{\bibfnamefont{I.}~\bibnamefont{Cholis}}, \bibnamefont{and}
  \bibinfo{author}{\bibfnamefont{J.}~\bibnamefont{Gelfand}},
  \bibinfo{journal}{Phys.Rev.} \textbf{\bibinfo{volume}{D80}},
  \bibinfo{pages}{063005} (\bibinfo{year}{2009}), \eprint{0903.1310}.

\bibitem[{\citenamefont{Faucher-Giguere and
  Kaspi}(2006)}]{FaucherGiguere:2005ny}
\bibinfo{author}{\bibfnamefont{C.-A.} \bibnamefont{Faucher-Giguere}}
  \bibnamefont{and} \bibinfo{author}{\bibfnamefont{V.~M.} \bibnamefont{Kaspi}},
  \bibinfo{journal}{Astrophys.J.} \textbf{\bibinfo{volume}{643}},
  \bibinfo{pages}{332} (\bibinfo{year}{2006}), \eprint{astro-ph/0512585}.

\bibitem[{\citenamefont{Cholis et~al.}(2012)\citenamefont{Cholis, Tavakoli,
  Evoli, Maccione, and Ullio}}]{Cholis:2011un}
\bibinfo{author}{\bibfnamefont{I.}~\bibnamefont{Cholis}},
  \bibinfo{author}{\bibfnamefont{M.}~\bibnamefont{Tavakoli}},
  \bibinfo{author}{\bibfnamefont{C.}~\bibnamefont{Evoli}},
  \bibinfo{author}{\bibfnamefont{L.}~\bibnamefont{Maccione}}, \bibnamefont{and}
  \bibinfo{author}{\bibfnamefont{P.}~\bibnamefont{Ullio}},
  \bibinfo{journal}{JCAP} \textbf{\bibinfo{volume}{1205}}, \bibinfo{pages}{004}
  (\bibinfo{year}{2012}), \eprint{1106.5073}.

\bibitem[{\citenamefont{Abdo et~al.}(2010{\natexlab{a}})}]{Abdo:2010nz}
\bibinfo{author}{\bibfnamefont{A.}~\bibnamefont{Abdo}} \bibnamefont{et~al.}
  (\bibinfo{collaboration}{Fermi-LAT collaboration}),
  \bibinfo{journal}{Phys.Rev.Lett.} \textbf{\bibinfo{volume}{104}},
  \bibinfo{pages}{101101} (\bibinfo{year}{2010}{\natexlab{a}}),
  \eprint{1002.3603}.

\bibitem[{\citenamefont{Tavakoli et~al.}(2011)\citenamefont{Tavakoli, Cholis,
  Evoli, and Ullio}}]{Tavakoli:2011wz}
\bibinfo{author}{\bibfnamefont{M.}~\bibnamefont{Tavakoli}},
  \bibinfo{author}{\bibfnamefont{I.}~\bibnamefont{Cholis}},
  \bibinfo{author}{\bibfnamefont{C.}~\bibnamefont{Evoli}}, \bibnamefont{and}
  \bibinfo{author}{\bibfnamefont{P.}~\bibnamefont{Ullio}}
  (\bibinfo{year}{2011}), \eprint{1110.5922}.

\bibitem[{\citenamefont{Dobler et~al.}(2010)\citenamefont{Dobler, Finkbeiner,
  Cholis, Slatyer, and Weiner}}]{Dobler:2009xz}
\bibinfo{author}{\bibfnamefont{G.}~\bibnamefont{Dobler}},
  \bibinfo{author}{\bibfnamefont{D.~P.} \bibnamefont{Finkbeiner}},
  \bibinfo{author}{\bibfnamefont{I.}~\bibnamefont{Cholis}},
  \bibinfo{author}{\bibfnamefont{T.~R.} \bibnamefont{Slatyer}},
  \bibnamefont{and} \bibinfo{author}{\bibfnamefont{N.}~\bibnamefont{Weiner}},
  \bibinfo{journal}{Astrophys.J.} \textbf{\bibinfo{volume}{717}},
  \bibinfo{pages}{825} (\bibinfo{year}{2010}), \eprint{0910.4583}.

\bibitem[{\citenamefont{Su et~al.}(2010)\citenamefont{Su, Slatyer, and
  Finkbeiner}}]{Su:2010qj}
\bibinfo{author}{\bibfnamefont{M.}~\bibnamefont{Su}},
  \bibinfo{author}{\bibfnamefont{T.~R.} \bibnamefont{Slatyer}},
  \bibnamefont{and} \bibinfo{author}{\bibfnamefont{D.~P.}
  \bibnamefont{Finkbeiner}}, \bibinfo{journal}{Astrophys. J.}
  \textbf{\bibinfo{volume}{724}}, \bibinfo{pages}{1044} (\bibinfo{year}{2010}),
  \eprint{1005.5480}.

\bibitem[{\citenamefont{Dobler et~al.}(2011)\citenamefont{Dobler, Cholis, and
  Weiner}}]{Dobler:2011mk}
\bibinfo{author}{\bibfnamefont{G.}~\bibnamefont{Dobler}},
  \bibinfo{author}{\bibfnamefont{I.}~\bibnamefont{Cholis}}, \bibnamefont{and}
  \bibinfo{author}{\bibfnamefont{N.}~\bibnamefont{Weiner}},
  \bibinfo{journal}{Astrophys.J.} \textbf{\bibinfo{volume}{741}},
  \bibinfo{pages}{25} (\bibinfo{year}{2011}), \eprint{1102.5095}.

\bibitem[{\citenamefont{Tavakoli et~al.}(2013)\citenamefont{Tavakoli, Cholis,
  Evoli, and Ullio}}]{Tavakoli:2013zva}
\bibinfo{author}{\bibfnamefont{M.}~\bibnamefont{Tavakoli}},
  \bibinfo{author}{\bibfnamefont{I.}~\bibnamefont{Cholis}},
  \bibinfo{author}{\bibfnamefont{C.}~\bibnamefont{Evoli}}, \bibnamefont{and}
  \bibinfo{author}{\bibfnamefont{P.}~\bibnamefont{Ullio}}
  (\bibinfo{year}{2013}), \eprint{1308.4135}.

\bibitem[{\citenamefont{Delahaye et~al.}(2011)\citenamefont{Delahaye, Fiasson,
  Pohl, and Salati}}]{Timur:2011vv}
\bibinfo{author}{\bibfnamefont{T.}~\bibnamefont{Delahaye}},
  \bibinfo{author}{\bibfnamefont{A.}~\bibnamefont{Fiasson}},
  \bibinfo{author}{\bibfnamefont{M.}~\bibnamefont{Pohl}}, \bibnamefont{and}
  \bibinfo{author}{\bibfnamefont{P.}~\bibnamefont{Salati}},
  \bibinfo{journal}{Astron.Astrophys.} \textbf{\bibinfo{volume}{531}},
  \bibinfo{pages}{A37} (\bibinfo{year}{2011}), \eprint{1102.0744}.

\bibitem[{\citenamefont{Abdo et~al.}(2010{\natexlab{b}})}]{Abdo:2010dk}
\bibinfo{author}{\bibfnamefont{A.~A.} \bibnamefont{Abdo}} \bibnamefont{et~al.}
  (\bibinfo{collaboration}{Fermi-LAT}), \bibinfo{journal}{JCAP}
  \textbf{\bibinfo{volume}{1004}}, \bibinfo{pages}{014}
  (\bibinfo{year}{2010}{\natexlab{b}}), \eprint{1002.4415}.

\bibitem[{\citenamefont{Hutsi et~al.}(2010)\citenamefont{Hutsi, Hektor, and
  Raidal}}]{Hutsi:2010ai}
\bibinfo{author}{\bibfnamefont{G.}~\bibnamefont{Hutsi}},
  \bibinfo{author}{\bibfnamefont{A.}~\bibnamefont{Hektor}}, \bibnamefont{and}
  \bibinfo{author}{\bibfnamefont{M.}~\bibnamefont{Raidal}},
  \bibinfo{journal}{JCAP} \textbf{\bibinfo{volume}{1007}}, \bibinfo{pages}{008}
  (\bibinfo{year}{2010}), \eprint{1004.2036}.

\bibitem[{\citenamefont{Fornasa et~al.}(2013)\citenamefont{Fornasa, Zavala,
  Sanchez-Conde, Siegal-Gaskins, Delahaye et~al.}}]{Fornasa:2012gu}
\bibinfo{author}{\bibfnamefont{M.}~\bibnamefont{Fornasa}},
  \bibinfo{author}{\bibfnamefont{J.}~\bibnamefont{Zavala}},
  \bibinfo{author}{\bibfnamefont{M.~A.} \bibnamefont{Sanchez-Conde}},
  \bibinfo{author}{\bibfnamefont{J.~M.} \bibnamefont{Siegal-Gaskins}},
  \bibinfo{author}{\bibfnamefont{T.}~\bibnamefont{Delahaye}},
  \bibnamefont{et~al.}, \bibinfo{journal}{MNRAS, 429,}
  \textbf{\bibinfo{volume}{1529}} (\bibinfo{year}{2013}), \eprint{1207.0502}.

\bibitem[{\citenamefont{Ando and Komatsu}(2013)}]{Ando:2013ff}
\bibinfo{author}{\bibfnamefont{S.}~\bibnamefont{Ando}} \bibnamefont{and}
  \bibinfo{author}{\bibfnamefont{E.}~\bibnamefont{Komatsu}},
  \bibinfo{journal}{Phys.Rev.} \textbf{\bibinfo{volume}{D87}},
  \bibinfo{pages}{123539} (\bibinfo{year}{2013}), \eprint{1301.5901}.

\bibitem[{\citenamefont{Donato et~al.}(2012)\citenamefont{Donato, Calore, and
  De~Romeri}}]{Donato:2011pe}
\bibinfo{author}{\bibfnamefont{F.}~\bibnamefont{Donato}},
  \bibinfo{author}{\bibfnamefont{F.}~\bibnamefont{Calore}}, \bibnamefont{and}
  \bibinfo{author}{\bibfnamefont{V.}~\bibnamefont{De~Romeri}},
  \bibinfo{journal}{J.Phys.Conf.Ser.} \textbf{\bibinfo{volume}{375}},
  \bibinfo{pages}{012039} (\bibinfo{year}{2012}), \eprint{1112.4171}.

\bibitem[{\citenamefont{Bringmann et~al.}(2013)\citenamefont{Bringmann, Calore,
  Di~Mauro, and Donato}}]{Calore:2013yia}
\bibinfo{author}{\bibfnamefont{T.}~\bibnamefont{Bringmann}},
  \bibinfo{author}{\bibfnamefont{F.}~\bibnamefont{Calore}},
  \bibinfo{author}{\bibfnamefont{M.}~\bibnamefont{Di~Mauro}}, \bibnamefont{and}
  \bibinfo{author}{\bibfnamefont{F.}~\bibnamefont{Donato}}
  (\bibinfo{year}{2013}), \eprint{1303.3284}.

\bibitem[{\citenamefont{Cholis et~al.}(2013)\citenamefont{Cholis, Hooper, and
  McDermott}}]{Cholis:2013ena}
\bibinfo{author}{\bibfnamefont{I.}~\bibnamefont{Cholis}},
  \bibinfo{author}{\bibfnamefont{D.}~\bibnamefont{Hooper}}, \bibnamefont{and}
  \bibinfo{author}{\bibfnamefont{S.~D.} \bibnamefont{McDermott}}
  (\bibinfo{year}{2013}), \eprint{1312.0608}.

\bibitem[{\citenamefont{Colafrancesco et~al.}(2006)\citenamefont{Colafrancesco,
  Profumo, and Ullio}}]{Colafrancesco:2005ji}
\bibinfo{author}{\bibfnamefont{S.}~\bibnamefont{Colafrancesco}},
  \bibinfo{author}{\bibfnamefont{S.}~\bibnamefont{Profumo}}, \bibnamefont{and}
  \bibinfo{author}{\bibfnamefont{P.}~\bibnamefont{Ullio}},
  \bibinfo{journal}{Astron.Astrophys.} \textbf{\bibinfo{volume}{455}},
  \bibinfo{pages}{21} (\bibinfo{year}{2006}), \eprint{astro-ph/0507575}.

\bibitem[{\citenamefont{Baltz et~al.}(2008)\citenamefont{Baltz, Berenji,
  Bertone, Bergstrom, Bloom et~al.}}]{Baltz:2008wd}
\bibinfo{author}{\bibfnamefont{E.}~\bibnamefont{Baltz}},
  \bibinfo{author}{\bibfnamefont{B.}~\bibnamefont{Berenji}},
  \bibinfo{author}{\bibfnamefont{G.}~\bibnamefont{Bertone}},
  \bibinfo{author}{\bibfnamefont{L.}~\bibnamefont{Bergstrom}},
  \bibinfo{author}{\bibfnamefont{E.}~\bibnamefont{Bloom}},
  \bibnamefont{et~al.}, \bibinfo{journal}{JCAP}
  \textbf{\bibinfo{volume}{0807}}, \bibinfo{pages}{013} (\bibinfo{year}{2008}),
  \eprint{0806.2911}.

\bibitem[{\citenamefont{Jeltema et~al.}(2009)\citenamefont{Jeltema, Kehayias,
  and Profumo}}]{Jeltema:2008vu}
\bibinfo{author}{\bibfnamefont{T.~E.} \bibnamefont{Jeltema}},
  \bibinfo{author}{\bibfnamefont{J.}~\bibnamefont{Kehayias}}, \bibnamefont{and}
  \bibinfo{author}{\bibfnamefont{S.}~\bibnamefont{Profumo}},
  \bibinfo{journal}{Phys.Rev.} \textbf{\bibinfo{volume}{D80}},
  \bibinfo{pages}{023005} (\bibinfo{year}{2009}), \eprint{0812.0597}.

\bibitem[{\citenamefont{Pinzke et~al.}(2009)\citenamefont{Pinzke, Pfrommer, and
  Bergstrom}}]{Pinzke:2009cp}
\bibinfo{author}{\bibfnamefont{A.}~\bibnamefont{Pinzke}},
  \bibinfo{author}{\bibfnamefont{C.}~\bibnamefont{Pfrommer}}, \bibnamefont{and}
  \bibinfo{author}{\bibfnamefont{L.}~\bibnamefont{Bergstrom}},
  \bibinfo{journal}{Phys.Rev.Lett.} \textbf{\bibinfo{volume}{103}},
  \bibinfo{pages}{181302} (\bibinfo{year}{2009}), \eprint{0905.1948}.

\bibitem[{\citenamefont{Pinzke et~al.}(2011)\citenamefont{Pinzke, Pfrommer, and
  Bergstrom}}]{Pinzke:2011ek}
\bibinfo{author}{\bibfnamefont{A.}~\bibnamefont{Pinzke}},
  \bibinfo{author}{\bibfnamefont{C.}~\bibnamefont{Pfrommer}}, \bibnamefont{and}
  \bibinfo{author}{\bibfnamefont{L.}~\bibnamefont{Bergstrom}},
  \bibinfo{journal}{Phys.Rev.} \textbf{\bibinfo{volume}{D84}},
  \bibinfo{pages}{123509} (\bibinfo{year}{2011}), \eprint{1105.3240}.

\bibitem[{\citenamefont{Ando and Nagai}(2012)}]{Ando:2012vu}
\bibinfo{author}{\bibfnamefont{S.}~\bibnamefont{Ando}} \bibnamefont{and}
  \bibinfo{author}{\bibfnamefont{D.}~\bibnamefont{Nagai}},
  \bibinfo{journal}{JCAP} \textbf{\bibinfo{volume}{1207}}, \bibinfo{pages}{017}
  (\bibinfo{year}{2012}), \eprint{1201.0753}.

\bibitem[{\citenamefont{Han et~al.}(2012)\citenamefont{Han, Frenk, Eke, Gao,
  and White}}]{Han:2012au}
\bibinfo{author}{\bibfnamefont{J.}~\bibnamefont{Han}},
  \bibinfo{author}{\bibfnamefont{C.~S.} \bibnamefont{Frenk}},
  \bibinfo{author}{\bibfnamefont{V.~R.} \bibnamefont{Eke}},
  \bibinfo{author}{\bibfnamefont{L.}~\bibnamefont{Gao}}, \bibnamefont{and}
  \bibinfo{author}{\bibfnamefont{S.~D.} \bibnamefont{White}}
  (\bibinfo{year}{2012}), \eprint{1201.1003}.

\bibitem[{\citenamefont{Hektor et~al.}(2013)\citenamefont{Hektor, Raidal, and
  Tempel}}]{Hektor:2012kc}
\bibinfo{author}{\bibfnamefont{A.}~\bibnamefont{Hektor}},
  \bibinfo{author}{\bibfnamefont{M.}~\bibnamefont{Raidal}}, \bibnamefont{and}
  \bibinfo{author}{\bibfnamefont{E.}~\bibnamefont{Tempel}},
  \bibinfo{journal}{Astrophys.J.} \textbf{\bibinfo{volume}{762}},
  \bibinfo{pages}{L22} (\bibinfo{year}{2013}), \eprint{1207.4466}.

\bibitem[{\citenamefont{Aharonian}(2009)}]{Aharonian:2009bc}
\bibinfo{author}{\bibfnamefont{F.}~\bibnamefont{Aharonian}}
  (\bibinfo{collaboration}{HESS Collaboration}) (\bibinfo{year}{2009}),
  \eprint{0907.0727}.

\bibitem[{\citenamefont{Aleksic et~al.}(2010)}]{Aleksic:2009ir}
\bibinfo{author}{\bibfnamefont{J.}~\bibnamefont{Aleksic}} \bibnamefont{et~al.}
  (\bibinfo{collaboration}{MAGIC Collaboration}),
  \bibinfo{journal}{Astrophys.J.} \textbf{\bibinfo{volume}{710}},
  \bibinfo{pages}{634} (\bibinfo{year}{2010}), \eprint{0909.3267}.

\bibitem[{\citenamefont{Ackermann
  et~al.}(2010{\natexlab{b}})\citenamefont{Ackermann, Ajello, Allafort,
  Baldini, Ballet et~al.}}]{Ackermann:2010rg}
\bibinfo{author}{\bibfnamefont{M.}~\bibnamefont{Ackermann}},
  \bibinfo{author}{\bibfnamefont{M.}~\bibnamefont{Ajello}},
  \bibinfo{author}{\bibfnamefont{A.}~\bibnamefont{Allafort}},
  \bibinfo{author}{\bibfnamefont{L.}~\bibnamefont{Baldini}},
  \bibinfo{author}{\bibfnamefont{J.}~\bibnamefont{Ballet}},
  \bibnamefont{et~al.}, \bibinfo{journal}{JCAP}
  \textbf{\bibinfo{volume}{1005}}, \bibinfo{pages}{025}
  (\bibinfo{year}{2010}{\natexlab{b}}), \eprint{1002.2239}.

\bibitem[{\citenamefont{Dugger et~al.}(2010)\citenamefont{Dugger, Jeltema, and
  Profumo}}]{Dugger:2010ys}
\bibinfo{author}{\bibfnamefont{L.}~\bibnamefont{Dugger}},
  \bibinfo{author}{\bibfnamefont{T.~E.} \bibnamefont{Jeltema}},
  \bibnamefont{and} \bibinfo{author}{\bibfnamefont{S.}~\bibnamefont{Profumo}},
  \bibinfo{journal}{JCAP} \textbf{\bibinfo{volume}{1012}}, \bibinfo{pages}{015}
  (\bibinfo{year}{2010}), \eprint{1009.5988}.

\bibitem[{\citenamefont{Zimmer et~al.}(2011)\citenamefont{Zimmer, Conrad, and
  Pinzke}}]{Zimmer:2011vy}
\bibinfo{author}{\bibfnamefont{S.}~\bibnamefont{Zimmer}},
  \bibinfo{author}{\bibfnamefont{J.}~\bibnamefont{Conrad}}, \bibnamefont{and}
  \bibinfo{author}{\bibfnamefont{A.}~\bibnamefont{Pinzke}}
  (\bibinfo{collaboration}{Fermi-LAT Collaboration}) (\bibinfo{year}{2011}),
  \eprint{1110.6863}.

\bibitem[{\citenamefont{Huang et~al.}(2012)\citenamefont{Huang, Vertongen, and
  Weniger}}]{Huang:2011xr}
\bibinfo{author}{\bibfnamefont{X.}~\bibnamefont{Huang}},
  \bibinfo{author}{\bibfnamefont{G.}~\bibnamefont{Vertongen}},
  \bibnamefont{and} \bibinfo{author}{\bibfnamefont{C.}~\bibnamefont{Weniger}},
  \bibinfo{journal}{JCAP} \textbf{\bibinfo{volume}{1201}}, \bibinfo{pages}{042}
  (\bibinfo{year}{2012}), \eprint{1110.1529}.

\bibitem[{\citenamefont{Walker}(2012)}]{Walker:2012td}
\bibinfo{author}{\bibfnamefont{M.~G.} \bibnamefont{Walker}}
  (\bibinfo{year}{2012}), \eprint{1205.0311}.

\bibitem[{\citenamefont{Evans et~al.}(2004)\citenamefont{Evans, Ferrer, and
  Sarkar}}]{Evans:2003sc}
\bibinfo{author}{\bibfnamefont{N.}~\bibnamefont{Evans}},
  \bibinfo{author}{\bibfnamefont{F.}~\bibnamefont{Ferrer}}, \bibnamefont{and}
  \bibinfo{author}{\bibfnamefont{S.}~\bibnamefont{Sarkar}},
  \bibinfo{journal}{Phys.Rev.} \textbf{\bibinfo{volume}{D69}},
  \bibinfo{pages}{123501} (\bibinfo{year}{2004}), \eprint{astro-ph/0311145}.

\bibitem[{\citenamefont{Colafrancesco et~al.}(2007)\citenamefont{Colafrancesco,
  Profumo, and Ullio}}]{Colafrancesco:2006he}
\bibinfo{author}{\bibfnamefont{S.}~\bibnamefont{Colafrancesco}},
  \bibinfo{author}{\bibfnamefont{S.}~\bibnamefont{Profumo}}, \bibnamefont{and}
  \bibinfo{author}{\bibfnamefont{P.}~\bibnamefont{Ullio}},
  \bibinfo{journal}{Phys.Rev.} \textbf{\bibinfo{volume}{D75}},
  \bibinfo{pages}{023513} (\bibinfo{year}{2007}), \eprint{astro-ph/0607073}.

\bibitem[{\citenamefont{Bovy}(2009)}]{Bovy:2009zs}
\bibinfo{author}{\bibfnamefont{J.}~\bibnamefont{Bovy}},
  \bibinfo{journal}{Phys.Rev.} \textbf{\bibinfo{volume}{D79}},
  \bibinfo{pages}{083539} (\bibinfo{year}{2009}), \eprint{0903.0413}.

\bibitem[{\citenamefont{Scott et~al.}(2010)\citenamefont{Scott, Conrad, Edsjo,
  Bergstrom, Farnier et~al.}}]{Scott:2009jn}
\bibinfo{author}{\bibfnamefont{P.}~\bibnamefont{Scott}},
  \bibinfo{author}{\bibfnamefont{J.}~\bibnamefont{Conrad}},
  \bibinfo{author}{\bibfnamefont{J.}~\bibnamefont{Edsjo}},
  \bibinfo{author}{\bibfnamefont{L.}~\bibnamefont{Bergstrom}},
  \bibinfo{author}{\bibfnamefont{C.}~\bibnamefont{Farnier}},
  \bibnamefont{et~al.}, \bibinfo{journal}{JCAP}
  \textbf{\bibinfo{volume}{1001}}, \bibinfo{pages}{031} (\bibinfo{year}{2010}),
  \eprint{0909.3300}.

\bibitem[{\citenamefont{Perelstein and Shakya}(2010)}]{Perelstein:2010at}
\bibinfo{author}{\bibfnamefont{M.}~\bibnamefont{Perelstein}} \bibnamefont{and}
  \bibinfo{author}{\bibfnamefont{B.}~\bibnamefont{Shakya}},
  \bibinfo{journal}{JCAP} \textbf{\bibinfo{volume}{1010}}, \bibinfo{pages}{016}
  (\bibinfo{year}{2010}), \eprint{1007.0018}.

\bibitem[{\citenamefont{Abdo et~al.}(2010{\natexlab{c}})\citenamefont{Abdo,
  Ackermann, Ajello, Atwood, Baldini et~al.}}]{Abdo:2010ex}
\bibinfo{author}{\bibfnamefont{A.}~\bibnamefont{Abdo}},
  \bibinfo{author}{\bibfnamefont{M.}~\bibnamefont{Ackermann}},
  \bibinfo{author}{\bibfnamefont{M.}~\bibnamefont{Ajello}},
  \bibinfo{author}{\bibfnamefont{W.}~\bibnamefont{Atwood}},
  \bibinfo{author}{\bibfnamefont{L.}~\bibnamefont{Baldini}},
  \bibnamefont{et~al.}, \bibinfo{journal}{Astrophys.J.}
  \textbf{\bibinfo{volume}{712}}, \bibinfo{pages}{147}
  (\bibinfo{year}{2010}{\natexlab{c}}), \eprint{1001.4531}.

\bibitem[{\citenamefont{Ackermann et~al.}(2011)}]{Ackermann:2011wa}
\bibinfo{author}{\bibfnamefont{M.}~\bibnamefont{Ackermann}}
  \bibnamefont{et~al.} (\bibinfo{collaboration}{Fermi-LAT collaboration}),
  \bibinfo{journal}{Phys.Rev.Lett.} \textbf{\bibinfo{volume}{107}},
  \bibinfo{pages}{241302} (\bibinfo{year}{2011}), \eprint{1108.3546}.

\bibitem[{\citenamefont{Ackermann et~al.}(2013)}]{Ackermann:2013yva}
\bibinfo{author}{\bibfnamefont{M.}~\bibnamefont{Ackermann}}
  \bibnamefont{et~al.} (\bibinfo{collaboration}{Fermi-LAT Collaboration})
  (\bibinfo{year}{2013}), \eprint{1310.0828}.

\bibitem[{\citenamefont{Cholis and Salucci}(2012)}]{Cholis:2012ve}
\bibinfo{author}{\bibfnamefont{I.}~\bibnamefont{Cholis}} \bibnamefont{and}
  \bibinfo{author}{\bibfnamefont{P.}~\bibnamefont{Salucci}}
  (\bibinfo{year}{2012}), \eprint{1203.2954},
  \urlprefix\url{http://arxiv.org/abs/1203.2954}.

\bibitem[{\citenamefont{Salucci et~al.}(2011)\citenamefont{Salucci, Wilkinson,
  Walker, Gilmore, Grebel et~al.}}]{Salucci:2011ee}
\bibinfo{author}{\bibfnamefont{P.}~\bibnamefont{Salucci}},
  \bibinfo{author}{\bibfnamefont{M.~I.} \bibnamefont{Wilkinson}},
  \bibinfo{author}{\bibfnamefont{M.~G.} \bibnamefont{Walker}},
  \bibinfo{author}{\bibfnamefont{G.~F.} \bibnamefont{Gilmore}},
  \bibinfo{author}{\bibfnamefont{E.~K.} \bibnamefont{Grebel}},
  \bibnamefont{et~al.} (\bibinfo{year}{2011}), \eprint{1111.1165}.

\bibitem[{\citenamefont{{Carrera} et~al.}(2002)\citenamefont{{Carrera},
  {Aparicio}, {Mart{\'{\i}}nez-Delgado}, and
  {Alonso-Garc{\'{\i}}a}}}]{carrera2002}
\bibinfo{author}{\bibfnamefont{R.}~\bibnamefont{{Carrera}}},
  \bibinfo{author}{\bibfnamefont{A.}~\bibnamefont{{Aparicio}}},
  \bibinfo{author}{\bibfnamefont{D.}~\bibnamefont{{Mart{\'{\i}}nez-Delgado}}},
  \bibnamefont{and}
  \bibinfo{author}{\bibfnamefont{J.}~\bibnamefont{{Alonso-Garc{\'{\i}}a}}},
  \bibinfo{journal}{\apj} \textbf{\bibinfo{volume}{123}}, \bibinfo{pages}{3199}
  (\bibinfo{year}{2002}), \eprint{astro-ph/0203300}.

\bibitem[{\citenamefont{{Bellazzini} et~al.}(2002)\citenamefont{{Bellazzini},
  {Ferraro}, {Origlia}, {Pancino}, {Monaco}, and {Oliva}}}]{Bellazzini2002}
\bibinfo{author}{\bibfnamefont{M.}~\bibnamefont{{Bellazzini}}},
  \bibinfo{author}{\bibfnamefont{F.~R.} \bibnamefont{{Ferraro}}},
  \bibinfo{author}{\bibfnamefont{L.}~\bibnamefont{{Origlia}}},
  \bibinfo{author}{\bibfnamefont{E.}~\bibnamefont{{Pancino}}},
  \bibinfo{author}{\bibfnamefont{L.}~\bibnamefont{{Monaco}}}, \bibnamefont{and}
  \bibinfo{author}{\bibfnamefont{E.}~\bibnamefont{{Oliva}}},
  \bibinfo{journal}{\apj} \textbf{\bibinfo{volume}{124}}, \bibinfo{pages}{3222}
  (\bibinfo{year}{2002}), \eprint{astro-ph/0209391}.

\bibitem[{\citenamefont{{Olszewski} and {Aaronson}}(1985)}]{Olszewski1985}
\bibinfo{author}{\bibfnamefont{E.~W.} \bibnamefont{{Olszewski}}}
  \bibnamefont{and}
  \bibinfo{author}{\bibfnamefont{M.}~\bibnamefont{{Aaronson}}},
  \bibinfo{journal}{\apj} \textbf{\bibinfo{volume}{90}}, \bibinfo{pages}{2221}
  (\bibinfo{year}{1985}).

\bibitem[{\citenamefont{{Cudworth} et~al.}(1986)\citenamefont{{Cudworth},
  {Olszewski}, and {Schommer}}}]{Cudworth1986}
\bibinfo{author}{\bibfnamefont{K.~M.} \bibnamefont{{Cudworth}}},
  \bibinfo{author}{\bibfnamefont{E.~W.} \bibnamefont{{Olszewski}}},
  \bibnamefont{and} \bibinfo{author}{\bibfnamefont{R.~A.}
  \bibnamefont{{Schommer}}}, \bibinfo{journal}{\apj}
  \textbf{\bibinfo{volume}{92}}, \bibinfo{pages}{766} (\bibinfo{year}{1986}).

\bibitem[{\citenamefont{{Piatek} et~al.}(2005)\citenamefont{{Piatek}, {Pryor},
  {Bristow}, {Olszewski}, {Harris}, {Mateo}, {Minniti}, and
  {Tinney}}}]{Piatek2005}
\bibinfo{author}{\bibfnamefont{S.}~\bibnamefont{{Piatek}}},
  \bibinfo{author}{\bibfnamefont{C.}~\bibnamefont{{Pryor}}},
  \bibinfo{author}{\bibfnamefont{P.}~\bibnamefont{{Bristow}}},
  \bibinfo{author}{\bibfnamefont{E.~W.} \bibnamefont{{Olszewski}}},
  \bibinfo{author}{\bibfnamefont{H.~C.} \bibnamefont{{Harris}}},
  \bibinfo{author}{\bibfnamefont{M.}~\bibnamefont{{Mateo}}},
  \bibinfo{author}{\bibfnamefont{D.}~\bibnamefont{{Minniti}}},
  \bibnamefont{and} \bibinfo{author}{\bibfnamefont{C.~G.}
  \bibnamefont{{Tinney}}}, \bibinfo{journal}{\apj}
  \textbf{\bibinfo{volume}{130}}, \bibinfo{pages}{95} (\bibinfo{year}{2005}),
  \eprint{astro-ph/0503620}.

\bibitem[{\citenamefont{{Nemec} et~al.}(1988)\citenamefont{{Nemec}, {Wehlau},
  and {Mendes de Oliveira}}}]{Nemec1988}
\bibinfo{author}{\bibfnamefont{J.~M.} \bibnamefont{{Nemec}}},
  \bibinfo{author}{\bibfnamefont{A.}~\bibnamefont{{Wehlau}}}, \bibnamefont{and}
  \bibinfo{author}{\bibfnamefont{C.}~\bibnamefont{{Mendes de Oliveira}}},
  \bibinfo{journal}{\apj} \textbf{\bibinfo{volume}{96}}, \bibinfo{pages}{528}
  (\bibinfo{year}{1988}).

\bibitem[{\citenamefont{{Mighell} and {Burke}}(1999)}]{Mighell1999}
\bibinfo{author}{\bibfnamefont{K.~J.} \bibnamefont{{Mighell}}}
  \bibnamefont{and} \bibinfo{author}{\bibfnamefont{C.~J.}
  \bibnamefont{{Burke}}}, \bibinfo{journal}{\apj}
  \textbf{\bibinfo{volume}{118}}, \bibinfo{pages}{366} (\bibinfo{year}{1999}),
  \eprint{astro-ph/9903065}.

\bibitem[{\citenamefont{{Lux} et~al.}(2010)\citenamefont{{Lux}, {Read}, and
  {Lake}}}]{lux2010}
\bibinfo{author}{\bibfnamefont{H.}~\bibnamefont{{Lux}}},
  \bibinfo{author}{\bibfnamefont{J.~I.} \bibnamefont{{Read}}},
  \bibnamefont{and} \bibinfo{author}{\bibfnamefont{G.}~\bibnamefont{{Lake}}},
  \bibinfo{journal}{\mnras} \textbf{\bibinfo{volume}{406}},
  \bibinfo{pages}{2312} (\bibinfo{year}{2010}), \eprint{1001.1731}.

\bibitem[{\citenamefont{{Mu{\~n}oz} et~al.}(2008)\citenamefont{{Mu{\~n}oz},
  {Majewski}, and {Johnston}}}]{munoz2008}
\bibinfo{author}{\bibfnamefont{R.~R.} \bibnamefont{{Mu{\~n}oz}}},
  \bibinfo{author}{\bibfnamefont{S.~R.} \bibnamefont{{Majewski}}},
  \bibnamefont{and} \bibinfo{author}{\bibfnamefont{K.~V.}
  \bibnamefont{{Johnston}}}, \bibinfo{journal}{\apj}
  \textbf{\bibinfo{volume}{679}}, \bibinfo{pages}{346} (\bibinfo{year}{2008}),
  \eprint{0712.4312}.

\bibitem[{\citenamefont{{Irwin} and {Hatzidimitriou}}(1995)}]{Irwin1995}
\bibinfo{author}{\bibfnamefont{M.}~\bibnamefont{{Irwin}}} \bibnamefont{and}
  \bibinfo{author}{\bibfnamefont{D.}~\bibnamefont{{Hatzidimitriou}}},
  \bibinfo{journal}{\mnras} \textbf{\bibinfo{volume}{277}},
  \bibinfo{pages}{1354} (\bibinfo{year}{1995}).

\bibitem[{\citenamefont{Geringer-Sameth and
  Koushiappas}(2011)}]{GeringerSameth:2011iw}
\bibinfo{author}{\bibfnamefont{A.}~\bibnamefont{Geringer-Sameth}}
  \bibnamefont{and} \bibinfo{author}{\bibfnamefont{S.~M.}
  \bibnamefont{Koushiappas}}, \bibinfo{journal}{Phys.Rev.Lett.}
  \textbf{\bibinfo{volume}{107}}, \bibinfo{pages}{241303}
  (\bibinfo{year}{2011}), \eprint{1108.2914}.

\bibitem[{\citenamefont{Bringmann and Weniger}(2012)}]{Bringmann:2012ez}
\bibinfo{author}{\bibfnamefont{T.}~\bibnamefont{Bringmann}} \bibnamefont{and}
  \bibinfo{author}{\bibfnamefont{C.}~\bibnamefont{Weniger}},
  \bibinfo{journal}{Phys.Dark Univ.} \textbf{\bibinfo{volume}{1}},
  \bibinfo{pages}{194} (\bibinfo{year}{2012}), \eprint{1208.5481}.

\bibitem[{1235384()}]{Fermi-LAT:2013uma}
1235384, \bibinfo{journal}{Physical Review D 88,}
  \textbf{\bibinfo{volume}{082002}} (\bibinfo{year}{2013}), \eprint{1305.5597}.

\bibitem[{\citenamefont{Abramowski et~al.}(2013)}]{Abramowski:2013ax}
\bibinfo{author}{\bibfnamefont{A.}~\bibnamefont{Abramowski}}
  \bibnamefont{et~al.} (\bibinfo{collaboration}{H.E.S.S. Collaboration}),
  \bibinfo{journal}{Phys.Rev.Lett.} \textbf{\bibinfo{volume}{110}},
  \bibinfo{pages}{041301} (\bibinfo{year}{2013}), \eprint{1301.1173}.

\bibitem[{\citenamefont{Cohen et~al.}(2013)\citenamefont{Cohen, Lisanti,
  Pierce, and Slatyer}}]{Cohen:2013ama}
\bibinfo{author}{\bibfnamefont{T.}~\bibnamefont{Cohen}},
  \bibinfo{author}{\bibfnamefont{M.}~\bibnamefont{Lisanti}},
  \bibinfo{author}{\bibfnamefont{A.}~\bibnamefont{Pierce}}, \bibnamefont{and}
  \bibinfo{author}{\bibfnamefont{T.~R.} \bibnamefont{Slatyer}},
  \bibinfo{journal}{JCAP} \textbf{\bibinfo{volume}{1310}}, \bibinfo{pages}{061}
  (\bibinfo{year}{2013}), \eprint{1307.4082}.

\bibitem[{\citenamefont{Fan and Reece}(2013)}]{Fan:2013faa}
\bibinfo{author}{\bibfnamefont{J.}~\bibnamefont{Fan}} \bibnamefont{and}
  \bibinfo{author}{\bibfnamefont{M.}~\bibnamefont{Reece}},
  \bibinfo{journal}{JHEP} \textbf{\bibinfo{volume}{1310}}, \bibinfo{pages}{124}
  (\bibinfo{year}{2013}), \eprint{1307.4400}.

\bibitem[{\citenamefont{Graham et~al.}(2006)\citenamefont{Graham, Merritt,
  Moore, Diemand, and Terzic}}]{Graham:2006ae}
\bibinfo{author}{\bibfnamefont{A.~W.} \bibnamefont{Graham}},
  \bibinfo{author}{\bibfnamefont{D.}~\bibnamefont{Merritt}},
  \bibinfo{author}{\bibfnamefont{B.}~\bibnamefont{Moore}},
  \bibinfo{author}{\bibfnamefont{J.}~\bibnamefont{Diemand}}, \bibnamefont{and}
  \bibinfo{author}{\bibfnamefont{B.}~\bibnamefont{Terzic}},
  \bibinfo{journal}{Astron.J.} \textbf{\bibinfo{volume}{132}},
  \bibinfo{pages}{2701} (\bibinfo{year}{2006}), \eprint{astro-ph/0608613}.

\bibitem[{\citenamefont{Burkert}(1996)}]{Burkert:1995yz}
\bibinfo{author}{\bibfnamefont{A.}~\bibnamefont{Burkert}},
  \bibinfo{journal}{IAU Symp.} \textbf{\bibinfo{volume}{171}},
  \bibinfo{pages}{175} (\bibinfo{year}{1996}), \eprint{astro-ph/9504041}.

\bibitem[{\citenamefont{Dasgupta and Laha}(2012)}]{Dasgupta:2012bd}
\bibinfo{author}{\bibfnamefont{B.}~\bibnamefont{Dasgupta}} \bibnamefont{and}
  \bibinfo{author}{\bibfnamefont{R.}~\bibnamefont{Laha}},
  \bibinfo{journal}{Phys.Rev.} \textbf{\bibinfo{volume}{D86}},
  \bibinfo{pages}{093001} (\bibinfo{year}{2012}), \eprint{1206.1322}.

\bibitem[{\citenamefont{Cholis}(2013)}]{Cholis:2012fr}
\bibinfo{author}{\bibfnamefont{I.}~\bibnamefont{Cholis}},
  \bibinfo{journal}{Phys. Rev. D 88,} \textbf{\bibinfo{volume}{063524}},
  \bibinfo{pages}{063524} (\bibinfo{year}{2013}), \eprint{1206.1607}.

\bibitem[{\citenamefont{Honda et~al.}(2007)\citenamefont{Honda, Kajita,
  Kasahara, Midorikawa, and Sanuki}}]{Honda:2006qj}
\bibinfo{author}{\bibfnamefont{M.}~\bibnamefont{Honda}},
  \bibinfo{author}{\bibfnamefont{T.}~\bibnamefont{Kajita}},
  \bibinfo{author}{\bibfnamefont{K.}~\bibnamefont{Kasahara}},
  \bibinfo{author}{\bibfnamefont{S.}~\bibnamefont{Midorikawa}},
  \bibnamefont{and} \bibinfo{author}{\bibfnamefont{T.}~\bibnamefont{Sanuki}},
  \bibinfo{journal}{Phys.Rev.} \textbf{\bibinfo{volume}{D75}},
  \bibinfo{pages}{043006} (\bibinfo{year}{2007}), \eprint{astro-ph/0611418}.

\bibitem[{\citenamefont{Aartsen et~al.}(2013)}]{Aartsen:2013jdh}
\bibinfo{author}{\bibfnamefont{M.}~\bibnamefont{Aartsen}} \bibnamefont{et~al.}
  (\bibinfo{collaboration}{IceCube}), \bibinfo{journal}{Science}
  \textbf{\bibinfo{volume}{342}}, \bibinfo{pages}{1242856}
  (\bibinfo{year}{2013}), \eprint{1311.5238}.

\bibitem[{\citenamefont{Anchordoqui et~al.}(2013)\citenamefont{Anchordoqui,
  Barger, Cholis, Goldberg, Hooper et~al.}}]{Anchordoqui:2013dnh}
\bibinfo{author}{\bibfnamefont{L.~A.} \bibnamefont{Anchordoqui}},
  \bibinfo{author}{\bibfnamefont{V.}~\bibnamefont{Barger}},
  \bibinfo{author}{\bibfnamefont{I.}~\bibnamefont{Cholis}},
  \bibinfo{author}{\bibfnamefont{H.}~\bibnamefont{Goldberg}},
  \bibinfo{author}{\bibfnamefont{D.}~\bibnamefont{Hooper}},
  \bibnamefont{et~al.} (\bibinfo{year}{2013}), \eprint{1312.6587}.

\bibitem[{\citenamefont{Tsirigotis et~al.}(2012)\citenamefont{Tsirigotis,
  Leisos, Tzamarias, and Consortium}}]{Tsirigotis:2012nr}
\bibinfo{author}{\bibfnamefont{A.}~\bibnamefont{Tsirigotis}},
  \bibinfo{author}{\bibfnamefont{A.}~\bibnamefont{Leisos}},
  \bibinfo{author}{\bibfnamefont{S.}~\bibnamefont{Tzamarias}},
  \bibnamefont{and} \bibinfo{author}{\bibfnamefont{o.~b. o. t.~K.}
  \bibnamefont{Consortium}} (\bibinfo{year}{2012}), \eprint{1201.5079}.

\bibitem[{\citenamefont{Galli et~al.}(2009)\citenamefont{Galli, Iocco, Bertone,
  and Melchiorri}}]{Galli:2009zc}
\bibinfo{author}{\bibfnamefont{S.}~\bibnamefont{Galli}},
  \bibinfo{author}{\bibfnamefont{F.}~\bibnamefont{Iocco}},
  \bibinfo{author}{\bibfnamefont{G.}~\bibnamefont{Bertone}}, \bibnamefont{and}
  \bibinfo{author}{\bibfnamefont{A.}~\bibnamefont{Melchiorri}},
  \bibinfo{journal}{Phys.Rev.} \textbf{\bibinfo{volume}{D80}},
  \bibinfo{pages}{023505} (\bibinfo{year}{2009}), \eprint{0905.0003}.

\bibitem[{\citenamefont{Slatyer et~al.}(2009)\citenamefont{Slatyer,
  Padmanabhan, and Finkbeiner}}]{Slatyer:2009yq}
\bibinfo{author}{\bibfnamefont{T.~R.} \bibnamefont{Slatyer}},
  \bibinfo{author}{\bibfnamefont{N.}~\bibnamefont{Padmanabhan}},
  \bibnamefont{and} \bibinfo{author}{\bibfnamefont{D.~P.}
  \bibnamefont{Finkbeiner}}, \bibinfo{journal}{Phys.Rev.}
  \textbf{\bibinfo{volume}{D80}}, \bibinfo{pages}{043526}
  (\bibinfo{year}{2009}), \eprint{0906.1197}.

\bibitem[{\citenamefont{Lopez-Honorez et~al.}(2013)\citenamefont{Lopez-Honorez,
  Mena, Palomares-Ruiz, and Vincent}}]{Lopez-Honorez:2013cua}
\bibinfo{author}{\bibfnamefont{L.}~\bibnamefont{Lopez-Honorez}},
  \bibinfo{author}{\bibfnamefont{O.}~\bibnamefont{Mena}},
  \bibinfo{author}{\bibfnamefont{S.}~\bibnamefont{Palomares-Ruiz}},
  \bibnamefont{and} \bibinfo{author}{\bibfnamefont{A.~C.}
  \bibnamefont{Vincent}}, \bibinfo{journal}{JCAP}
  \textbf{\bibinfo{volume}{1307}}, \bibinfo{pages}{046} (\bibinfo{year}{2013}),
  \eprint{1303.5094}.

\bibitem[{\citenamefont{Galli et~al.}(2013)\citenamefont{Galli, Slatyer,
  Valdes, and Iocco}}]{Galli:2013dna}
\bibinfo{author}{\bibfnamefont{S.}~\bibnamefont{Galli}},
  \bibinfo{author}{\bibfnamefont{T.~R.} \bibnamefont{Slatyer}},
  \bibinfo{author}{\bibfnamefont{M.}~\bibnamefont{Valdes}}, \bibnamefont{and}
  \bibinfo{author}{\bibfnamefont{F.}~\bibnamefont{Iocco}}
  (\bibinfo{year}{2013}), \eprint{1306.0563}.

\bibitem[{\citenamefont{Komatsu et~al.}(2009)}]{Komatsu:2008hk}
\bibinfo{author}{\bibfnamefont{E.}~\bibnamefont{Komatsu}} \bibnamefont{et~al.}
  (\bibinfo{collaboration}{WMAP Collaboration}),
  \bibinfo{journal}{Astrophys.J.Suppl.} \textbf{\bibinfo{volume}{180}},
  \bibinfo{pages}{330} (\bibinfo{year}{2009}), \eprint{0803.0547}.

\bibitem[{\citenamefont{Sievers et~al.}(2013)}]{Sievers:2013ica}
\bibinfo{author}{\bibfnamefont{J.~L.} \bibnamefont{Sievers}}
  \bibnamefont{et~al.} (\bibinfo{collaboration}{Atacama Cosmology Telescope}),
  \bibinfo{journal}{JCAP} \textbf{\bibinfo{volume}{1310}}, \bibinfo{pages}{060}
  (\bibinfo{year}{2013}), \eprint{1301.0824}.

\bibitem[{\citenamefont{Story et~al.}(2013)\citenamefont{Story, Reichardt, Hou,
  Keisler, Aird et~al.}}]{Story:2012wx}
\bibinfo{author}{\bibfnamefont{K.}~\bibnamefont{Story}},
  \bibinfo{author}{\bibfnamefont{C.}~\bibnamefont{Reichardt}},
  \bibinfo{author}{\bibfnamefont{Z.}~\bibnamefont{Hou}},
  \bibinfo{author}{\bibfnamefont{R.}~\bibnamefont{Keisler}},
  \bibinfo{author}{\bibfnamefont{K.}~\bibnamefont{Aird}}, \bibnamefont{et~al.},
  \bibinfo{journal}{Astrophys.J.} \textbf{\bibinfo{volume}{779}},
  \bibinfo{pages}{86} (\bibinfo{year}{2013}), \eprint{1210.7231}.

\bibitem[{\citenamefont{Donato et~al.}(2000)\citenamefont{Donato, Fornengo, and
  Salati}}]{Donato:1999gy}
\bibinfo{author}{\bibfnamefont{F.}~\bibnamefont{Donato}},
  \bibinfo{author}{\bibfnamefont{N.}~\bibnamefont{Fornengo}}, \bibnamefont{and}
  \bibinfo{author}{\bibfnamefont{P.}~\bibnamefont{Salati}},
  \bibinfo{journal}{Phys.Rev.} \textbf{\bibinfo{volume}{D62}},
  \bibinfo{pages}{043003} (\bibinfo{year}{2000}), \eprint{hep-ph/9904481}.

\bibitem[{\citenamefont{Duperray et~al.}(2005)\citenamefont{Duperray, Baret,
  Maurin, Boudoul, Barrau et~al.}}]{Duperray:2005si}
\bibinfo{author}{\bibfnamefont{R.}~\bibnamefont{Duperray}},
  \bibinfo{author}{\bibfnamefont{B.}~\bibnamefont{Baret}},
  \bibinfo{author}{\bibfnamefont{D.}~\bibnamefont{Maurin}},
  \bibinfo{author}{\bibfnamefont{G.}~\bibnamefont{Boudoul}},
  \bibinfo{author}{\bibfnamefont{A.}~\bibnamefont{Barrau}},
  \bibnamefont{et~al.}, \bibinfo{journal}{Phys.Rev.}
  \textbf{\bibinfo{volume}{D71}}, \bibinfo{pages}{083013}
  (\bibinfo{year}{2005}), \eprint{astro-ph/0503544}.

\bibitem[{\citenamefont{Donato et~al.}(2008)\citenamefont{Donato, Fornengo, and
  Maurin}}]{Donato:2008uq}
\bibinfo{author}{\bibfnamefont{F.}~\bibnamefont{Donato}},
  \bibinfo{author}{\bibfnamefont{N.}~\bibnamefont{Fornengo}}, \bibnamefont{and}
  \bibinfo{author}{\bibfnamefont{D.}~\bibnamefont{Maurin}}
  (\bibinfo{year}{2008}), \eprint{0803.2640v1},
  \urlprefix\url{http://arxiv.org/abs/0803.2640v1}.

\bibitem[{\citenamefont{Braeuninger and Cirelli}(2009)}]{Braeuninger:2009bh}
\bibinfo{author}{\bibfnamefont{C.~B.} \bibnamefont{Braeuninger}}
  \bibnamefont{and} \bibinfo{author}{\bibfnamefont{M.}~\bibnamefont{Cirelli}}
  (\bibinfo{year}{2009}), \eprint{0904.1165v1},
  \urlprefix\url{http://arxiv.org/abs/0904.1165v1}.

\bibitem[{\citenamefont{Kadastik et~al.}(2009)\citenamefont{Kadastik, Raidal,
  and Strumia}}]{Kadastik:2009fk}
\bibinfo{author}{\bibfnamefont{M.}~\bibnamefont{Kadastik}},
  \bibinfo{author}{\bibfnamefont{M.}~\bibnamefont{Raidal}}, \bibnamefont{and}
  \bibinfo{author}{\bibfnamefont{A.}~\bibnamefont{Strumia}}
  (\bibinfo{year}{2009}), \eprint{0908.1578v2},
  \urlprefix\url{http://arxiv.org/abs/0908.1578v2}.

\bibitem[{\citenamefont{Cui et~al.}(2010)\citenamefont{Cui, Mason, and
  Randall}}]{Cui:2010ud}
\bibinfo{author}{\bibfnamefont{Y.}~\bibnamefont{Cui}},
  \bibinfo{author}{\bibfnamefont{J.~D.} \bibnamefont{Mason}}, \bibnamefont{and}
  \bibinfo{author}{\bibfnamefont{L.}~\bibnamefont{Randall}},
  \bibinfo{journal}{JHEP} \textbf{\bibinfo{volume}{1011}}, \bibinfo{pages}{017}
  (\bibinfo{year}{2010}), \eprint{1006.0983}.

\bibitem[{\citenamefont{Ibarra and Wild}(2013)}]{Ibarra:2013qt}
\bibinfo{author}{\bibfnamefont{A.}~\bibnamefont{Ibarra}} \bibnamefont{and}
  \bibinfo{author}{\bibfnamefont{S.}~\bibnamefont{Wild}},
  \bibinfo{journal}{Phys.Rev.} \textbf{\bibinfo{volume}{D88}},
  \bibinfo{pages}{023014} (\bibinfo{year}{2013}), \eprint{1301.3820}.

\bibitem[{\citenamefont{Fornengo et~al.}(2013)\citenamefont{Fornengo, Maccione,
  and Vittino}}]{Fornengo:2013osa}
\bibinfo{author}{\bibfnamefont{N.}~\bibnamefont{Fornengo}},
  \bibinfo{author}{\bibfnamefont{L.}~\bibnamefont{Maccione}}, \bibnamefont{and}
  \bibinfo{author}{\bibfnamefont{A.}~\bibnamefont{Vittino}},
  \bibinfo{journal}{JCAP} \textbf{\bibinfo{volume}{1309}}, \bibinfo{pages}{031}
  (\bibinfo{year}{2013}), \eprint{1306.4171}.

\bibitem[{\citenamefont{Dal and Kachelriess}(2012)}]{Dal:2012my}
\bibinfo{author}{\bibfnamefont{L.}~\bibnamefont{Dal}} \bibnamefont{and}
  \bibinfo{author}{\bibfnamefont{M.}~\bibnamefont{Kachelriess}}
  (\bibinfo{year}{2012}), \eprint{1207.4560}.

\bibitem[{\citenamefont{von Doetinchem}(2012)}]{GAPS}
\bibinfo{author}{\bibfnamefont{P.}~\bibnamefont{von Doetinchem}},
  \emph{\bibinfo{title}{The general antiparticle spectrometer (gaps) - hunt for
  dark matter using low energy antideuterons}}, \bibinfo{howpublished}{Talk at
  the 9th International Conference Identification of Dark Matter, Chicago, USA}
  (\bibinfo{year}{2012}).

\bibitem[{\citenamefont{Fuke et~al.}(2005)\citenamefont{Fuke, Maeno, Abe,
  Haino, Makida et~al.}}]{Fuke:2005it}
\bibinfo{author}{\bibfnamefont{H.}~\bibnamefont{Fuke}},
  \bibinfo{author}{\bibfnamefont{T.}~\bibnamefont{Maeno}},
  \bibinfo{author}{\bibfnamefont{K.}~\bibnamefont{Abe}},
  \bibinfo{author}{\bibfnamefont{S.}~\bibnamefont{Haino}},
  \bibinfo{author}{\bibfnamefont{Y.}~\bibnamefont{Makida}},
  \bibnamefont{et~al.}, \bibinfo{journal}{Phys.Rev.Lett.}
  \textbf{\bibinfo{volume}{95}}, \bibinfo{pages}{081101}
  (\bibinfo{year}{2005}), \eprint{astro-ph/0504361}.

\bibitem[{\citenamefont{Choutko and Giovacchini}(2007)}]{amslimit}
\bibinfo{author}{\bibfnamefont{V.}~\bibnamefont{Choutko}} \bibnamefont{and}
  \bibinfo{author}{\bibfnamefont{F.}~\bibnamefont{Giovacchini}}
  (\bibinfo{year}{2007}).

\bibitem[{\citenamefont{Maccione}(2013)}]{Maccione:2012cu}
\bibinfo{author}{\bibfnamefont{L.}~\bibnamefont{Maccione}},
  \bibinfo{journal}{Phys.Rev.Lett.} \textbf{\bibinfo{volume}{110}},
  \bibinfo{pages}{081101} (\bibinfo{year}{2013}), \eprint{1211.6905}.

\end{thebibliography}
\bibliographystyle{apsrev}
\end{document}